\pdfoutput=1

\documentclass[usenatbib]{mnras}

\usepackage{amssymb}
\usepackage{amsmath}
\usepackage{graphicx}
\usepackage[autolanguage]{numprint}
\usepackage{subfigure}

\usepackage{enumitem}
\setlist{noitemsep}

\graphicspath{{./figs/}} 

\newcommand{\mscript}[1]{{\mbox{\scriptsize #1}}}
\newcommand{\mtiny}[1]{{\mbox{\tiny #1}}}

\title[Evidences against cuspy dark matter halos in large galaxies]{Evidences against cuspy dark matter halos in large galaxies}
\author[D.C. Rodrigues et al]{Davi C. Rodrigues,$^{1}$\thanks{E-mail:davi.rodrigues@cosmo-ufes.org}
 Antonino del Popolo$^{2,3}$\thanks{E-mail:adelpopolo@oact.inaf.it} Valerio Marra,$^{1}$\thanks{E-mail:valerio.marra@cosmo-ufes.org} \newauthor Paulo L. C. de Oliveira,$^{1}$\thanks{E-mail:paulo.oliveira@cosmo-ufes.org}\\ 
$^{1}$Departamento de F\'{\i}sica, Universidade Federal do Esp\'{\i}rito Santo, Av. F.Ferrari, 514, 29075-910, Vit\'oria, Brazil.\\
$^{2}$Dipartamento di Fisica e Astronomia, Universit\`a di Catania, Viale Andrea Doria 6, 95125 Catania, Italy.\\
$^{3}$INFN sezione di Catania, Via S. Sofia 64, 95123 Catania, Italy} 
\begin{document}

\date{}

\pagerange{\pageref{firstpage}--\pageref{lastpage}} 

\maketitle

\label{firstpage}

\begin{abstract} 
\noindent
We develop and apply new techniques in order to  {uncover} galaxy rotation curves (RC) systematics. Considering that an ideal dark matter (DM) profile should yield RCs that have no bias towards any particular radius, we find that the Burkert DM profile satisfies the test, while the Navarro-Frenk-While (NFW) profile has a  tendency of better fitting the region between one and two disc scale lengths than the inner disc scale length region. Our sample indicates that this behaviour happens to more than 75\% of the galaxies fitted with  {an} NFW halo.  Also, this tendency does not weaken by considering ``large'' galaxies, for instance those with $M_*\gtrsim 10^{10} M_\odot$.  Besides the tests on the homogeneity of the fits, we also use a sample of 62 galaxies of diverse types to perform tests on the quality of the overall fit of each galaxy, and to search for correlations with stellar mass, gas mass and the disc scale length. In particular, we find that only 13 galaxies are better fitted by the NFW halo; and that even for the galaxies with $M_* \gtrsim 10^{10} M_\odot$ the Burkert profile either fits as good as, or better than, the NFW profile. This result is relevant since different baryonic effects important for the smaller galaxies, like supernova feedback and dynamical friction from baryonic clumps, indicate that at such large stellar masses the NFW profile should be preferred over the Burkert profile. Hence, our results  either suggest a new baryonic effect or a change of the dark matter physics.
\end{abstract}


\section{Introduction}

According to the $\Lambda$CDM model, our Universe is mainly composed  {of} non-baryonic matter.  This model is very  successful in describing the early universe state, the formation and evolution of cosmic structures, and the abundance of  {the} matter-energy content of the Universe \citep[e.g.,][]{2011PhRvL.107b1301D, 2012PhRvL.109d1101H, Hinshaw:2012aka, Ade:2015xua}, for reviews see \citet{0521857937, Popolo:2012bna, DelPopolo:2013qba}. However, it has several issues on small scales \citep[e.g.,][]{Moore:1994yx, 1994ApJ...427L...1F, Gilmore:2007fy, 2009NJPh...11j5029P,deBlok:2009sp, Weinberg:2013aya, Pawlowski:2015qta, Onorbe:2015ija}, see  \citet{DelPopolo:2016emo} for a recent review.

The most persistent of the quoted problems is the so-called cusp-core problem \citep{Moore:1994yx, 1994ApJ...427L...1F} concerning the discrepancy between the cuspy profiles obtained in N-body simulations \citep[e.g., the NFW profile,][]{Navarro:1996bv, Navarro:1996gj, 2010MNRAS.402...21N} and the profiles inferred from the observed dwarf and low surface brightness (LSB) galaxies, which show cored profiles.

The N-body dark matter (DM)  cosmological simulations find inner DM density profiles of virialized halos sharply increasing towards  their centres (the cusp of the DM profiles). In the case of the NFW profile, the inner DM halo slope is $\rho \propto r^{-1}$, while in more recent simulations, or semi-analytical models, the inner slope decreases towards the centre, reaching  $\rho \propto r^{-0.8}$ at $\sim 100$ pc from the centre \citep{Stadel:2008pn, 2010MNRAS.402...21N, Taylor:2001bq, DelPopolo:2011zz}.\footnote{This profile is dubbed Einasto profile \citep[see][]{Gao:2007gh}.} To be more precise, we should recall that several authors, considering dark matter only simulations or semi-analytical results, found a correlation between the inner slope and the mass of the object considered \linebreak \citep[e.g.,][]{Ricotti:2002qu, Ricotti:2007jh, DelPopolo:2010rj, DelPopolo:2012de, 2014MNRAS.437..415D}, such that the inner slope could be either a bit above or below -1, depending on the system mass. 

Contrary to the simulation results, the profiles of real galaxies, and in particular that of the dwarf and low surface brightness (LSB) galaxies, are  usually better described by cored DM profiles (whose density is about constant at the centre), like the pseudo-isothermal or the Burkert profiles \citep{BlaisOuellette:2000ma,Borriello:2000rv, 2001AJ....122.2396D, deBlok:2001fe, Swaters:2002rx,  Gentile:2004tb, 2005ApJ...634L.145G,  2011AJ....142...24O}. Hence there is a conflict between the  {DM-only simulation} results and the DM profiles that are observationally favoured. This conflict is well known in the context of dwarf and LSB galaxies, which should have a cuspy profile (with slope $\alpha \lesssim -1$) according to DM-only simulations, while observational data favour cored profiles ($\alpha \sim 0$). The previous tendency is not valid for all galaxies. \citet{2008AJ....136.2648D} found that in the THINGS sample  larger galaxies ($M_B<-19$) are described equally well by cuspy (NFW) or cored profiles (pseudo-isothermal), while smaller ones ($M_B >- 19$) are better described by the pseudo-isothermal profile.\footnote{Also, there are some observational results that do not favour any universal profile \citep[e.g.,][]{Simon:2004sr}, which may be related to the environment and the different ways the galaxies formed \citep[][]{DelPopolo:2011cj}.} 

The situation with the most massive disc galaxies is not so clear, since the inner parts of these galaxies are usually  baryon dominated. Nonetheless, \citet{Spano:2007nt} using 36 disc galaxies of diverse types found that only 4 of the 36 galaxies yielded fits that were clearly better with the NFW profile, while 18 yielded fits that were clearly better with the pseudo-isothermal profile. They could not find a morphological trend on a possible preference between the NFW profile or the pseudo-isothermal one. Also, it is suggested that the comparison of $\chi^2$ values limited to the central regions could  clarify further their results. In the present work we aim to re-evaluate this issue with a larger sample and new techniques, which also make use of $\chi^2$ analyses limited to the central regions of galaxies.

Apart from considerations on  alternative approaches to DM, like self-interacting DM \citep{Spergel:1999mh, Rocha:2012jg}, change of the spectrum at small scales \citep{Bode:2000gq, Zentner:2003yd, 2013MNRAS.428..882M}, or modified gravity \citep[e.g.,][]{vandenBosch:1999dz, Zlosnik:2006zu, Rodrigues:2009vf, Famaey:2011kh, Rodrigues:2014xka, deAlmeida:2016chs, 2016MNRAS.462.3918S},  different proposals on how to solve this disagreement between simulations and observational data consider that baryonic effects may play a relevant role.  Within the latter picture, interactions of baryons with DM through gravity could ``heat'' the DM component giving rise to flatter inner profiles \citep{DelPopolo:2009df,2010Natur.463..203G,DelPopolo:2011cj, Pontzen:2011ty, Governato:2012fa,DelPopolo:2014yta}.

Independently of the precise dominant baryonic mechanism (which includes supernova feedback, and baryonic clumps with dynamical friction), the transformation from a cusp to a core would depend on the baryonic content of each galaxy, and would be more efficient on some galaxies than in others. All the cited approaches agree that, for the largest galaxies, one should not find a cored profile. In particular, and in accordance with \citet{2014MNRAS.437..415D} and \citet{DelPopolo:2015nda}, this transformation of the central cusp into a core correlates with the galaxy stellar mass ($M_*$), such that galaxies with $ M_* \sim 10^{8.5} M_\odot$ have DM profiles that are  close to a cored profile, while the largest galaxies (i.e., those with stellar masses about or above $10^{9.5} M_\odot$) are better described by a cuspy DM halo with  central slope about -1, or even lower. This behaviour would be a consequence of the fact that the ratio between stellar mass to halo mass is higher in the largest galaxies and that the central regions of these galaxies are dominated by baryons. The large amount of baryonic matter deepens the Newtonian potential  more than what happens in dwarf galaxies and, consequently, the outflows generated by the supernovae, or by the dynamical friction from baryonic clumps, are not able to drag away enough DM and flatten the DM profile.

This work aims to develop new approaches to evaluate galaxy fits, which will be used to re-evaluate the cusp-core issue. Several galaxies of diverse types are considered here, but focus is given to the largest galaxies, since the approaches that indicate that baryonic physics can transform the cusp into a core usually also state that this transformation happens for ``small'' galaxies, while the same baryonic mechanism cannot remove the cusp for galaxies with $M_* \gtrsim 10^{10} M_\odot$ \citep[e.g.,][]{DelPopolo:2009df, 2010Natur.463..203G, deSouza:2011pb, 2011MNRAS.418.2527I, Governato:2012fa, 2014MNRAS.437..415D, DelPopolo:2014eta, 2016MNRAS.456.3542T, DelPopolo:2015nda}. Actually, the baryonic physics in such large galaxies is expected to lead to DM profiles whose central slope becomes more negative than -1. Here, we look for possible systematics that could favor, or disfavour, the presence of  DM cusps in large galaxies. This is an important issue since it could indicate poor understanding of the baryonic physics, or issues with the standard DM model.

The paper is organized  as follows: in the next section we present the technique for evaluating the homogeneity of  galaxy RC fits. This technique is based on approaches developed in \citet{deBlok:2002tg, Rodrigues:2014xka}. Sections \ref{sec:profiles} and \ref{sec:samples} explain, respectively, the DM profiles and the galaxy samples that are here used. In Section \ref{sec:results} we present our main results, which include the application of the technique introduced in Sec.~\ref{sec:xizeta}. Section \ref{sec:conclusions} is devoted to our conclusions and discussions, while the Appendices \ref{app:Fdist} and \ref{app:xizeta} clarify assumptions and results from the Sections \ref{sec:xizeta} and \ref{sec:results}, respectively.

\section{Testing the uniformity of fits and data: the quantities $\xi$, $\zeta$ and $\Delta \xi$} \label{sec:xizeta}
 
\citet{Rodrigues:2014xka} generalized the approach proposed by \cite{deBlok:2002tg}, which will be further developed here. Hence, first we will briefly review the quantities $\chi^2_\mscript{inn}$ and $\chi^2_\mscript{out}$ which were introduced in the latter reference. After the minimum value of $\chi^2$ is found ($\chi^2_\mscript{min}$), one considers two quantities, the inner and the outer values of $\chi^2$, and these are denoted by $\chi^2_\mscript{inn}$, $\chi^2_\mscript{out}$. Let $R_\mscript{max}$ be the largest radius of the observational RC. The value of $\chi^2_\mscript{inn}$  is found from $\chi^2_\mscript{min}$ but considering only the observational data from the galaxy centre to $R_\mscript{max}/2$, while $\chi^2_\mscript{out}$ considers the radii from $R_\mscript{max}/2$ to $R_\mscript{max}$. \citet{deBlok:2002tg}  found that the pseudo-isothermal halo leads to better fits than the NFW halo for most of the cases of their sample (this step is just a straightforward comparison of $\chi^2_\mscript{min}$). And, by using the quantities $\chi^2_\mscript{inn}$ and $\chi^2_\mscript{out}$, they could point out that the main problem with the NFW fits were clearly in the inner region.

In order to further explore the inner radii dynamics, \citet{Rodrigues:2014xka} consider three reference radii, and these are not based on $R_\mscript{max}$, which is not directly related to the inner dynamics, but to the disc scale length ($h$). These reference radii lead  to the definition of the quantities: $\chi^2_{h/2}$, $\chi^2_{h}$ and $\chi^2_{2 h}$. These three quantities are given by $\chi^2_\mscript{min}$ but considering only radii either up to $h/2$,  $h$, or $2h$, respectively. 

To introduce a proper notation, we write the quantity $\chi^2$ as
\begin{equation}
	    \chi^2(p_j) \equiv \sum_{i=1}^{N} \left( \frac{V_\mtiny{model}(R_i, p_j) - V_i}{\sigma_i} \right)^2,
\end{equation}
where $V_i$ and $\sigma_i$ are the observed RC velocity and its corresponding error at the radius $R_i$, $N$ is the number of observational data points of the RC (i.e., $R_N = R_\mscript{max}$), and $V_\mtiny{model}(R_i,p_j)$ is the theoretical circular velocity at the radius $R_i$  with the model parameters $p_j$. Using this notation, the quantity $\chi^2_h$, for example, can be written as,
\begin{equation}\label{eq:chi2h}
    \chi^2_h \equiv \sum_{i=1}^{N(h)} \left( \frac{V_\mtiny{model}(R_i, \bar p_j) - V_i}{\sigma_i} \right)^2,
\end{equation}
where $\bar p_j$ are the parameters values that minimize $\chi^2$ (i.e., $\chi^2(\bar p_i) = \chi^2_\mscript{min}$). The number $N(h)$ is the largest natural number such that $R_{N(h)} \leq h$. Equivalently, $N(h)$ is the number of RC data points at $0\leq R \leq h$. Analogous definitions are used for $\chi^2_{h/2}$ and $\chi^2_{2h}$.

In order to evaluate the uniformity of the fits along the galaxy radius, we introduce the  quantity
\begin{equation} \label{defxi}
	\xi(m,n)\equiv\frac{\chi^2_{m  h}}{\chi^2_{n  h}} ,
\end{equation} 
in a similar way as done by \citet{Rodrigues:2014xka}, where $m$ and $n$ are real dimensionless numbers. The quantity $\chi^2_{m h}$ is defined as in eq.~(\ref{eq:chi2h}), but with $N(h)$ replaced by $N(m h)$. 

 For an ideal set of galaxies whose observational data is homogeneously distributed along their radius, and for an ideal model with no bias towards any radius, on average one should find 
\begin{equation} \label{eq:xiaverage}
	\langle \xi (m,n)\rangle \approx \frac mn,
\end{equation}
where $\langle \; \; \rangle$ stands for a certain average, which will be detailed afterwards.

It is important to select a suitable average for the problem. Since the quantity $\xi(m,n)$, when applied to real galaxies, sometimes changes by more than one order of magnitude from one galaxy to another, the arithmetic mean becomes easily dominated by a few outliers. Instead of developing an algorithm to define and eliminate the outliers, we simply use -- as in \citet{Rodrigues:2014xka} -- the median as a robust estimator for the average. Doing so we consider the complete data, without discarding any ``outlier''.
Moreover, Appendix~\ref{app:Fdist} describes in detail a particular case, in contact with the procedures here used, in which eq.~\eqref{eq:xiaverage} holds exactly if the median is employed. One of the conditions for the latter result is that $m = 2n$, and this relation will be used in Sec. \ref{sec:results}. Unless otherwise stated, all the averages in this work are performed using the median.

Apart from notation changes, the framework presented above for testing the homogeneity of galaxy fits  was proposed in \citet{Rodrigues:2014xka}. In particular it was found that the fits derived from the NFW halo had a tendency of better fitting the region $2h> R > h$ than the region $R<h$. It should be emphasised that this test is not a comparison between two different models, it is a consistent test. It compares the fit yielded by certain model at certain radius to the fit of the same model at a different radius.

Even using the median as the average and a perfect model with no bias towards any galaxy radius, eq.~(\ref{eq:xiaverage}) may fail to hold as observational data are not, in general, uniformly distributed and with constant error.
In order to quantify the non-uniformity of RC data, we extend the approach of \cite{Rodrigues:2014xka} and introduce here the quantity $\zeta(m,n)$. This quantity is supposed to extend eq.~(\ref{eq:xiaverage}) to the case of real galaxies. That is, it should be such that for a model without a significative bias towards any particular radius, 
\begin{equation} \label{eq:xizeta}
	\langle \xi (m,n)\rangle \approx \langle \zeta(m,n) \rangle.
\end{equation}

 If a given RC has constant error bars, then $\zeta$ should only depend on the number of data points with radius $R \leq m h$ (i.e., $N(m h)$) and $R \leq n h$ (i.e., $N(n h)$). Hence, in this context a natural definition for $\zeta$ would be $\zeta(m,n) = N(m h)/N(n h)$. If the data points are evenly spaced, then $N(m h)/N(n h) = m/n$ and one recovers  eq.~(\ref{eq:xiaverage}).

Non-constant error bars are another source of non-uniformity along the galaxy radius. Since $\chi^2$ depends on the sum of the inverse of $\sigma_i^2$,  the following quantity will be useful
\begin{equation}
	\Sigma (m h) \equiv \sum_{i=1}^{N(m h)} \frac{1}{\sigma_i^2}.	
\end{equation}
For a RC whose error bars have  the same magnitude, one finds that $\Sigma (m h) / \Sigma (n h) = N(m h) /N(n h)$, thus finding the previous case. This quantity already depends on both the magnitude of the error bars and the number of data points, it is also directly related to the definition of $\chi^2$ and generalizes previous considerations. Hence, considering eq.~(\ref{eq:xizeta}), we  define $\zeta$ as,
\begin{equation}
	\zeta(m,n) \equiv   \frac{\Sigma(m h)}{\Sigma(n h)}.
\end{equation}

For ideal models without bias towards any radii, one should also expect that the dispersions of $\xi$ and $\zeta$ should  be similar. To quantify the dispersion we introduce the quantities  $\sigma_{50\%}^{\pm}$ and $\sigma_{25\%}^{\pm}$. The first one, applied to some set of numerical data $\{X\}$ whose median value is $\langle X \rangle$,  is defined as
\begin{eqnarray}
		\sigma^+_{50\%}(X) &=& \langle \{X \; | \;  X \geq \langle X\rangle \}\rangle,	 \\[.1in]
		\sigma^-_{50\%}(X) &=& \langle \{X \; | \;  X \leq \langle X\rangle \}\rangle.		
\end{eqnarray}
In other words, $\sigma^+_{50\%}(X)$ is the median of the subsample of $\{X\}$ composed by the $X$ values that are larger or equal to $\langle X \rangle$.

Since, from the definition of the median, about half of the members of a set $\{ X \}$ will be above its median, and half below it, one sees that about half of set $\{ X \}$ will be in the range  $\sigma^-_{50\%}(X) \leq X \leq \sigma^+_{50\%}(X)$.

The quantity $\sigma_{25\%}^{\pm}$ subdivides further the set $\{X\}$. It  fixes a range that includes the median and in which about $25\%$ of the sample elements are present, namely,
\begin{eqnarray}
		\sigma^+_{25\%}(X) &=& \langle \{X \; | \; \sigma^+_{50\%}(X) \geq X \geq \langle X\rangle \}\rangle,	  \\[.1in]
		\sigma^-_{25\%}(X) &=& \langle \{X \; | \; \sigma^-_{50\%}(X)  \leq X \leq \langle X\rangle \}\rangle.		
\end{eqnarray}

If the sample is sufficiently representative, the  above quantities can be probabilistically interpreted in the following ways: i) the probability for a random galaxy to lie inside the region between $\sigma^{-}_{k\%}$ and $\sigma^{+}_{k\%}$ is $k\%$; ii) The probability of finding a member of the sample $X$  that is above the corresponding $\sigma^{+}_{50\%}(X)$ is $25\%$.; iii) and thus the probability of finding an element $X$ that is below $\sigma^{+}_{50\%}(X)$ is 75\%.

At last, to further clarify and simplify the analysis, we also introduce the quantity
\begin{equation}
	\Delta \xi(m,n) \equiv \xi(m,n) - \zeta(m,n),
\end{equation}
whose average, for an ideal model, should yield,
\begin{equation} \label{eq:deltaxi}
	\langle \Delta \xi(m,n) \rangle \approx 0.
\end{equation}
For an arbitrary sample of data neither eq.~(\ref{eq:deltaxi}) implies eq.~(\ref{eq:xizeta}) nor the contrary, but both are expected to hold if the sample is sufficiently large.

\section{Dark matter profiles} \label{sec:profiles}

As said in the Introduction, there are different approaches that try to solve the cusp-core problem by flattening the central DM profile of dwarf and LSB galaxies. These mechanisms are not expected to alter the cuspy DM profile of the largest disc galaxies. It is the purpose of this work to use traditional tests in order to compare different DM halo proposals and, also, to apply the methodology presented in the previous section. The main motivation is to look for new evidences against or in favor of the existence of cusps in the DM profiles of the large galaxies.

We consider here two DM profiles that only differ on their behaviour close to the galactic centre, the Navarro-Frenk-White (NFW) profile \citep{1996ApJ...462..563N,Navarro:1996gj, 2010MNRAS.402...21N},
\begin{equation} \label{eq:nfwDef}
	\rho_\mtiny{NFW}(r) = \frac{\rho_s}{\frac r{r_s}{\left( 1 + \frac r{r_s}\right )^2} },
\end{equation}
which depends on two parameters, $r_s$ and $\rho_s$, and the {Burkert} profile \citep{1995ApJ...447L..25B},
\begin{equation}
		\rho_\mtiny{B}(r) = \frac{\rho_c}{\left(1+\frac r{r_c} \right)\left( 1 + \frac {r^2}{r_c^2}\right )},
\end{equation}
which also depends on two parameters, $r_c$ (the core radius) and $\rho_c$. 

The Burkert profile is a cored profile that is well known for its phenomenological success\footnote{Another well known cored profile is the pseudo-isothermal profile \citep{1991MNRAS.249..523B}, nonetheless this profile differs from the NFW one at both small and large radius.} \citep[e.g.,][]{2005ApJ...634L.145G, Gentile:2004tb, Gentile:2006hv, Salucci:2007tm}, and it is such that for small radii it has a constant density, and for large radii it decays just like the NFW profile, that is, with $r^{-3}$. 

According to \citet{2014MNRAS.437..415D, DelPopolo:2015nda, 2016MNRAS.456.3542T}, galaxies with stellar to DM mass ratio $ {M_*}/{M_{\mtiny{DM}}} \gtrsim 10^{-1.7}$ (or, equivalently, using the \citet{Moster:2012fv} relation, $M_* \gtrsim 10^{9.5} M_\odot$) have inner slope $\alpha \leq -0.6 $; while for ${M_*}/{M_{\mtiny{DM}}} \gtrsim 10^{-1.5}$ (or $M_* \gtrsim 10^{10.0} M_\odot $) the inner slope is $\alpha \leq -1.0$. Since the NFW and Burkert profiles' inner slopes are respectively $-1$ and 0, while their outer slopes are both $-3$, it is expected that for galaxies with stellar mass about or above $10^{9.5} M_\odot$ one should find that the NFW halo leads to better fits than the Burkert halo.

Although the NFW profile, as defined in eq.~(\ref{eq:nfwDef}), depends on two parameters, several simulations assert that there is a correlation between these parameters (the correlation is usually parameterised with the concentration $c$ and $M_{200}$) \citep[e.g.,][]{Maccio:2008pcd}. Some works use this correlation to write one parameter as a function of the other \cite[e.g.,][]{2005ApJ...634L.145G}, thus arriving on a one-parameter NFW halo. Since there is significative dispersion on such correlations (including differences between different works), here both the parameters are fitted without constraints, which implies that the NFW results used in this work are the best possible fits with this profile. 

The present work uses the (two-parameter) NFW fits from \cite{Rodrigues:2014xka}, where further details (including the correlation between $c$ and $M_{200}$ from the observational data) can be found. For the Burkert fits, all the fits are done here and  constitute part of the results of this work. Some of the galaxies that we consider here were previously fitted with the Burkert profile; nonetheless, to assure uniformity on all the conventions, we fitted all the galaxies with the Burkert profile using precisely the same procedures that we used for the NFW fits.

\section{Samples} \label{sec:samples}
  
\begin{table}
\caption{The five data samples considered in \citet{Rodrigues:2014xka}. Unless otherwise stated, we refer to different baryonic models as different galaxies. There are different ways of removing these repetitions, but neither has lead to significant systematic effects that could change any of the conclusions.}
\label{tab:samples}
\begin{tabular}{@{}lcc}
\hline
Sample & Fitted galaxies &   Main Refs. \cr
\hline
A & 18  &\cite{2008AJ....136.2648D}\cr
B & 05  & \cite{Gentile:2004tb} \cr
C & 13  &\cite{deBlok:2002tg}   \cr
D & 08  & \cite{2001AJ....122.2396D}\cr
E & 18 & \cite{Swaters:2011yq}\cr
\hline 
Total & 62 & \hspace*{-0.50in} {\footnotesize different baryonic models for galaxies} \cr
	& 53 & \hspace*{-1.53in}{\footnotesize different galaxies}\cr
\hline 
\end{tabular}
\end{table}

\begin{table} \begin{center}
\caption{The samples and the corresponding number of galaxies that have one or more RC data points at $R<h/2$, $R<h$, and $R<2h$. These are respectively denoted by $N_G(h/2)$, $N_G(h)$, and $N_G(2h)$. The samples ${\cal S}_\mscript{*1}$ and ${\cal S}_\mscript{*2}$ are the only ones whose number of members is model dependent, hence their $N_G$ values are stated in the form NFW/Burkert. Below, $M_\mscript{*}$ is the total stellar mass (bulge and disc), $h$ is the disc scale length and $M_\mscript{gas}$ is the gas mass (it includes hydrogen and helium contributions, and follows  the prescriptions from the corresponding original reference).}
\label{tab:samplenumbers}
\begin{tabular}{@{}llrrr}
\hline
Sample & Sample criterion & $N_G(h/2)$ &  $N_G(h)$ & $N_G(2h)$ \cr
\hline
A &- & 14&17&18\\[.1cm]
B & - &5&5&5\\[.1cm]
C &-& 13&13&13\\[.1cm]
D &-& 7&8&8\\[.1cm]
E &-& 12&18&18 \\[.1cm]
${\cal S}$ &All galaxies& 51 & 61 & 62 \\[.1cm]
${\cal S}_\mscript{*1}$&$M_\mscript{*}> 10^9 M_\odot$ &29/32&34/39&35/40\\[.1cm]
${\cal S}_\mscript{*2}$&$M_\mscript{*}>  10^{10} M_\odot$&13/12&16/16&17/17\\[.1cm]
${\cal S}_\mscript{g1}$&$M_\mscript{gas}> 10^9 M_\odot$ &39&48&49\\[.1cm]
${\cal S}_\mscript{g2}$&$M_\mscript{gas}> 5 \times 10^9 M_\odot$&14&17&18\\[.1cm]
${\cal S}_\mscript{h1}$&$h > 1.5$ kpc&42&47&48 \\[.1cm]
${\cal S}_\mscript{h2}$&$h>3.0$ kpc&17&19&19 \\[.1cm]
\hline 
\end{tabular} \end{center}
\end{table}

\nprounddigits{2} 
\begin{table*} 
\caption{Values of $N$ and $\Sigma$ for all the galaxies. These figures are directly derived from the observational data, and hence are model independent.}.
\label{tab:xidata}
\begin{tabular}{@{}llrrrrn{1}{2}n{1}{2}n{1}{2}n{2}{2}}
\hline
S & Galaxy & \multicolumn{1}{c}{$N(h/2)$} &  \multicolumn{1}{c}{$N(h)$} & \multicolumn{1}{c}{$N(2h)$} & \multicolumn{1}{c}{$N(R_\mscript{max})$}&  \multicolumn{1}{c}{$\Sigma(h/2)$} &  \multicolumn{1}{c}{$\Sigma(h)$} & \multicolumn{1}{c}{$\Sigma(2h)$} &\multicolumn{1}{c}{$\Sigma(R_\mscript{max})$ } \cr
\hline
A&DDO 154&3&7&14&60&1.85&2.56&4.19&17.36\\ 
A&NGC 2403 1D&14&28&57&287&0.53&1.10&2.17&16.25\\ 
A&NGC 2403 2D&14&28&57&287&0.53&1.10&2.17&16.25\\ 
A&NGC 2841&0&2&14&140&0.00&0.02&0.49&2.71\\ 
A&NGC 2903&0&0&6&86&0.00&0.00&0.10&2.91\\ 
A&NGC 2976&13&27&42&41&2.28&3.53&4.63&4.55\\ 
A&NGC 3031&0&5&31&116&0.00&0.20&2.23&4.16\\ 
A&NGC 3198 1D&3&7&15&93&0.08&0.18&0.48&5.31\\ 
A&NGC 3198 2D&3&7&15&93&0.08&0.18&0.48&5.31\\ 
A&NGC 3521&20&41&83&99&0.61&0.92&1.00&1.06\\ 
A&NGC 3621&6&12&24&122&0.39&0.74&1.88&8.11\\ 
A&NGC 4736&5&14&31&81&0.16&0.43&1.01&2.76\\ 
A&NGC 5055&4&9&19&198&0.05&0.27&0.89&4.66\\ 
A&NGC 6946&2&19&54&206&0.10&0.44&1.64&5.86\\ 
A&NGC 7331&0&12&38&104&0.00&0.17&0.44&1.41\\ 
A&NGC 7793&7&14&28&67&1.27&2.65&3.96&6.22\\ 
A&NGC 7793 R&7&14&28&41&1.27&2.65&3.96&4.87\\ 
A&NGC 925&8&18&38&95&0.19&0.81&1.52&3.16\\ 
B&ESO 116-G12&1&3&5&14&0.08&0.27&0.48&1.82\\ 
B&ESO 287-G13&3&6&12&25&0.12&0.34&0.61&2.11\\ 
B&ESO 79-G14&3&5&9&14&0.03&0.10&0.21&0.94\\ 
B&NGC 1090&3&3&6&23&0.08&0.08&0.21&2.14\\ 
B&NGC 7339&2&4&9&14&0.09&0.17&0.86&1.40\\ 
C&F 563-1&2&3&3&7&0.01&0.02&0.02&0.08\\ 
C&UGC 1230&2&3&6&10&0.02&0.03&0.05&0.08\\ 
C&UGC 3060&7&14&29&58&1.75&3.50&7.25&19.43\\ 
C&UGC 3371&3&7&12&17&0.03&0.06&0.09&0.24\\ 
C&UGC 3851&8&15&18&27&0.31&0.60&0.64&1.02\\ 
C&UGC 4173&3&6&10&12&0.06&0.12&0.23&0.28\\ 
C&UGC 4325&3&5&11&15&0.04&0.09&0.23&0.26\\ 
C&UGC 5005&1&3&6&10&0.02&0.02&0.07&0.10\\ 
C&UGC 5721&1&3&5&22&0.05&0.12&0.20&0.97\\ 
C&UGC 7524&11&23&41&54&0.30&0.57&1.05&1.47\\ 
C&UGC 7603&2&4&7&19&0.12&0.24&0.42&1.14\\ 
C&UGC 8837&3&3&8&7&0.18&0.18&0.43&0.39\\ 
C&UGC 9211&1&2&4&10&0.02&0.03&0.05&0.19\\ 
D&F 563-1&0&1&2&9&0.00&0.00&0.00&0.05\\ 
D&F 568-3&3&5&8&10&0.07&0.10&0.16&0.19\\ 
D&F 571-8&3&4&9&12&0.15&0.21&0.39&0.57\\ 
D&F 579-V1&3&6&11&13&0.03&0.06&0.10&0.12\\ 
D&F 583-1&2&5&9&16&0.03&0.08&0.22&0.36\\ 
D&F 583-4&3&3&6&8&0.12&0.12&0.25&0.33\\ 
D&UGC 5750&2&4&7&10&0.06&0.08&0.23&0.25\\ 
D&UGC 6614&3&3&9&14&0.03&0.03&0.10&0.13\\ 
E&UGC 11707&1&3&7&12&0.01&0.05&0.51&1.09\\ 
E&UGC 12060&0&1&3&8&0.00&0.05&0.15&0.41\\ 
E&UGC 12632&2&5&10&16&0.06&0.38&0.81&1.54\\ 
E&UGC 12732&1&2&4&15&0.09&0.14&0.24&1.17\\ 
E&UGC 3371&1&3&6&10&0.09&0.36&0.75&1.20\\ 
E&UGC 4325&1&2&4&7&0.11&0.21&0.43&0.75\\ 
E&UGC 4499&0&1&3&8&0.00&0.06&0.28&0.94\\ 
E&UGC 5414&1&2&4&5&0.16&0.33&0.66&0.83\\ 
E&UGC 6446&1&2&4&10&0.14&0.30&0.60&1.49\\ 
E&UGC 731&1&2&5&11&0.18&0.35&0.67&1.53\\ 
E&UGC 7323&1&3&7&9&0.07&0.20&0.47&0.60\\ 
E&UGC 7399&0&1&2&17&0.00&0.13&0.22&2.22\\ 
E&UGC 7524&5&10&20&30&0.44&1.05&1.71&2.68\\ 
E&UGC 7559&1&2&5&8&0.10&0.19&0.48&0.76\\ 
E&UGC 7577&1&3&6&8&0.10&0.30&0.60&0.79\\ 
E&UGC 7603&0&1&3&11&0.00&0.12&0.36&1.32\\ 
E&UGC 8490&0&1&3&29&0.00&0.07&0.22&2.13\\ 
E&UGC 9211&0&1&2&8&0.00&0.06&0.12&0.48\\
\hline 
\end{tabular} 
\end{table*}

Table \ref{tab:samples} lists the five galaxy data samples that were studied in \citet{Rodrigues:2014xka} and their corresponding main references. We refer to the latter reference for a table with the galaxy global parameters  (including luminosity, distance and disc scale length).

The complete sample contains precisely 53 different galaxies and 62 different baryonic models for galaxies. For instance, in the Sample A two different models for the galaxy NGC 3198 can be found (one with a bulge and the other without), and the galaxy F 563-1 can be found in both the samples C and D. We do not try to advocate which of these baryonic models is to be preferred, and we use all the 62 galaxy data. There are different strategies to eliminate duplicate galaxies, some of them were explicitly tested and neither has lead to significant systematic effects that could change our conclusions (which is in part expected since the median is a robust type of average).

The Total Sample (${\cal S}$) is composed by the union of the samples A, B, C, D and E. The subsamples of ${\cal S}$ composed by all the galaxies with stellar mass (bulge plus disc stellar masses) above $10^9 M_\odot$ or $10^{10} M_\odot$ constitute respectively the samples named ${\cal S}_\mscript{*1}$ and ${\cal S}_\mscript{*2}$. The subsamples of ${\cal S}$ composed by all the galaxies with gas mass above $10^9 M_\odot$ or $5 \times 10^9 M_\odot$ constitute respectively the samples named ${\cal S}_\mscript{g1}$ and ${\cal S}_\mscript{g2}$. The subsamples of ${\cal S}$ composed by all the galaxies with disc scale length above 1.5 kpc or 3.0 kpc  constitute respectively the samples named ${\cal S}_\mscript{h1}$ and ${\cal S}_\mscript{h2}$. Further details on these samples  are shown in Table \ref{tab:samplenumbers}.

Table \ref{tab:xidata} shows the values of $N(n h)$ and $\Sigma(n h)$ for each of the galaxies.

The galaxy samples used here are well known and used as part of several different tests \citep[e.g., for some recent examples, see][]{Rodrigues:2014xka, Saburova:2014opa,  Oman:2015xda, Sanchez-Salcedo:2016bof, 2016MNRAS.460.3610O, deAlmeida:2016chs, 2016MNRAS.456.3542T, 2017MNRAS.465.4703K}. Sample A \citep{2008AJ....136.2648D} is the original THINGS sample that includes large and massive spirals, its 21 cm data was presented in \citet{2008AJ....136.2563W} and  it uses different infrared bands for modeling the stellar part, including 3.6 $\mu$m from Spitzer. Sample B \citep{Gentile:2004tb} is a small sample of galaxies with dynamical masses from $\sim 10^{10} M_\odot$ to $\sim 10^{11} M_\odot$ that was carefully modeled to study the core-cusp issue with combined HI and H$\alpha$ data, it uses the infrared I-band to model the stellar part. Samples C \citep{deBlok:2002tg} and D \citep{2001AJ....122.2396D} are classic references on LSB galaxies and on the cusp-core problem. The Sample E \citep{Swaters:2011yq} is a sample with dwarf and LSB galaxies whose RC were derived from both HI and H$\alpha$ observations. This sample is a selection of the 18 highest quality RC data from the 62 galaxies of \citet{2009A&A...493..871S}.

Recently, a new large catalogue on 175 disc galaxies was compiled, the SPARC sample \citep{2016AJ....152..157L}. There is a significant intersection between  the galaxies of that catalogue and the galaxies that are used in this work, namely, there are 10 galaxies from the SPARC sample that also appear in Sample A, 4 galaxies from Sample B, 4 from Sample C, 3 from Sample D, and 8 from Sample E.  On the other hand, there is also a significant amount of galaxies that appear in the latter five samples and do not appear in SPARC. The differences between the galaxy data and baryonic models that appear in more then one sample is commonly small, and some features are identical (e.g., most of the RC data are identical). Among the differences, perhaps unexpectedly, some galaxies that are part of the THINGS sample appear in SPARC, but with RC data from older references. The reason for this choice is detailed in \citet{2016AJ....152..157L}. The most relevant difference comes from the indication that all the galaxies may share a fixed stellar mass-to-light ratio ($\Upsilon_*$) at the 3.6 $\mu$m wave length. In this work we do not consider the latter  as a starting point, we follow one of the standard approaches to the subject, and find $\Upsilon_*$ for each galaxy from a best fit. In the Appendix \ref{app:expectedY} this issue is discussed in detail, and our results on $\Upsilon_*$ are compared to the expectations posed by  \citet{2016AJ....152..157L}.

\section{Results} \label{sec:results}

\nprounddigits{2} 
\begin{table*}
\caption{Best-fit results for our sample of 62 galaxies using the Burkert dark matter profile. This table extends Table 4 of \citet{Rodrigues:2014xka}. Col.~(1): sample. Col.~(3): minimum $\chi^2$. Col.~(4): reduced $\chi^2$. Cols.~(5)-(7): see Sec.~\ref{sec:xizeta} for their definitions. Cols.~(8)-(9): disc and bulge stellar mass-to-light ratios in the appropriate band for each sample. Col.~(10): $r_c$ (kpc). An ``$\infty$'' means that the resulting $r_c$ from the fits is larger than 1 Mpc. Col.~(11): $\rho_c$ ($M_\odot$/kpc$^3$). }
\begin{tabular}{cln{3}{2}n{1}{2}n{3}{2}n{3}{2}n{3}{2}n{1}{2}n{1}{2}n{2}{2}n{1}{2}}
 \hline
{S} & \multicolumn{1}{c}{Galaxy}   & \multicolumn{1}{c}{$\chi^2_\mscript{min}$}   & \multicolumn{1}{c}{$\chi_\mscript{red}^2$} & \multicolumn{1}{c}{$\chi^2_{2h}$} & \multicolumn{1}{c}{$\chi^2_{h}$} & \multicolumn{1}{c}{$\chi^2_{h/2}$} & \multicolumn{1}{c}{$\Upsilon_{*D}$} & \multicolumn{1}{c}{$\Upsilon_{*B}$} & \multicolumn{1}{c}{$r_c$} & \multicolumn{1}{c}{$\rho_c$}  \\

{(1)} &   \multicolumn{1}{c}{(2)}   &  \multicolumn{1}{c}{(3)}    &   \multicolumn{1}{c}{(4)}  &  \multicolumn{1}{c}{ (5)}   &  \multicolumn{1}{c}{(6)}   & \multicolumn{1}{c}{(7)} & \multicolumn{1}{c}{(8)} & \multicolumn{1}{c}{(9)} & \multicolumn{1}{c}{(10)}& \multicolumn{1}{c}{(11)}   \\
\hline
A&DDO 154&15.578&0.269&2.961&2.520&1.798&3.45& \multicolumn{1}{c}{---} &4.31&1.033E7\\
A&NGC 2403 1D&163.768&0.575&35.216&13.368&9.103&0.68& \multicolumn{1}{c}{---} &7.24&2.554E7\\ 
A&NGC 2403 2D&162.182&0.571&26.518&12.956&9.586&0.59&1.07&6.82&2.865E7\\ 
A&NGC 2841&33.230&0.243&6.380&2.089&0.000&0.96&1.58&13.91&2.525E7\\ 
A&NGC 2903&20.474&0.247&0.176&0.000&0.000&1.63&2.45&6.80&4.782E7\\ 
A&NGC 2976&17.176&0.440&17.176&11.612&9.299&0.25& \multicolumn{1}{c}{---} &2.38&1.103E8\\ 
A&NGC 3031&369.135&3.267&113.266&8.464&0.000&0.92&0.26&5.03&2.834E7\\ 
A&NGC 3198 1D&34.689&0.381&2.925&0.595&0.362&0.12& \multicolumn{1}{c}{---} &4.34&9.736E7\\ 
A&NGC 3198 2D&34.272&0.381&2.803&0.299&0.154&0.07&0.08&4.21&1.059E8\\ 
A&NGC 3521&130.603&1.346&127.220&114.225&113.699&0.00& \multicolumn{1}{c}{---} &2.14&1.005E9\\ 
A&NGC 3621&86.585&0.722&23.460&11.923&8.546&0.61& \multicolumn{1}{c}{---} &12.04&1.049E7\\ 
A&NGC 4736&111.520&1.430&61.805&19.909&3.187&0.41&0.33&0.84&9.831E8\\ 
A&NGC 5055&142.333&0.730&71.637&15.148&4.410&0.50&0.38&13.71&1.041E7\\ 
A&NGC 6946&193.554&0.953&85.301&23.860&12.695&0.61&0.68&16.91&1.023E7\\ 
A&NGC 7331&27.986&0.277&8.460&4.934&0.000&0.56&0.68&18.20&8.745E6\\ 
A&NGC 7793&38.334&1.009&33.969&12.852&10.527&0.45& \multicolumn{1}{c}{---} &\multicolumn{1}{c}{$\infty$}&2.497E7\\ 
A&NGC 7793 R&39.524&1.040&34.684&17.364&15.885&0.44& \multicolumn{1}{c}{---} &\multicolumn{1}{c}{$\infty$}&2.539E7\\ 
A&NGC 925&61.217&0.658&28.655&22.984&19.593&0.15& \multicolumn{1}{c}{---} &8.46&1.605E7\\ 
B&ESO 116-G12&9.356&0.780&4.083&3.729&2.569&0.43& \multicolumn{1}{c}{---} &4.39&4.653E7\\ 
B&ESO 287-G13&28.635&1.245&22.337&17.347&15.977&1.96& \multicolumn{1}{c}{---} &27.59&4.539E6\\ 
B&ESO 79-G14&7.404&0.617&5.042&4.260&1.446&0.75& \multicolumn{1}{c}{---} &7.96&3.451E7\\ 
B&NGC 1090&13.337&0.635&6.325&0.408&0.408&1.47& \multicolumn{1}{c}{---} &8.97&1.846E7\\ 
B&NGC 7339&13.107&1.092&6.350&3.896&0.318&1.82& \multicolumn{1}{c}{---} &5.54&5.417E7\\ 
C&F563-1&2.360&0.472&2.280&2.280&0.839&8.48& \multicolumn{1}{c}{---} &19.59&3.532E6\\ 
C&UGC 1230&2.114&0.264&1.797&0.944&0.801&0.00& \multicolumn{1}{c}{---} &3.53&7.770E7\\ 
C&UGC 3060&119.633&2.136&76.217&42.786&14.662&4.25& \multicolumn{1}{c}{---}&13.47&6.658E6\\ 
C&UGC 3371&0.233&0.016&0.134&0.122&0.112&0.00& \multicolumn{1}{c}{---} &5.55&2.082E7\\ 
C&UGC 3851&25.678&1.027&24.665&24.504&9.529&0.00& \multicolumn{1}{c}{---} &1.06&1.732E8\\ 
C&UGC 4173&0.433&0.043&0.396&0.313&0.183&0.00& \multicolumn{1}{c}{---} &4.12&8.883E6\\ 
C&UGC 4325&0.095&0.007&0.075&0.027&0.019&0.46& \multicolumn{1}{c}{---} &4.32&1.036E8\\ 
C&UGC 5005&0.223&0.028&0.167&0.121&0.004&2.56& \multicolumn{1}{c}{---} &11.66&5.314E6\\ 
C&UGC 5721&8.700&0.435&1.876&0.760&0.101&1.99& \multicolumn{1}{c}{---} &1.24&3.073E8\\ 
C&UGC 7524&24.474&0.471&22.074&8.181&4.446&6.67& \multicolumn{1}{c}{---} &0.68&1.668E8\\ 
C&UGC 7603&4.007&0.236&0.698&0.446&0.206&1.28& \multicolumn{1}{c}{---} &3.57&2.811E7\\ 
C&UGC 8837&6.322&1.264&6.322&0.596&0.596&0.00& \multicolumn{1}{c}{---} &\multicolumn{1}{c}{$\infty$}&1.911E7\\ 
C&UGC 9211&0.291&0.036&0.199&0.156&0.145&0.00& \multicolumn{1}{c}{---} &1.74&1.000E8\\ 
D&F563-1&0.829&0.118&0.644&0.223&0.000&10.46& \multicolumn{1}{c}{---} &16.23&3.498E6\\ 
D&F568-3&4.775&0.597&4.233&2.285&2.143&0.00& \multicolumn{1}{c}{---} &4.42&4.356E7\\ 
D&F578-1&1.157&0.129&1.051&0.526&0.337&0.00&0.46&5.30&6.423E7\\ 
D&F579-V1&1.042&0.095&0.378&0.147&0.129&5.01& \multicolumn{1}{c}{---}&0.93&6.397E8\\ 
D&F583-1&0.314&0.022&0.171&0.101&0.019&0.00& \multicolumn{1}{c}{---} &3.77&3.868E7\\ 
D&F583-4&1.320&0.220&0.624&0.268&0.268&9.84& \multicolumn{1}{c}{---}&0.42&1.138E8\\ 
D&UGC 5750&0.935&0.117&0.499&0.354&0.244&0.00& \multicolumn{1}{c}{---} &6.73&1.150E7\\ 
D&UGC 6614&15.912&1.447&15.815&14.844&14.844&0.01&2.48&12.96&1.873E7\\ 
E&UGC 11707&10.348&1.035&3.283&0.674&0.194&9.24&\multicolumn{1}{c}{---} &\multicolumn{1}{c}{$\infty$}&6.917E5\\ 
E&UGC 12060&0.349&0.058&0.113&0.044&0.000&7.74& \multicolumn{1}{c}{---} &23.55&1.083E6\\ 
E&UGC 12632&14.603&1.043&8.658&6.100&1.720&14.08& \multicolumn{1}{c}{---} &\multicolumn{1}{c}{$\infty$}&1.172E6\\ 
E&UGC 12732&2.059&0.158&0.440&0.106&0.083&6.14& \multicolumn{1}{c}{---} &12.51&4.235E6\\ 
E&UGC 3371&3.785&0.473&1.454&0.810&0.582&10.04& \multicolumn{1}{c}{---} &10.76&3.874E6\\ 
E&UGC 4325&2.361&0.472&2.100&0.908&0.900&0.16& \multicolumn{1}{c}{---} &1.45&3.085E8\\ 
E&UGC 4499&0.706&0.118&0.197&0.007&0.000&0.00& \multicolumn{1}{c}{---} &2.52&5.908E7\\ 
E&UGC 5414&0.478&0.159&0.356&0.251&0.107&2.76& \multicolumn{1}{c}{---} &5.51&9.170E6\\ 
E&UGC 6446&1.726&0.216&0.923&0.796&0.509&3.21& \multicolumn{1}{c}{---} &4.53&1.531E7\\ 
E&UGC 731&0.826&0.092&0.474&0.224&0.009&12.59& \multicolumn{1}{c}{---} &5.86&6.874E6\\ 
E&UGC 7323&0.902&0.129&0.854&0.427&0.273&1.96& \multicolumn{1}{c}{---} &6.91&1.302E7\\ 
E&UGC 7399&20.720&1.381&2.302&2.017&0.000&6.11& \multicolumn{1}{c}{---} &3.97&5.199E7\\ 
E&UGC 7524&2.432&0.087&0.847&0.394&0.291&4.72& \multicolumn{1}{c}{---}&3.59&1.866E7\\ 
E&UGC 7559&0.360&0.060&0.269&0.059&0.000&0.00& \multicolumn{1}{c}{---}&0.88&1.060E8\\ 
E&UGC 7577&0.648&0.108&0.465&0.294&0.023&0.40& \multicolumn{1}{c}{---} &\multicolumn{1}{c}{$\infty$}&8.251E5\\ 
E&UGC 7603&1.993&0.221&0.413&0.043&0.000&0.66& \multicolumn{1}{c}{---} &1.94&7.830E7\\ 
E&UGC 8490&4.199&0.156&2.691&1.408&0.000&3.63& \multicolumn{1}{c}{---} &2.88&5.073E7\\ 
E&UGC 9211&0.225&0.038&0.023&0.011&0.000&2.53& \multicolumn{1}{c}{---} &2.36&5.187E7\\
\hline
\end{tabular}
\label{tab:results1galaxy}
\end{table*}

\nprounddigits{2} 
\begin{table*}
\caption{The medians of the quantities $\chi^2_\mscript{red}$, $\chi^2$, $\chi^2_\mscript{2h}$, $\chi^2_\mscript{h}$ and $\chi^2_\mscript{h/2}$. For all of these quantities, and for all the samples and subsamples, the Burkert profile  yields lower median results than the NFW profile.}
\begin{tabular}{lln{1}{2}n{3}{2}n{2}{2}n{2}{2}n{2}{2}}
\hline
\multicolumn{1}{c}{S}&Model& {$\langle \chi^2_\mscript{red}\rangle$} & {$\langle \chi^2\rangle$} & {$\langle \chi^2_{2h}\rangle$} & {$\langle \chi^2_{h}\rangle$} & {$\langle \chi^2_{h/2}\rangle$} \\
\multicolumn{1}{c}{(1)} &   \multicolumn{1}{c}{(2)}   &  \multicolumn{1}{c}{(3)}    &   \multicolumn{1}{c}{(4)}  &  \multicolumn{1}{c}{ (5)}   &  \multicolumn{1}{c}{(6)}   & \multicolumn{1}{c}{(7)}\\
\hline
A&Burkert&0.62&50.37&27.59&12.85&9.20\\ 
&NFW&0.92&106.45&43.89&22.69&16.55\\ 
B&Burkert&0.78&13.11&6.32&3.90&1.45\\ 
&NFW&1.58&31.15&21.18&13.39&3.11\\ 
C&Burkert&0.26&2.36&1.80&0.60&0.21\\ 
&NFW&0.54&7.32&5.35&3.45&1.89\\ 
D&Burkert&0.12&1.10&0.63&0.31&0.27\\ 
&NFW&0.97&10.03&6.73&5.24&2.57\\ 
E&Burkert&0.16&1.86&0.66&0.34&0.23\\ 
&NFW&0.42&4.11&2.00&1.37&0.61\\ 
$\cal S$&Burkert&0.38&6.86&2.50&0.81&0.58\\ 
&NFW&0.67&14.72&6.11&4.42&2.93\\ 
${\cal S}_\mscript{*1}$&Burkert&0.47&14.60&4.08&2.18&0.84\\ 
&NFW&0.71&22.87&10.69&6.64&4.18\\ 
${\cal S}_\mscript{*2}$&Burkert&0.73&24.23&7.42&4.93&1.72\\ 
&NFW&1.27&31.97&21.18&10.41&7.91\\
${\cal S}_\mscript{g1}$&Burkert&0.47&10.35&2.93&1.71&0.90\\ 
&NFW&0.68&20.64&7.51&6.05&4.08\\ 
${\cal S}_\mscript{g2}$&Burkert&0.43&28.31&6.35&2.28&2.62\\ 
&NFW&0.62&29.24&13.85&8.29&7.35\\ 
${\cal S}_\mscript{h1}$&Burkert&0.47&11.73&3.68&2.09&0.82\\ 
&NFW&0.57&19.04&7.80&5.94&3.19\\ 
${\cal S}_\mscript{h2}$&Burkert&0.38&7.40&2.93&0.81&0.41\\ 
&NFW&0.57&17.44&7.51&4.42&4.24\\ 
\hline
\end{tabular}
\label{tab:resultsmedians}
\end{table*}

\begin{figure}
        \includegraphics[trim = 0cm 0cm 0cm 0cm, clip=true, width=0.45\textwidth]{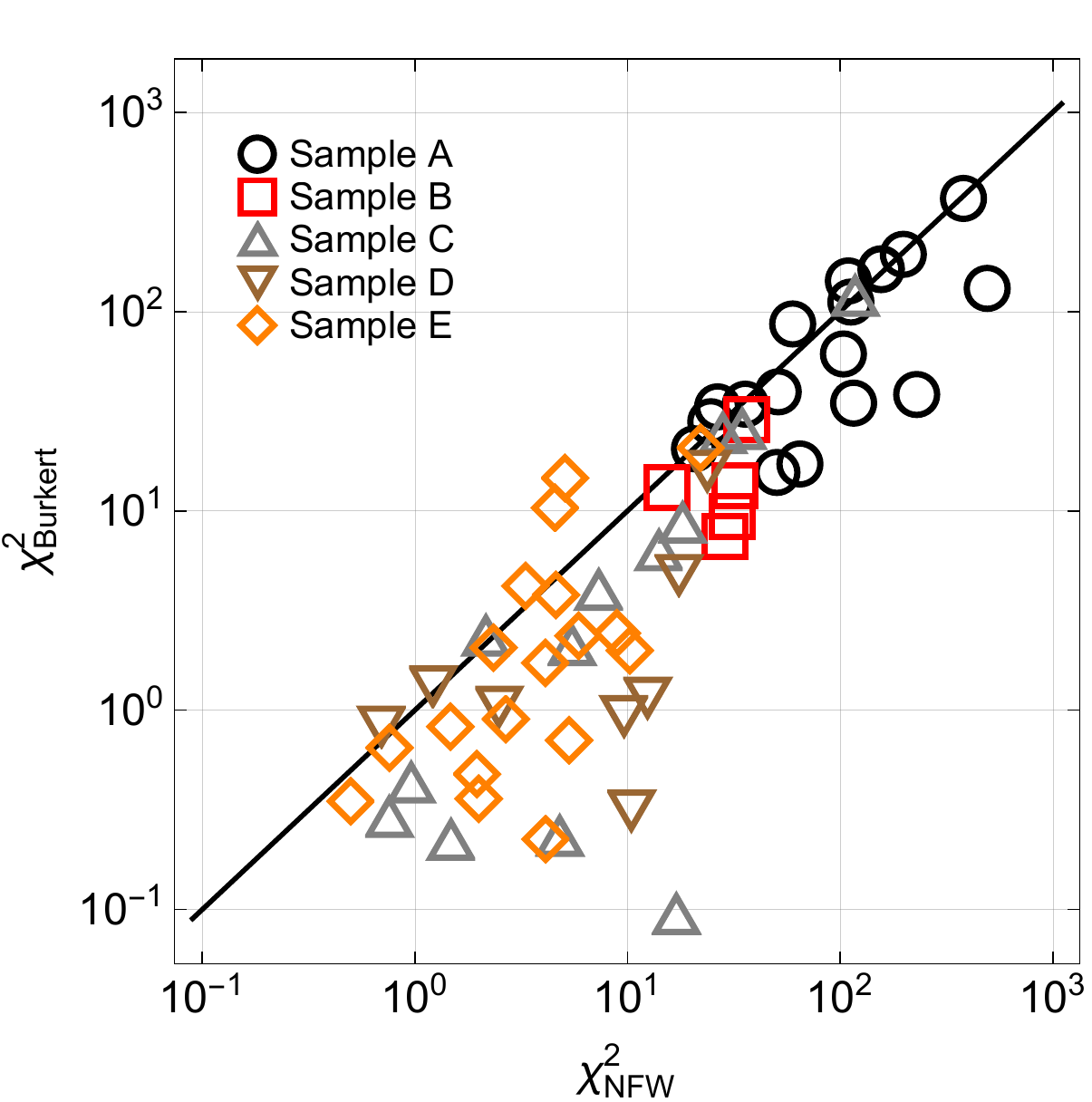}
	\caption{Comparison between the Burkert and the NFW fits considering the minimum $\chi^2$ (which are respectively denoted by $\chi^2_\mscript{Burkert}$ and $\chi^2_\mscript{NFW}$). The black line is the the straight line where $\chi^2_\mscript{NFW} = \chi^2_\mscript{Burkert}$. Among our sample of 62 galaxies, only 13 are fitted better with the NFW dark matter halo than with the Burkert one (i.e., they satisfy $\chi^2_\mscript{NFW} < \chi^2_\mscript{Burkert}$).} \label{fig:chiBurkertxNFW}
\end{figure}

\begin{figure*}
    \begin{subfigure}
     	\centering 
        \includegraphics[trim = 0cm 0cm 0cm 0.6cm, clip=true, height=0.35\textwidth]{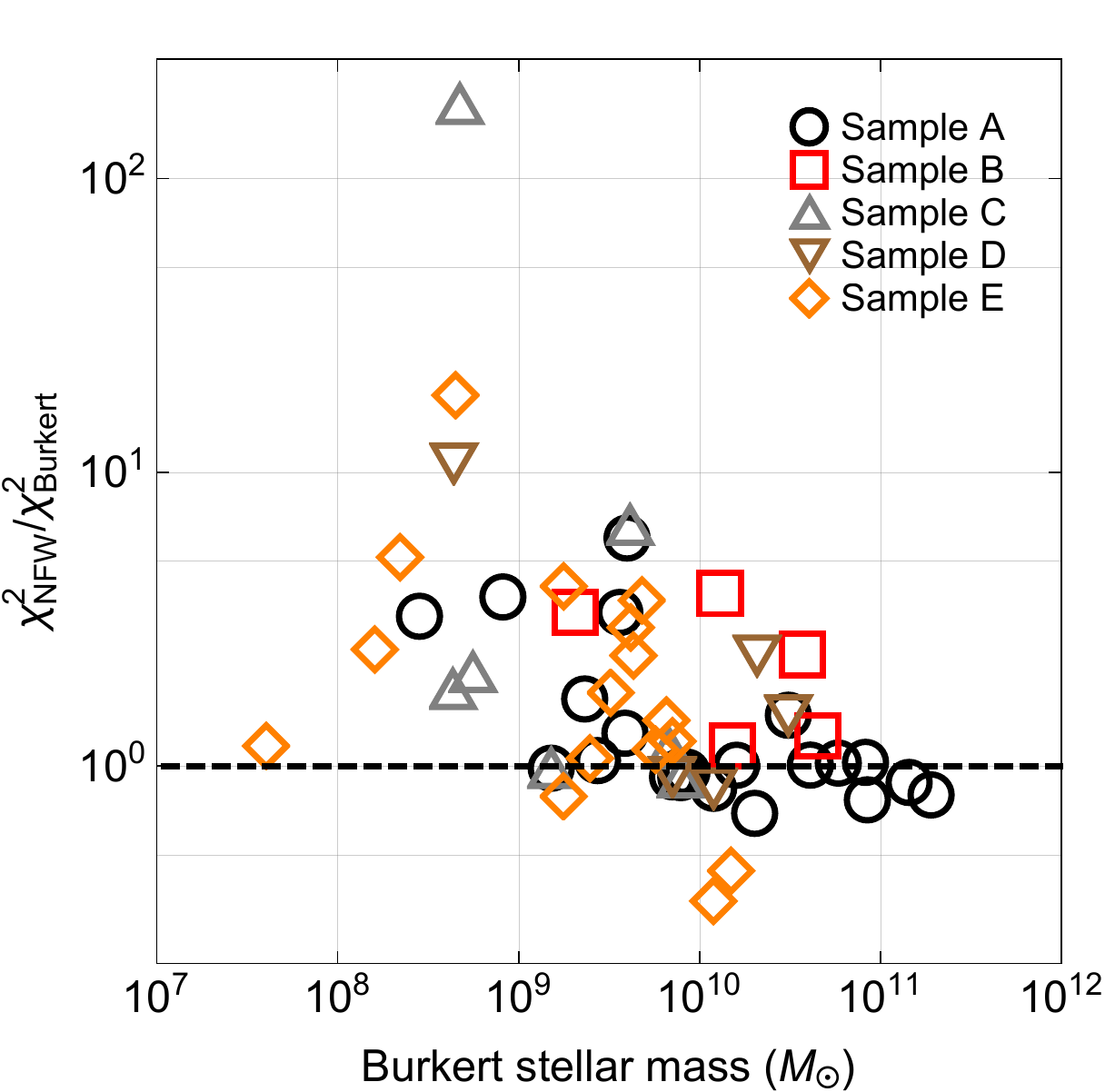}
    \end{subfigure} \hspace*{-0.35cm}
    \begin{subfigure}
     	\centering 
        \includegraphics[trim = 0cm 0cm 0cm 0.6cm, clip=true, height=0.35\textwidth]{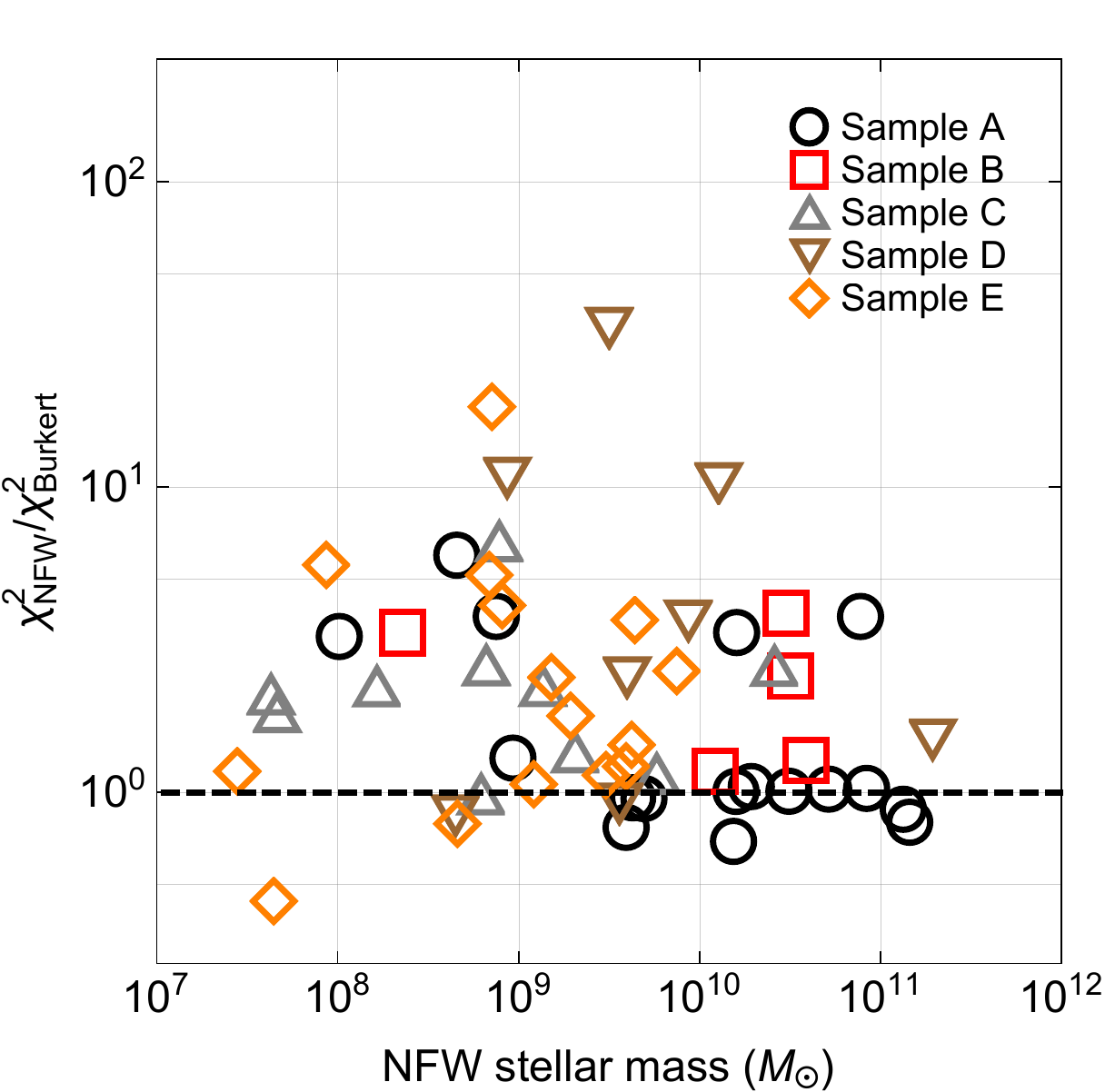}
    \end{subfigure} \hspace*{-0.35cm}
    \begin{subfigure}
     	\centering 
        \hspace*{0.4cm} \includegraphics[trim = 0cm 0cm 0.05cm 0cm, clip=true, height=0.345\textwidth]{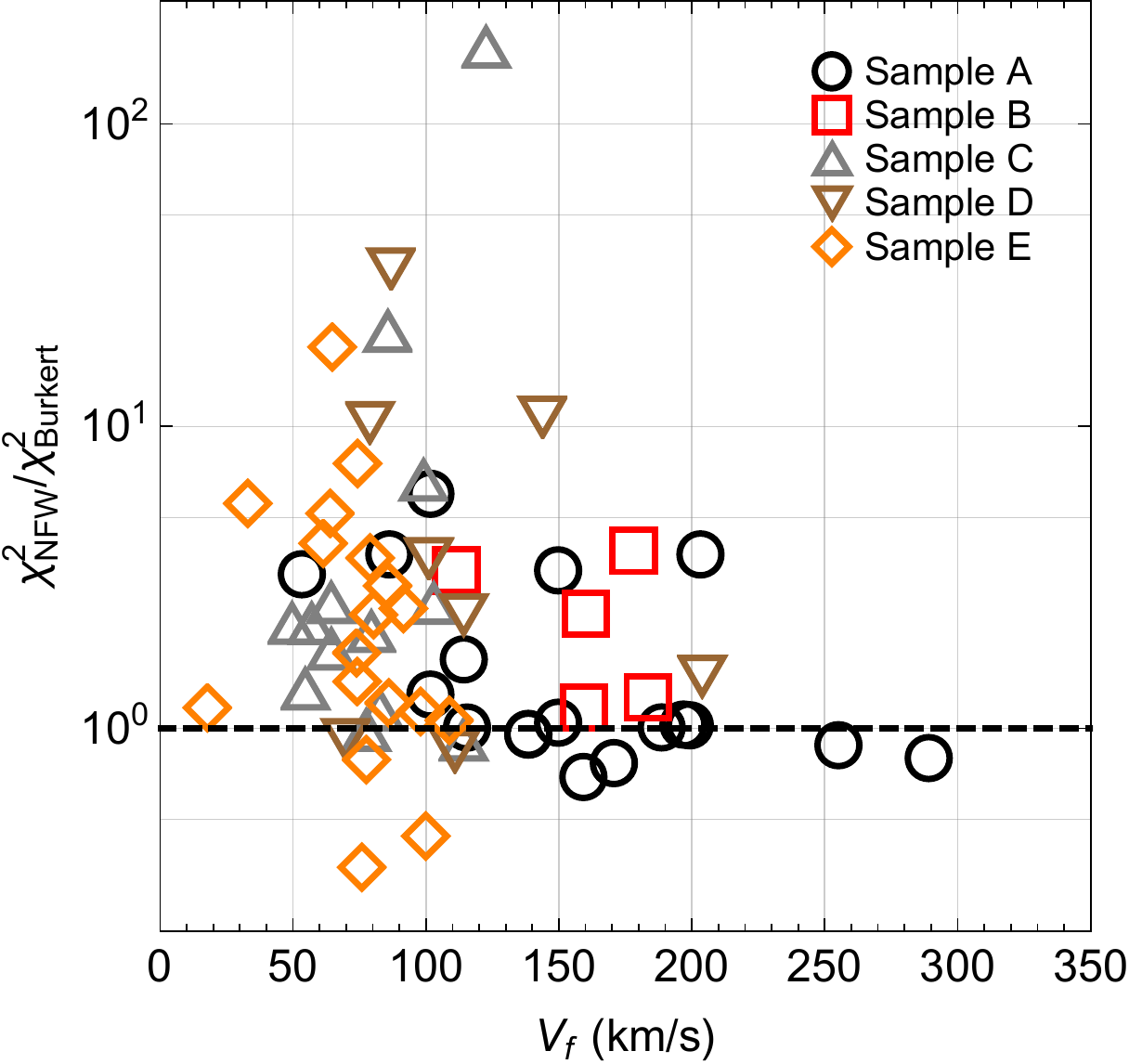}
    \end{subfigure} 
        \begin{subfigure}
     	\centering 
        \includegraphics[trim = 0cm 0cm 0cm 0.35cm, clip=true, height=0.355\textwidth]{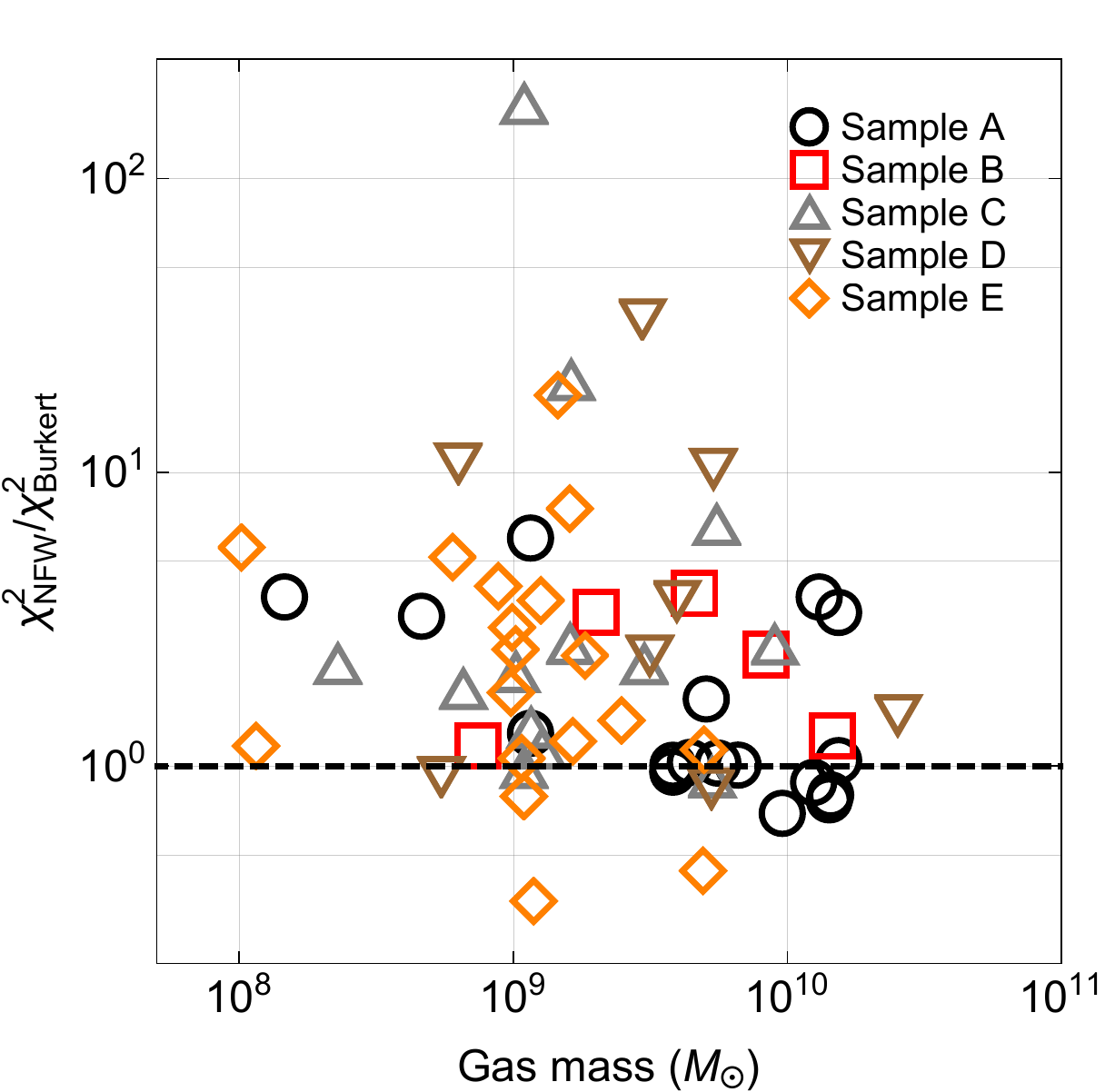}
    \end{subfigure} 
	\caption{Each plot shows the relation between the ratio $\chi^2_\mscript{NFW}/\chi^2_\mscript{Burkert}$ and the following parameters: i) (top left) the total stellar mass (disc and bulge) derived from the fits that use the Burkert profile, ii) (top right) same as the previous case, but using the NFW profile, iii) (bottom left) the final velocity $V_f$ (see Appendix \ref{app:expectedY} for further details), and iv) (bottom right) the total gas mass. The first two plots show a trend such that, for galaxies with stellar mass above $\sim 10^{9.5} M_\odot$, the higher is the stellar mass the lower is the dispersion on the plane  $\chi^2_\mscript{NFW}/\chi^2_\mscript{Burkert} \times M_*$, and the closer the data are to $\chi^2_\mscript{NFW}/\chi^2_\mscript{Burkert}\sim 1$.  Qualitatively similar trends can also be seen in the other plots above.}\label{fig:chiBNcorrelations}
\end{figure*}

\begin{figure*}
    \begin{subfigure}
     	\centering 
        \includegraphics[trim = 0cm 0.80cm 0cm 0cm, clip=true, width=0.70\textwidth]{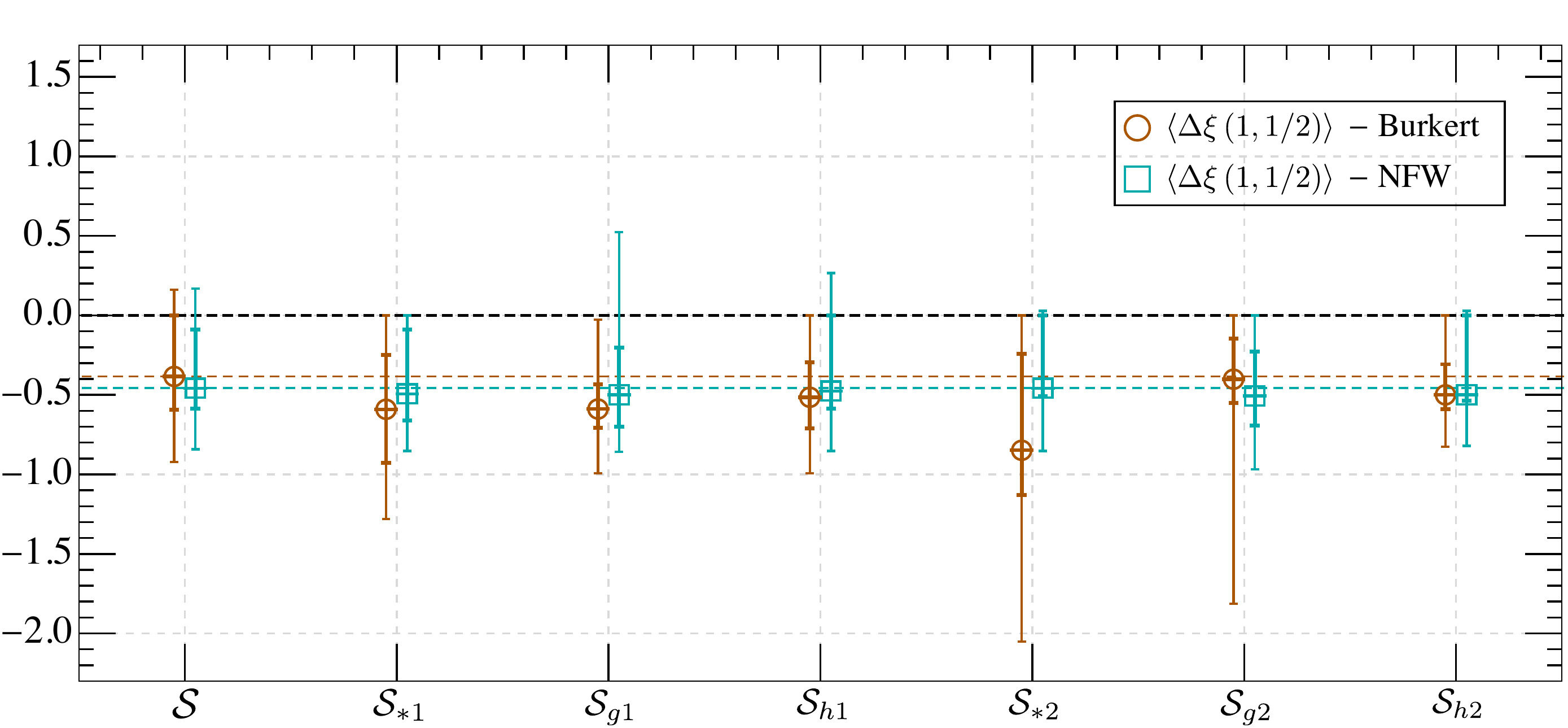}
    \end{subfigure}\hspace*{0.1cm} \vspace*{-0.1cm}
    \begin{subfigure}
     	\centering 
        \includegraphics[trim = 0cm 0cm 0cm 0.80cm, clip=true, width=0.70\textwidth]{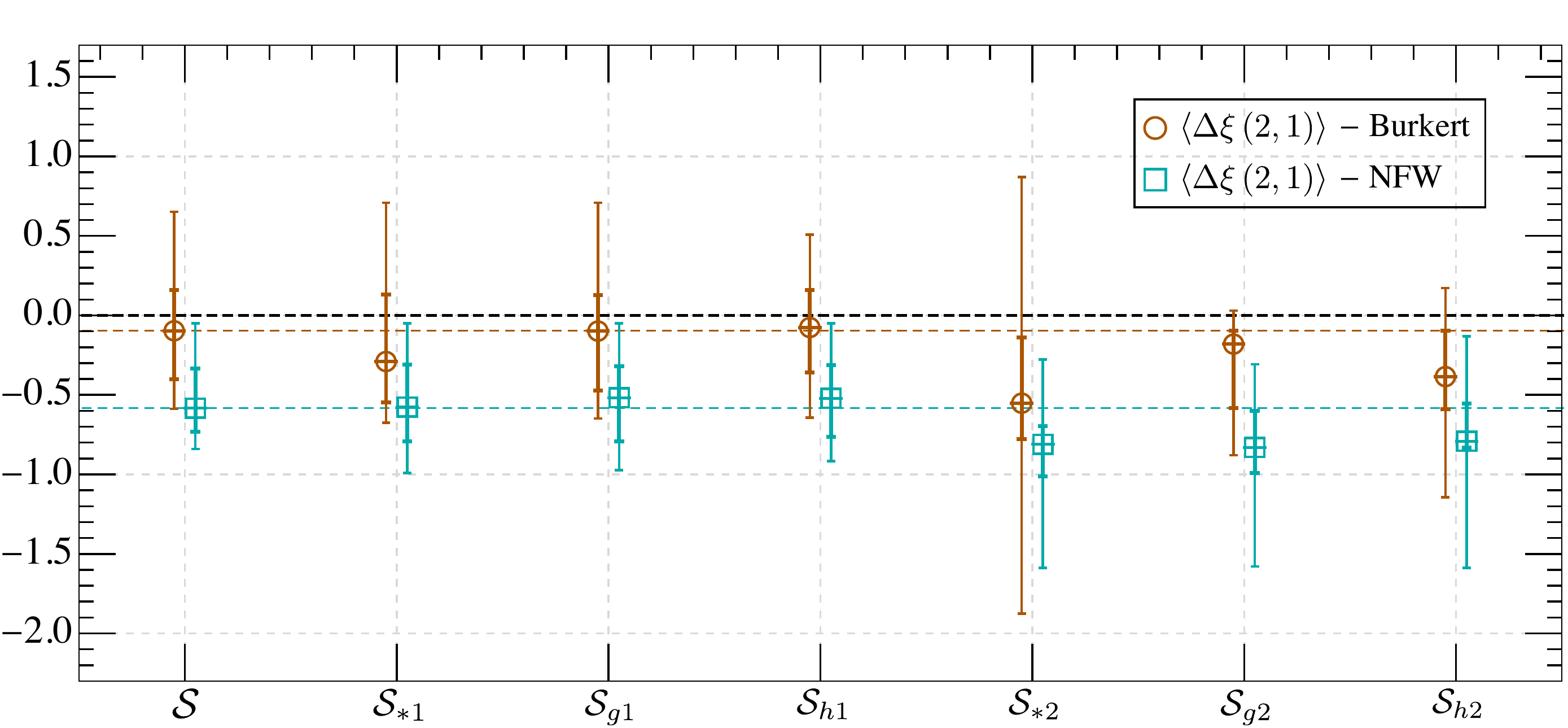}
    \end{subfigure} 
	\caption{Results for the medians and dispersions of $\Delta \xi (1,1/2)$ and $\Delta \xi (2,1)$, considering  the complete sample ${\cal S}$ and the six subsamples whose definitions can be found in Table \ref{tab:samplenumbers}. The medians are  denoted by a  circle, for the Burkert profile, and with an open square, for the NFW profile.  Each of these medians have two error bars, one, the most interior one, for the dispersion evaluated using $\sigma_\mscript{$25\%$}$, while the other is computed from $\sigma_\mscript{$50\%$}$. The thick black dashed line indicates the expected value of $\langle\Delta \xi\rangle$ for an ideal model whose fits are homogeneous along the galaxy radius, which is zero. The two thinner dashed lines indicate the values of $\langle\Delta \xi\rangle$ computed for the complete ${\cal S}$ sample and associated either to the Burkert profile (with brown color), or to the NFW profile (with cyan color). These results are discussed in Secs. \ref{sec:results} and \ref{sec:conclusions}.} \label{fig:plotDxiSamples}
\end{figure*}

Our results can be grouped as follows:
\begin{enumerate}[leftmargin=*,label=\arabic*)]

\item Burkert fits of individual galaxies, see Table \ref{tab:results1galaxy}.

\item Analyses of the $\chi^2$ values for each galaxy, comparing Burkert and NFW profiles.
See Figs.~\ref{fig:chiBurkertxNFW}-\ref{fig:chiBNcorrelations}.

\item Medians of the quantities $\chi^2, \chi^2_\mscript{red}$ and $\chi^2_{m h}$, see  Table \ref{tab:resultsmedians}.

\item Analyses of the quantities $\xi$, $\zeta$ and $\Delta \xi$. 
Fig.~\ref{fig:plotDxiSamples} summarizes the detailed results shown in Appendix \ref{app:xizeta}.

\end{enumerate}

Figure \ref{fig:chiBurkertxNFW} compares the minimum $\chi^2$ derived from the Burkert and  NFW profiles. There is a clear preference for the Burkert profile since among our sample of 62 galaxies only 13 have better fits when using the NFW profile. Moreover, those that are better fitted with the NFW profile only slightly favor the latter.

Figure \ref{fig:chiBurkertxNFW} also shows that some samples have larger $\chi^2$ values than others. This is expected since the $\chi^2$ values depend on the number of RC data points, and the latter depend on both the angular resolution of the 21 cm data and on the size and distance of the observed galaxies.  For example, Sample A includes several  large nearby galaxies and features 21 cm observations with the highest angular resolution, thus it is expected to yield the highest values for $\chi^2$. For the reduced $\chi^2$ results of Sample A, one can see from Table \ref{tab:resultsmedians} that there is no  discrepancy in regard to other samples.

Figure \ref{fig:chiBNcorrelations} shows plots whose purpose is to analyse correlations between the fraction  $\chi^2_\mscript{NFW}/\chi^2_\mscript{Burkert}$ and certain galaxy parameters, namely: the stellar mass, gas mass and the final circular velocity $V_f$. It is not shown but correlations with the disc scale length were also tested, and they lead to qualitatively similar results, but with a dispersion about the same or higher. It can be noted from the upper plots of Fig.~\ref{fig:chiBNcorrelations} that the values of $\chi^2_\mscript{NFW}/\chi^2_\mscript{Burkert}$ have larger dispersion at about $M_* \sim 10^8 M_\odot$ or $M_* \sim 10^9 M_\odot$, and that the dispersion decreases and the fraction $\chi^2_\mscript{NFW}/\chi^2_\mscript{Burkert}$ approaches 1 as one considers larger stellar masses. It was not possible to find that galaxies with $10^{9.5} M_\odot$ or higher stellar masses favor the NFW profile (i.e., $\chi^2_\mscript{Burkert} > \chi^2_\mscript{NFW}$).\footnote{We have included the bulge in our analyses, but no significative change is observed if the bulge is not considered.} The analyses with the disc scale length ($h$) and the gas mass lead to similar results, but with a less clear correlation related to the fraction  $\chi^2_\mscript{NFW}/\chi^2_\mscript{Burkert}$.

In Table \ref{tab:resultsmedians}, medians of $\chi^2$-related quantities are displayed for the various samples. For all the samples, even those that select the largest galaxies (i.e., ${\cal S}_{*2}, {\cal S}_{g2}$ and ${\cal S}_{h2}$), all the $\chi^2$-related quantities have lower values when the dark matter halo profile is the Burkert one.\footnote{Some care is necessary on the issue of $\chi^2_\mscript{red}$, since a large fraction of the found values have very low values of $\chi^2_\mscript{red}$. Supposing that the error bars of all galaxies were properly evaluated, one is to expect that $\langle \chi^2_\mscript{red}\rangle \approx 1$.  To properly consider all the diverse systematical errors in external galaxies is not an easy task, and a reliable and feasible procedure is probably  currently unknown. Likewise in many other papers on the subject \citep[e.g.,][]{deBlok:2002tg, 2008AJ....136.2648D, Gentile:2010xt} we use $\chi^2$ or $\chi^2_\mscript{red}$ to compare fits relative to different models and not to obtain an absolute goodness-of-fit.}

We now discuss our results regarding the quantities $\xi$, $\zeta$ and $\Delta \xi$.
With the values of $\chi^2_\mscript{h/2}, \chi^2_\mscript{h}$ and $\chi^2_\mscript{2h}$ for each galaxy, essentially two different $\xi$ quantities, as introduced in Sec.~\ref{sec:xizeta}, can be evaluated: $\xi(1, 1/2)$ and $\xi(2,1)$. The quantity $\xi(2,1/2)$ is a combination of the previous two. 
Considering the median results for the sample ${\cal S}$, the upper plot of Fig.~\ref{fig:plotDxiSamples} shows that both the profiles have about the same behaviour, and both display a tendency to better fit  the region $h/2 < R < h$ than the region $0 < R < h/2$.\footnote{The fits are on average about $25\%$ better in the region $h/2 < R < h$, since $\langle \Delta \xi(1,1/2)\rangle \approx -0.5$, and since 0.5 is $25\%$ of 2 $\approx \langle \zeta(1,1/2) \rangle$.} Considering the inferred dispersions, one sees that the expected value of $\langle \Delta \xi(1,1/2)\rangle $, which is zero, is close to the upper limit of $\sigma_{25\%}$ (i.e., $\sigma_{25\%}^+$) for both of the profiles.\footnote{If $\sigma_{25\%}^+$ of some quantity $X$ is accurately determined, then the probability of a value of $X$ to be smaller than $\sigma_{25\%}^+(X)$ is  $62.5\%$ (i.e., $P(X < \sigma_{25\%}^+(X))= 0.5 + 0.25/2 = 0.625$).} One sees, from considering only the largest galaxies (i.e., the other six samples), that the above  ``tension'' has a small tendency to increase. In case further analyses confirm and enlarge this tension for both of the profiles, a possible interpretation is that a systematic issue with the central part of the stellar profiles is being uncovered, see also Sec.~\ref{sec:conclusions}. In particular, it may be related to disc and bulge decomposition issues, non-circular motions or differential dust opacity \citep[see e.g.,][]{Courteau:2013cjm}.

The results associated to $\langle \Delta \xi (2,1)\rangle$ display stronger differences between the profile results.  As it can be seen in the bottom plot of Fig.~\ref{fig:plotDxiSamples}, the $\cal S$ sample results indicate the existence of a good agreement between the Burkert value of $\langle \Delta \xi (2,1)\rangle$ and the expected value of zero. The expected value is clearly well inside the $\sigma_\mscript{25\%}$ error bars of the Burkert profile. On the other hand, for the NFW profile results, the expected value is outside the $\sigma_\mscript{50\%}$ error bars, hence more than 75\% of the galaxies fitted with NFW are in tension with a homogeneous fit.\footnote{Since $P(X < \sigma_{50\%}^+(X))= 0.5 + 0.50/2 = 0.75$.}

Considering the sample $\cal S$ results, the plot at the bottom of Fig.~\ref{fig:plotDxiSamples} shows that the Burkert profile provides RC fits that are homogeneous with respect to the regions $0<R<h$ and $h<R<2h$, while the NFW profile has a clear tension with homogeneity,  fitting on average the region $h<R<2h$ better than the region $0<R<h$. Upon considering the six subsamples that select the largest galaxies, both the models lead essentially to the same results, with a small tendency towards more negative $\langle \Delta \xi (2,1) \rangle$ values for the three most restrictive subsamples (${\cal S}_\mscript{*2}, {\cal S}_\mscript{g2}$ and ${\cal S}_\mscript{h2}$). Perhaps the best DM profile is neither one of these two, but clearly the Burkert profile results are better than the NFW results, and this tendency persists even considering only the largest galaxies (i.e., using the subsamples ${\cal S}_*$, ${\cal S}_g$, ${\cal S}_h$). This is one of the main results of this work.

For the subsamples $S_*, S_g$, and $S_h$, the Burkert profile results are essentially the same, with a small tendency towards better fitting the region $h < R < 2h$ than the region $R < h$ for the three most stringent subsamples. On the other hand, the NFW profile is clearly worse for these subsamples. The restriction to such large galaxies actually worsens the NFW situation instead of improving it, as it can be seen from Fig.~\ref{fig:plotDxiSamples} and also, in more detail, from Figs.~\ref{fig:bluered1}, \ref{fig:bluered2}, \ref{fig:yellow1}, \ref{fig:yellow2}.

\section{Conclusions and discussion} \label{sec:conclusions}

Here we use observational data of 62 galaxies fitted with both the NFW profile \citep[whose fits come from][]{Rodrigues:2014xka} and the Burkert profile (which are new results presented here, see Table \ref{tab:results1galaxy}).
We perform four different comparisons between the NFW and Burkert profiles, namely: i)~a straightforward test that compares the values of the minimum $\chi^2$ for each galaxy and each profile (Fig.~\ref{fig:chiBurkertxNFW}, see also Table \ref{tab:resultsmedians}); ii)~correlations between quality of the fits (i.e., minimum $\chi^2$) and global galaxy parameters  (stellar mass, disc scale length, final velocity $V_f$ and gas mass, see Fig.~\ref{fig:chiBNcorrelations}); iii)~evaluations on the homogeneity of the fits along the galaxy radius for the whole sample by using the quantities $\xi$ and $\zeta$ that were introduced in Sec.~\ref{sec:xizeta}, and whose results can be seen in the first plots of Figs. \ref{fig:bluered1} and \ref{fig:yellow1}; iv)~evaluation of trends on the evolution of homogeneity using different criteria to specify ``large'' galaxies (Fig.~\ref{fig:plotDxiSamples} summarizes the results, and the details are given in Appendix \ref{app:xizeta}).

Considering the four analyses above, we find that: i) among the 62 galaxies, only 13 are better fitted by the NFW halo profile with respect to the Burkert profile; ii) we found evidence for a trend such that for larger galaxies the NFW profile has a systematic tendency towards improving its fits in comparison with the Burkert one, but it does not fit better than the Burkert profile for  $M_* \lesssim 10^{10.5} M_\odot$. The NFW profile may be the best profile for $M_* \gtrsim 10^{11}$, but these are very massive galaxies, and the sample that we use in this work only has a few of them. iii) The homogeneity tests show that the Burkert profile results are consistent with homogeneity (considering the quantity $\Delta \xi(2,1)$), while the NFW fits have a tendency towards better fitting the region between $h$ and $2 h$ than the region between the galaxy centre and $h$, where $h$ is the disc scale length. iv) By restricting the galaxy sample to the subsamples that select the largest galaxies according to different criteria, we find that the results on the homogeneity tests with $\xi$ and $\zeta$ are essentially the same, and hence the NFW profile still leads to non-homogeneous fits  considering only the galaxies with $M_* > 10^9 M_\odot$, or even $M_* > 10^{10} M_*$. Therefore, we confirm the results of \citet{Spano:2007nt} that a cored profile -- the Burkert profile in this work -- can on average lead to significantly better results than the  NFW profile, even for large, very massive, galaxies.\footnote{On the other hand, there is the possibility that an important aspect of baryonic physics is not being properly modeled by the observational data analysis. If this is the case, then the results relative to the largest galaxies are more prone to significative changes than the results relative to the smaller ones.}

 If the DM content of real galaxies follows a universal profile, the above result  states that such universal profile should be closer to the Burkert profile than the NFW one. This interpretation is in accordance with the much debated existence of a universal constant dark matter halo surface density   \citep{2004IAUS..220..377K, Salucci:2007tm, 2009MNRAS.397.1169D,Gentile:2009bw, 2016ApJ...817...84K}, see, however, \citet{2013MNRAS.429.1080D, Saburova:2014opa}. On the other hand,  it is also important to stress that our results do not imply the existence of a universal DM profile, since there may exist a significative amount of galaxies that evolve naturally towards cuspy DM profiles. For instance, our results are not in conflict with those of \citet{Simon:2004sr}.

If the trends that we find here persist once the sample is enlarged, the derived results would be in conflict with certain expectations from the most well known mechanisms able to flatten the DM cusp, namely, supernova feedback and dynamical friction generated by baryonic clumps. They have different predictions for low mass galaxies, like for the dwarf spheroidals \citep{DelPopolo:2016emo}, but both of them are especially effective at $M_*\sim 10^{8.5} M_\odot$, and both lead to DM halos that are well described by a NFW profile when $M_*\sim 10^{10} M_\odot$. From Fig. \ref{fig:chiBNcorrelations} it is possible to see that there is a trend such that, for the most massive galaxies, the internal dynamics reduces its strong preference for the cored profile in favour of the cuspy NFW profile, qualitatively as expected from the simulations and the two mechanisms just cited. The problem comes from the details, since a clear preference for the NFW cannot be spotted as even for the galaxies with $M_* > 10^{10} M_\odot$ the data still favour the Burkert profile. For such massive galaxies, these two effects are not expected to be effective on flattening the central profile, hence it may be a sign that an additional baryonic effect is taking place. For instance, for the largest galaxies  considered here, AGN feedback is perhaps relevant, and it may be responsible for the DM profile flattening of many of the largest disc galaxies \citep{2016arXiv161109922P}  \citep[at cluster scales, see, e.g.,][]{2012MNRAS.424...38D, 2013MNRAS.432.1947M}. Another possible interpretation is that the baryonic physics modeling is correct, but the  DM physics must be changed (e.g., some kind of  self-interacting DM, or modified gravity).

At last, concerning the new technique presented here, we tested the quantities $\xi (2,1)$, $\xi(1,1/2)$ and related quantities ($\zeta$ and $\Delta \xi$). We found that the values of $\langle \xi(2,1) \rangle$ are compatible with homogeneous fits if the Burkert profile is used, while homogeneity is not achieved by using the NFW profile (see Fig. \ref{fig:plotDxiSamples}). This tension with the NFW profile is not reduced by selecting only the largest galaxies from our sample. For the quantity $\langle \xi(1,1/2) \rangle$, both the profiles yielded similar results, with both of them being marginally compatible with homogeneous fits. The latter small tension for both profiles either stays the same or increases when considering the largest galaxies. This behaviour suggests the presence of  a systematic issue with the stellar profile close to the galaxy centres. Nonetheless, further investigation is necessary to confirm the latter issue, which we plan to do in a future work.

\section*{Acknowledgements}

We thank Luciano Casarini for discussions on hydrodynamical simulations and Nicola Napolitano for discussions on the stellar mass-to-light ratios. DCR and VM thank CNPq (Brazil) and FAPES (Brazil) for partial financial support. PLCO thanks CAPES for financial support. AP thanks CNPq (Brazil) for partial financial support during his stay at UFES.
\appendix

\vspace{.5cm}

\section{Distribution of $\xi$}
\label{app:Fdist}

\begin{figure}
        \includegraphics[ width=0.47\textwidth]{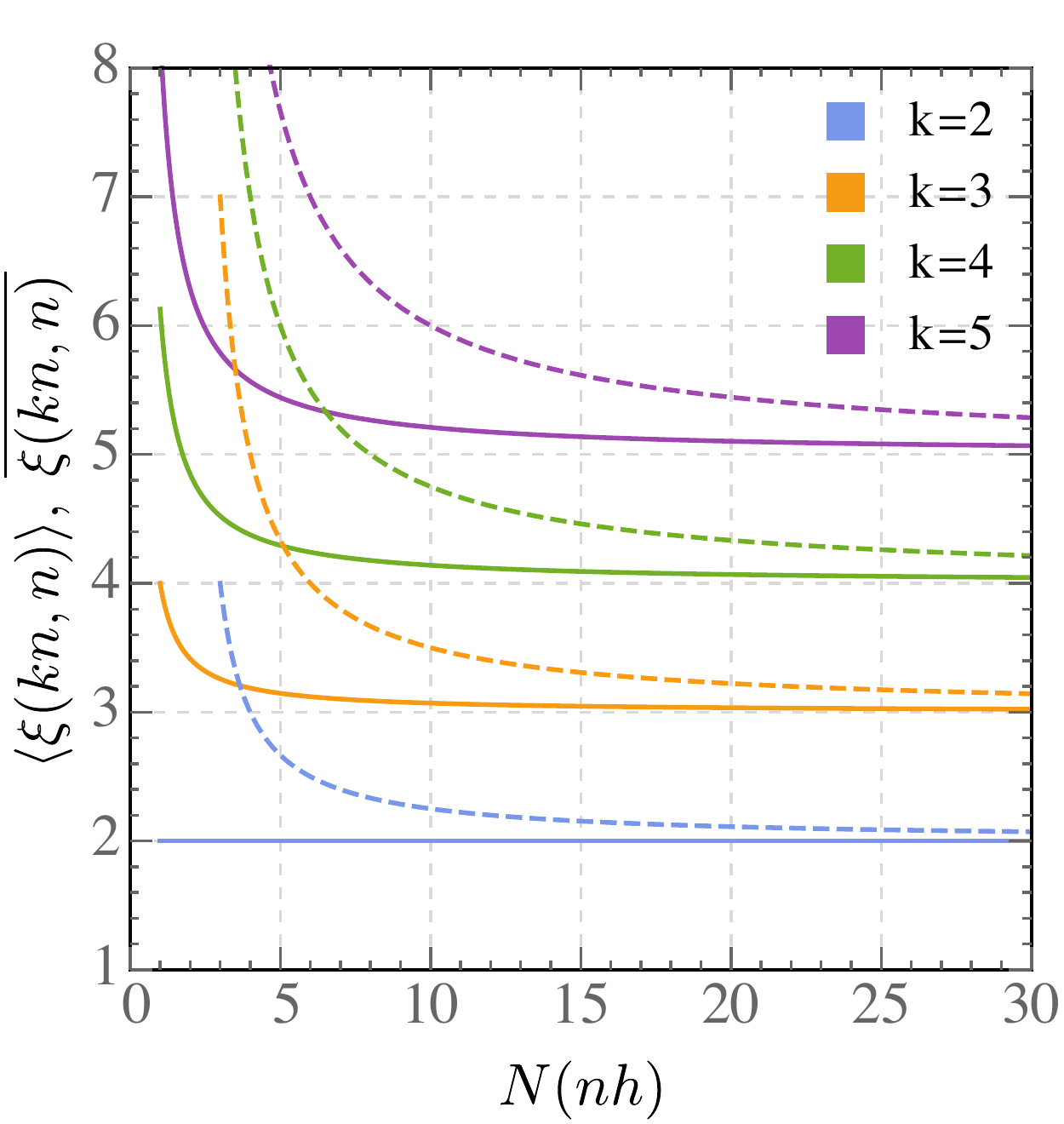}
	\caption{The median (solid line, plotted for $N(nh)\geq 1$) and the mean (dashed line, plotted for $N(nh)\geq 3)$) of $\xi(k n,n)$ as functions of $N(nh)$. From the plot above, the mean shows a much slower convergence than the median. The case $k=2$ is a special case, it is the only case in which the median is independent of $N(nh)$ (within the assumptions of this appendix). All the $\xi$ quantities computed from galaxies data in this paper use $k=2$, and the plot above motivates this choice.} \label{fig:xidist}
\end{figure} 

To derive the quantity $\xi$, as defined in eq.~\eqref{defxi}, one first minimizes the $\chi^2$ relative to the full sample of $N$ points and then takes the ratio of the two pieces of  $\chi^2$ with number of data points given by $N(n h)$ and $N(m h)$, respectively, where $1\le  N(mh) \le N $ and $1\le  N(nh) \le N$.

In order to understand the $\xi$ statistics, we start by assuming that the data are homogeneously distributed and dense enough such that $N(m  h)/N(n  h)=m/n$. To clarify the analyses we introduce here the following quantity, which is similar to $\chi^2_h$ (see eq.~\ref{eq:chi2h}),
\begin{equation}
    \chi^2_{m h,n h} \equiv \sum_{i=N(n h)+1}^{N(m h)} \left( \frac{V_\mtiny{model}(R_i, \bar p_j) - V_i}{\sigma_i} \right)^2,
\end{equation}
so that one can define (with $m>n$),
\begin{equation} \label{eq:xiind}
\xi_{\mtiny{ind}}(m,n)\equiv \frac{\chi^2_{m h,n h}}{\chi^2_{n  h}}  = \xi(m,n)  -1 \,.
\end{equation}

Although its relation to $\xi$ is simple, the quantity  $\xi_{\mtiny{ind}}$ is useful since it clearly only depends on independent data points. To simplify the analysis, we assume that  $N(n h) \gg N_p $,  where $N_p$ is the number of parameters $p_j$. Then, one sees from eq.~\eqref{eq:xiind} that $\xi_{\mtiny{ind}}$ is distributed according to a scaled F-distribution with $\{ N(mh)-N(n h), N(n h) \}$ degrees of freedom. Consequently, its median and its mean can be derived as follows
\begin{align}
 \langle \xi_{\mtiny{ind}}(k n ,n) \rangle &= \frac{1}{I^{-1}_{\left(1,-\frac{1}{2}\right)}\left ( \frac{N(nh)}{2},(k-1)\frac{N(nh)}{2} \right)} -1 \,, \\[.1in]
 \overline{\xi_{\mtiny{ind}}(k n,n)} &= (k-1)\frac{ N(n h)}{N(n h)-2} \,, 	
\end{align}
where we used $kn$ in place of $m$, $\langle \;\; \rangle$ denotes the median, a bar over a quantity denotes its mean value, the result for the mean is valid for $N(nh) \geq  3$, and  $I^{-1}$ is the inverse of the generalized regularized incomplete beta function.\footnote{That is, $I_{(z_0, z_1)} (a,b) = B(z_0,z_1, a,b)/B(a,b)$, where $B(a,b)$ is the beta function and $B(z_0,z_1, a,b) \equiv \int_{z_0}^{z_1} t^{a-1}(1-t)^{b-1}dt$ is the generalized incomplete beta function.} For $N(n h)$ sufficiently large, one finds that  $\langle \xi_{\mtiny{ind}} \rangle \approx \bar \xi_{\mtiny{ind}} \approx k-1$, which is equivalent to eq.~\eqref{eq:xiaverage}.

For the particular case $k=2$, changing the variable back to $\xi$, in place of $\xi_{\mtiny{ind}}$, we find,
\begin{align}
 \langle \xi(2n,n) \rangle &= 2 \,, \\
 \overline{\xi(2n,n)} &=2 \, \frac{N(n h)-1}{N(n h)-2}  \,.
\end{align}

This shows that -- within the assumption of this section -- eq.~\eqref{eq:xiaverage} holds exactly if the average is the median and if $m=2n$. For other values of $m$ and $n$, the same equation still holds, but under an additional approximation. 

Besides the important issue with outliers, commented in Sec.~\ref{sec:xizeta}, the median has an additional convenience, since the convergence of the median of the F-distribution to the value given by eq.~\eqref{eq:xiaverage} is much faster than the convergence of the mean. This can be seen in Fig.~\ref{fig:xidist}.

The main purpose of this appendix is to further clarify and motivate the use of $\xi(2n,n)$ and related quantities that we used in this paper. Some assumptions used in this appendix were evoked for simplicity  and are too restrictive considering the data that we use here. Further analyses, either with more data from galaxies, or theoretical developments on the statistics  will be purpose of a future work.

In Sec.~\ref{sec:xizeta} we agued in favour of the existence of some kind of average that would be compatible with eq.~\eqref{eq:xiaverage}, and also be compatible with the type of data that we deal with galaxies, namely data with a significative number of outliers. The above results confirm that the median is suited for describing the average \eqref{eq:xiaverage}, and favour the use of $k=2$.

\section{Plots of $\xi$, $\zeta$ and $\Delta \xi$} \label{app:xizeta}
\begin{figure*}
    \begin{subfigure}
	    \centering 
        \includegraphics[trim = 0cm 0cm 0cm 0cm, clip=true, width=0.45\textwidth]{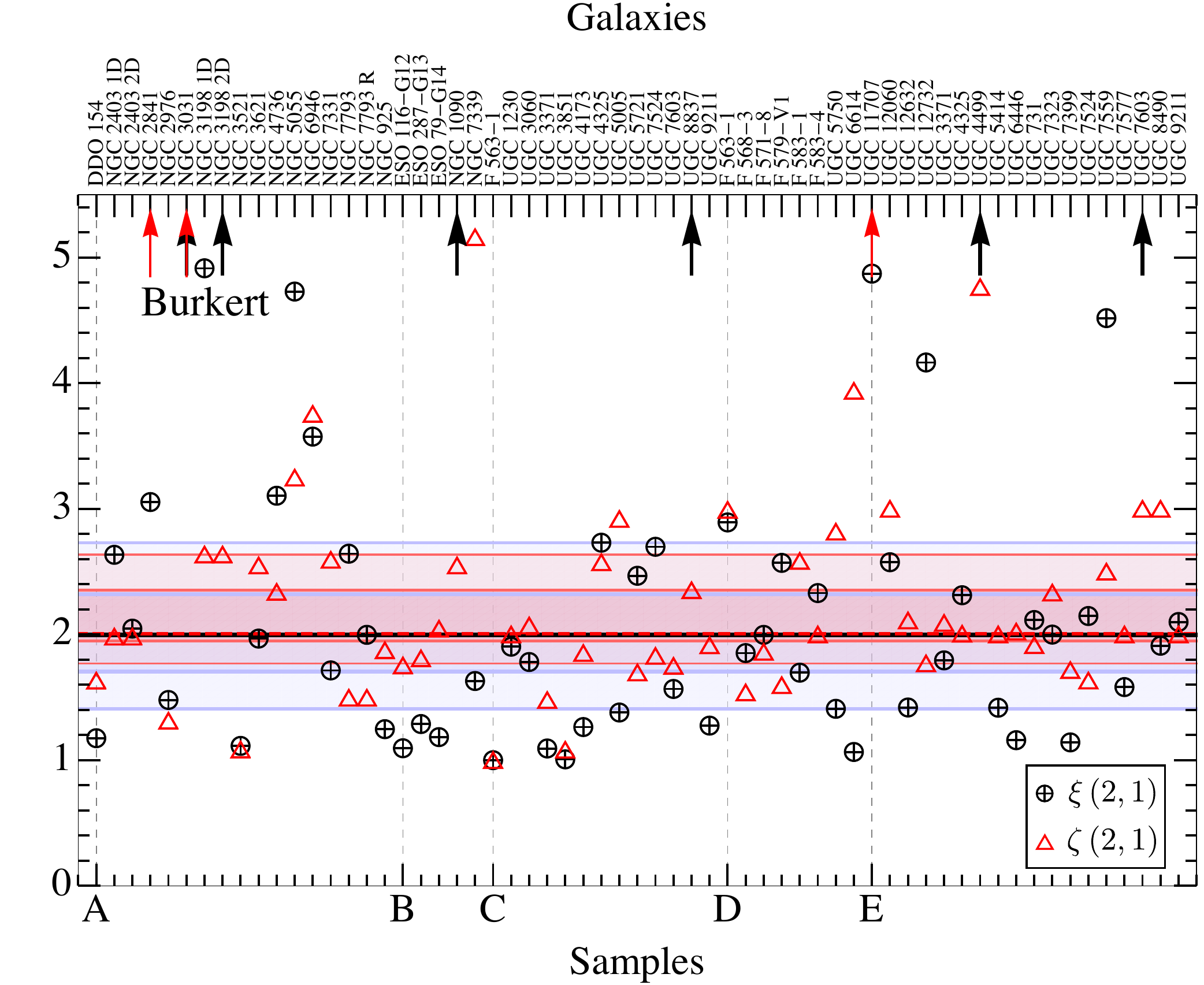}
   \end{subfigure} \hspace*{-0.75cm}
    \begin{subfigure}
	    \centering 
        \includegraphics[width=0.45\textwidth]{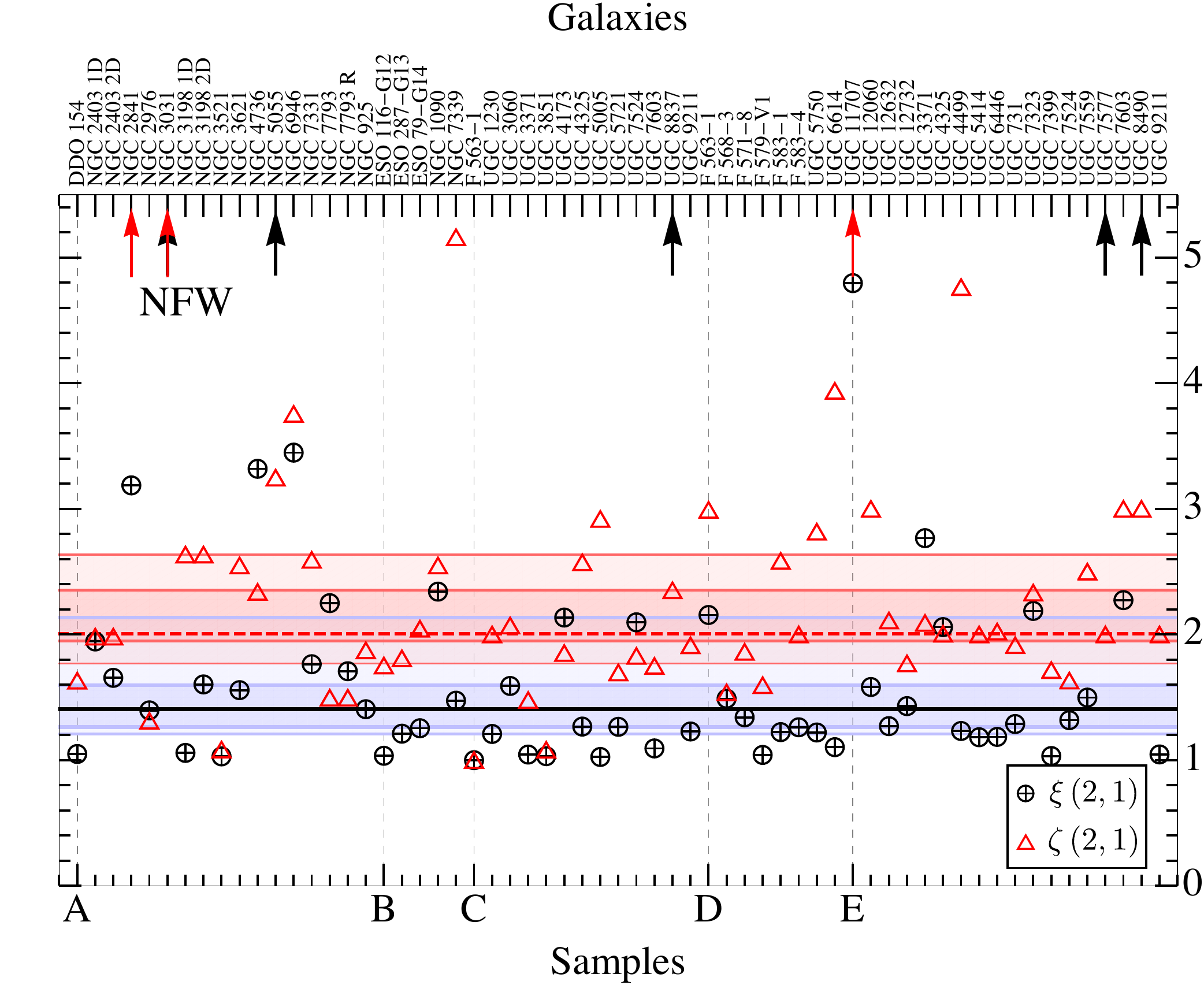}
    \end{subfigure}
    \begin{subfigure}
	    \centering 
        \includegraphics[trim = 0cm 0cm 0cm 0cm, clip=true, width=0.45\textwidth]{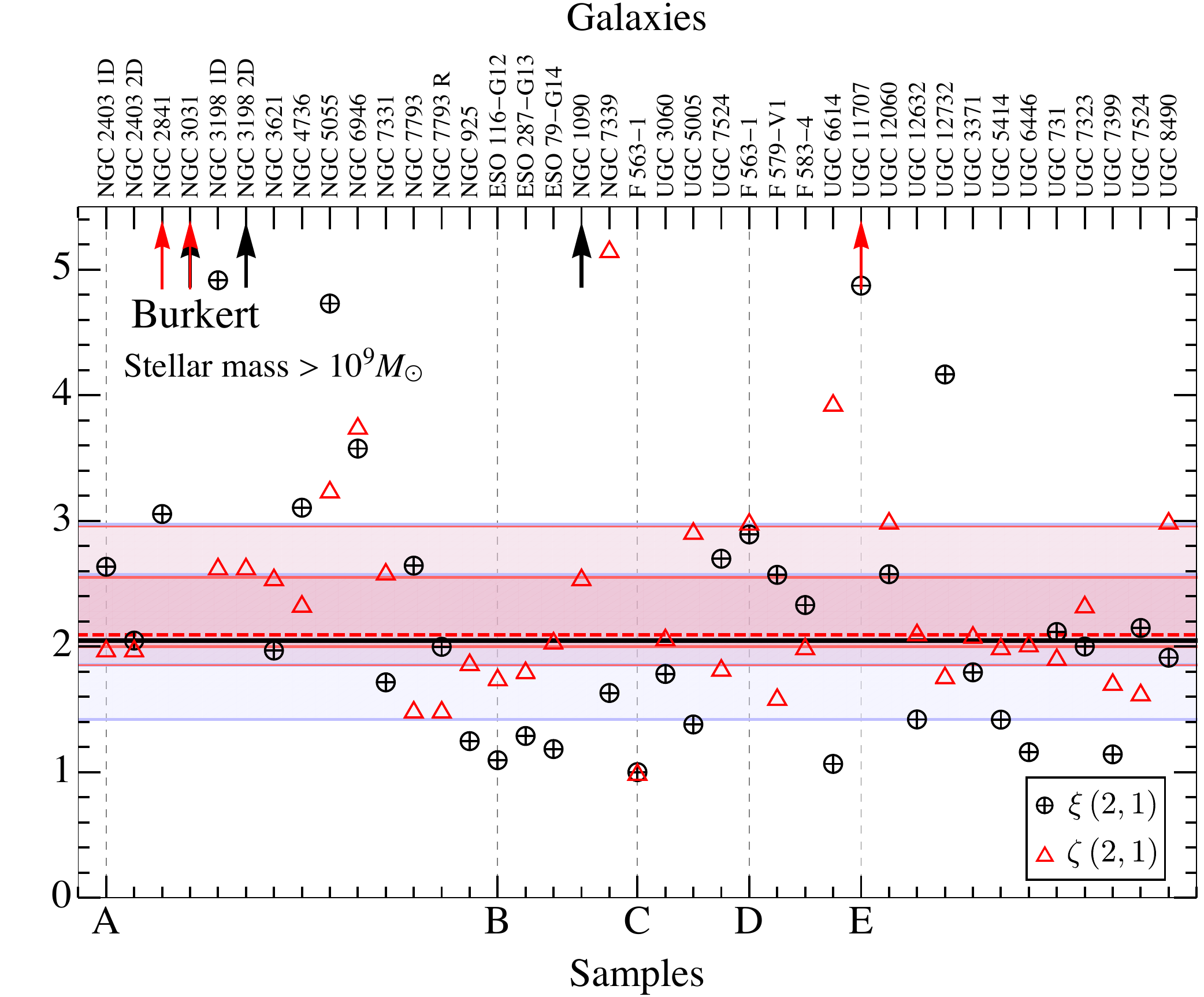}
    \end{subfigure} \hspace*{-0.75cm}
    \begin{subfigure}
	    \centering 
        \includegraphics[width=0.45\textwidth]{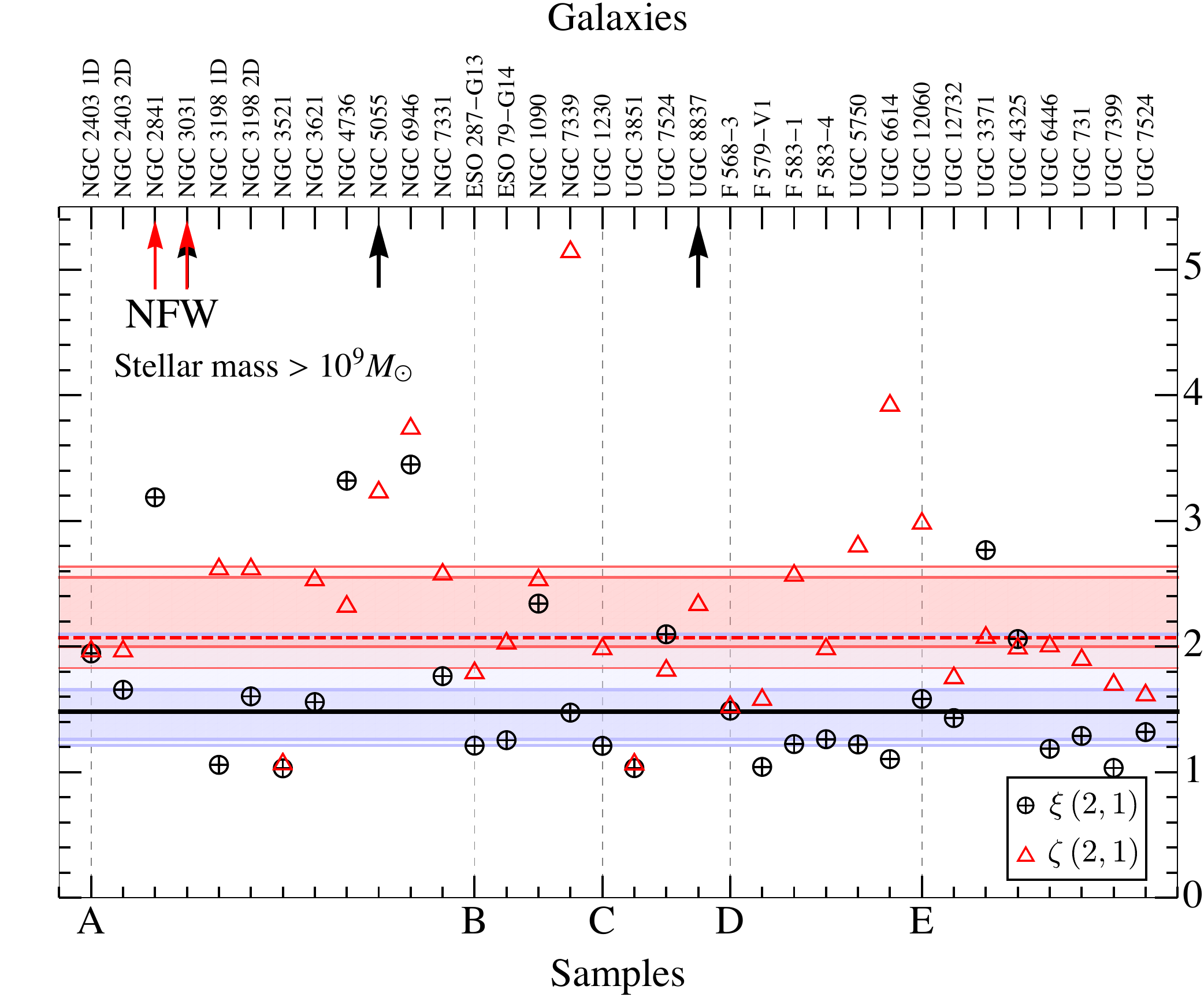}
    \end{subfigure}
	\hspace*{0.2cm}
    \begin{subfigure}
	    \centering 
        \includegraphics[trim = 0cm 0cm 0cm 0cm, clip=true, width=0.45\textwidth]{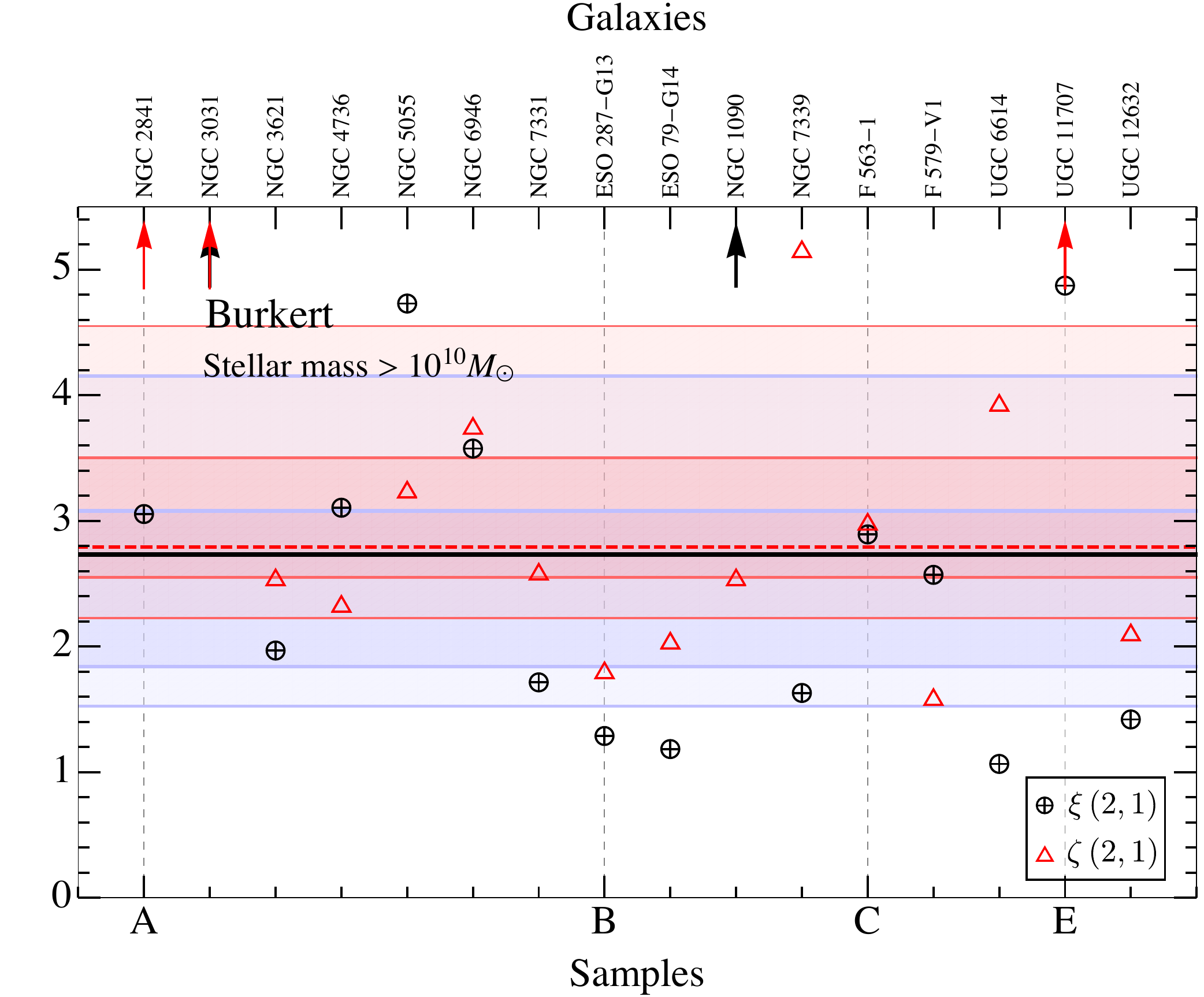}
    \end{subfigure} \hspace*{-0.75cm}
    \begin{subfigure}
	    \centering 
        \includegraphics[width=0.45\textwidth]{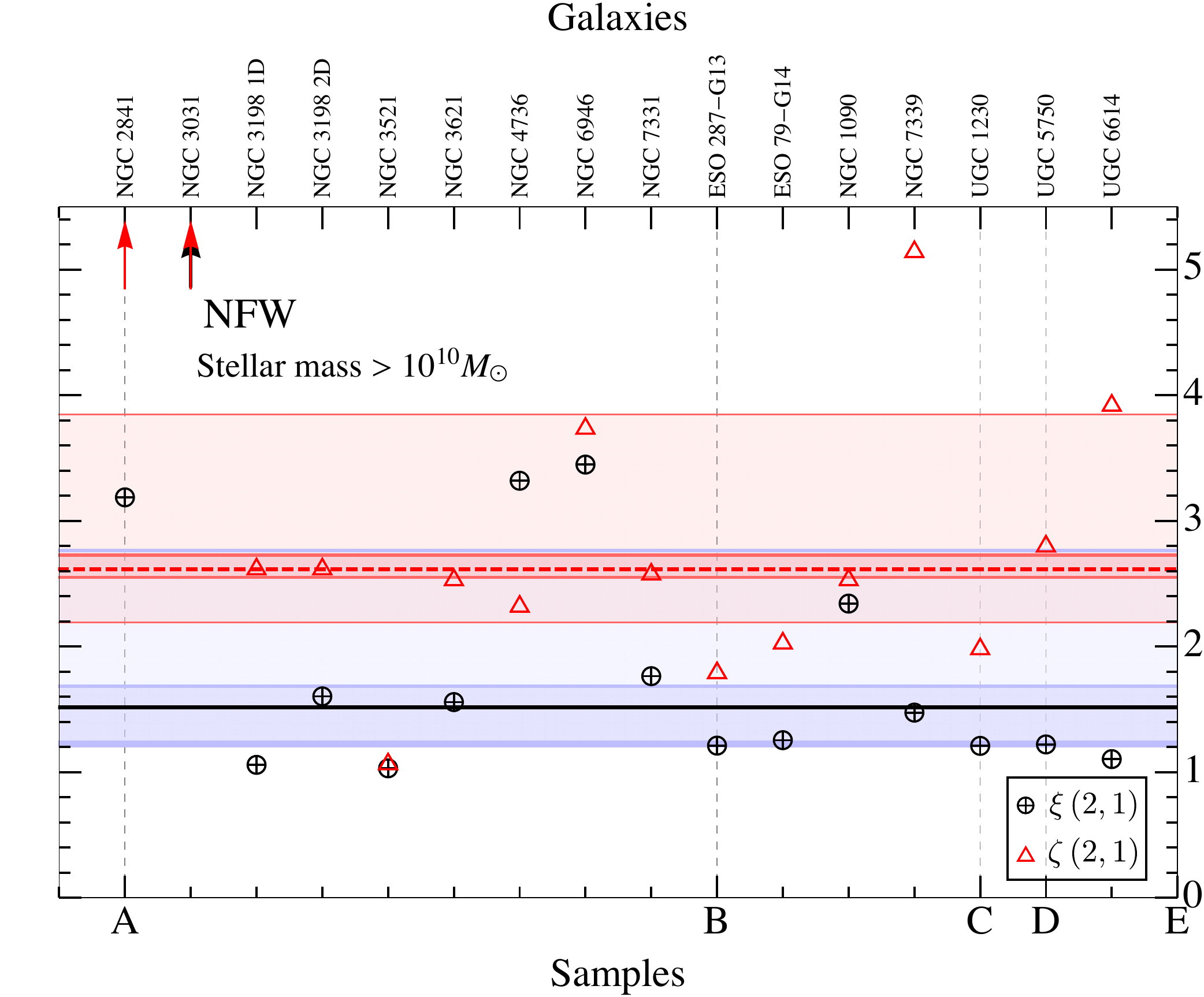}
    \end{subfigure}
    \caption{Plots that show the values of $\xi(2,1), \zeta(2,1)$, their medians and their dispersions. The dashed red and the solid black lines show respectively the values of $\langle \zeta(2,1) \rangle$ and  $\langle \xi(2,1) \rangle$. The lighter and darker red regions are respectively  the regions between  $\sigma^-_{50\%}(\zeta(2,1))$ and $\sigma^+_{50\%}(\zeta(2,1))$, and between $\sigma^-_{25\%}(\zeta(2,1))$ and $\sigma^+_{25\%}(\zeta(2,1))$. The darker and lighter blue regions follow analogously, but for $\xi(2,1)$. The two plots in the first line refer to the total sample $\cal S$, and those in the second and third lines refer respectively to the samples ${\cal S}_{*1}$ and ${\cal S}_{*2}$. The arrows indicate data whose corresponding values are outside the plotted region. The plots above show that $\langle \xi(2,1) \rangle \approx \langle \zeta(2,1) \rangle $ for the Burkert fits, while $\langle \xi(2,1) \rangle < \langle \zeta(2,1) \rangle $ for the NFW fits.} \label{fig:bluered1}
\end{figure*}

\begin{figure*}
    \begin{subfigure}
	    \centering 
        \includegraphics[trim = 0cm 0cm 0cm 0cm, clip=true, width=0.39\textwidth]{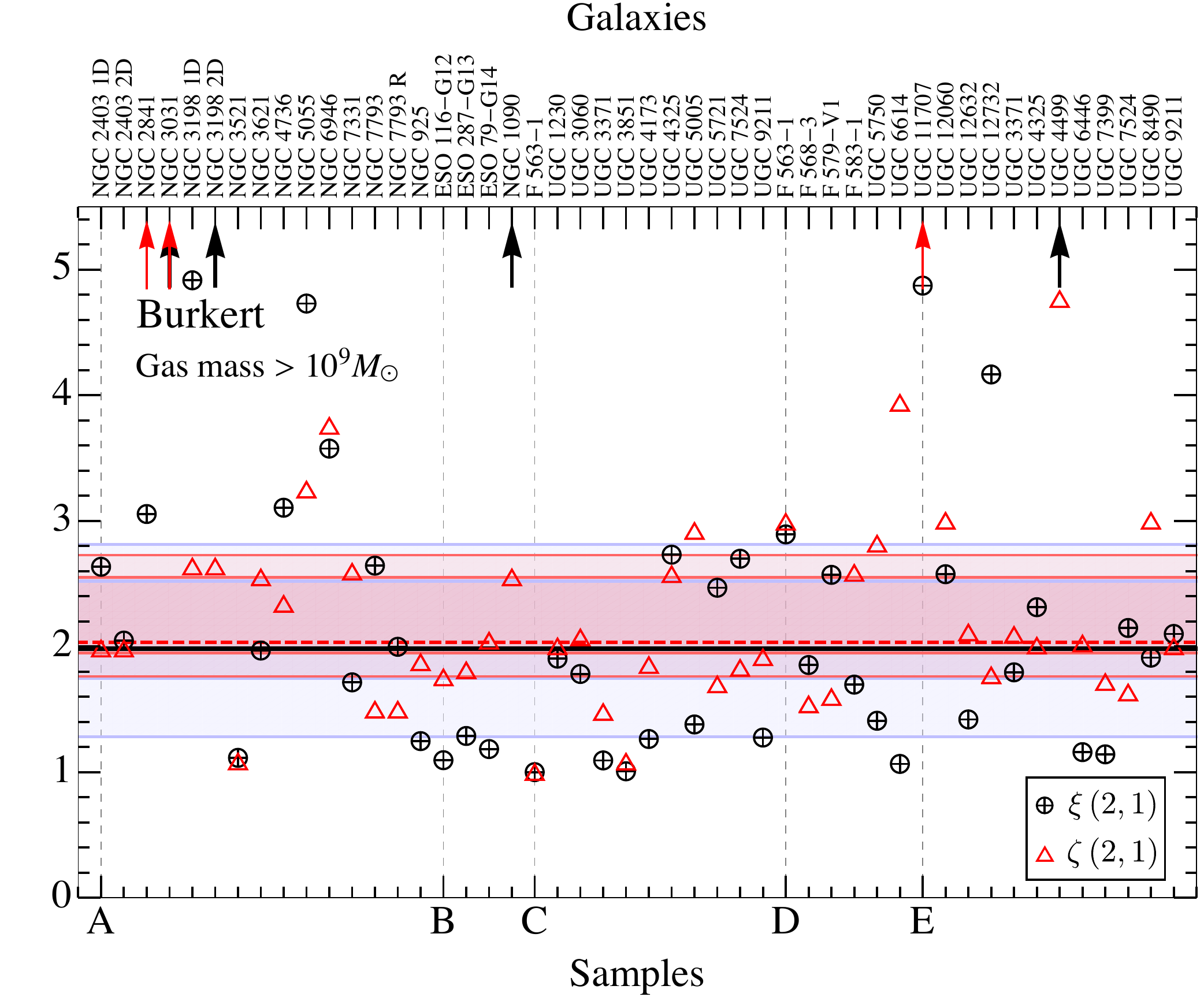}
   \end{subfigure} \hspace*{-0.66cm}
    \begin{subfigure}
	    \centering 
        \includegraphics[width=0.39\textwidth]{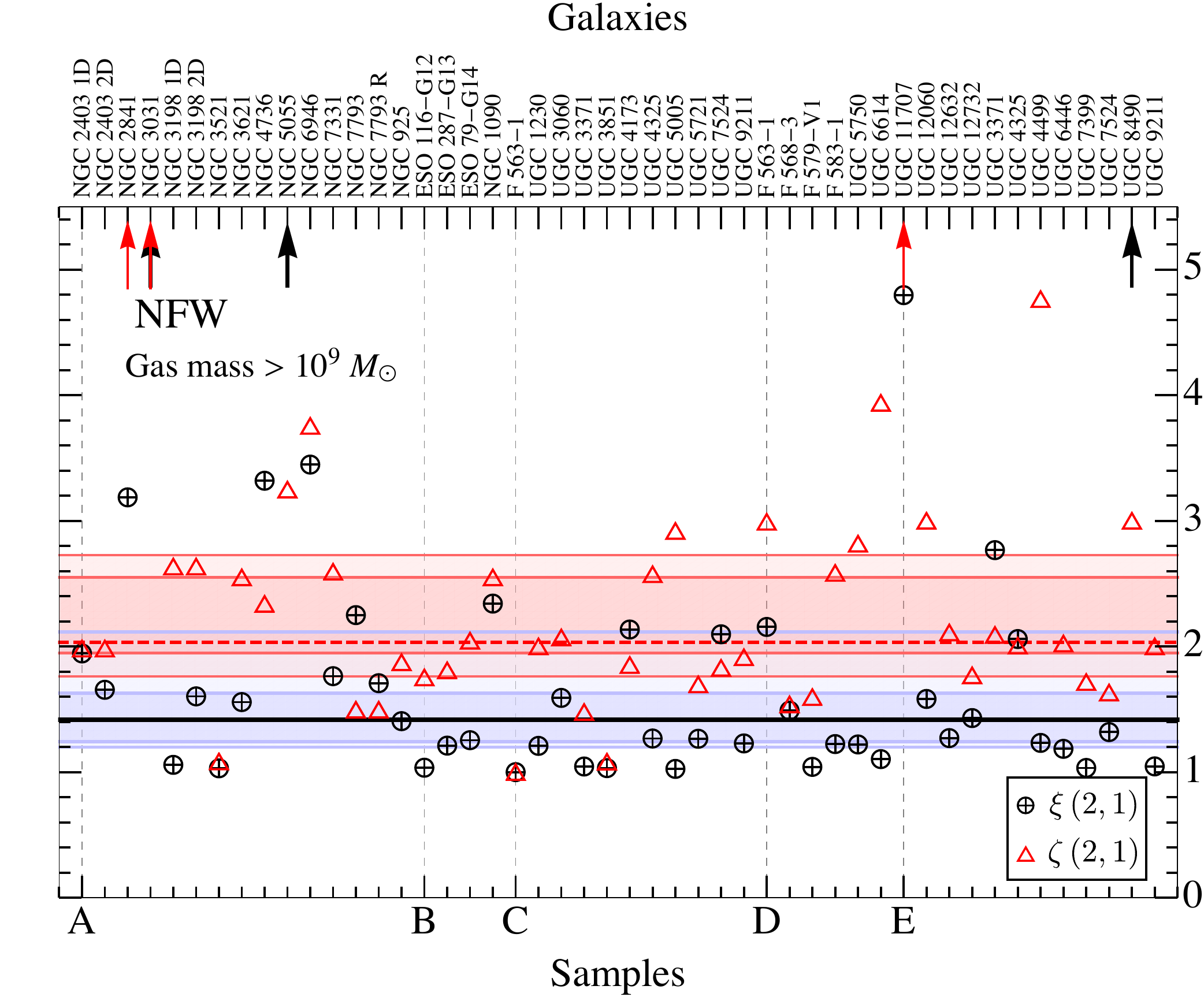}
    \end{subfigure}
    \begin{subfigure}
	    \centering 
        \includegraphics[trim = 0cm 0cm 0cm 0cm, clip=true, width=0.39\textwidth]{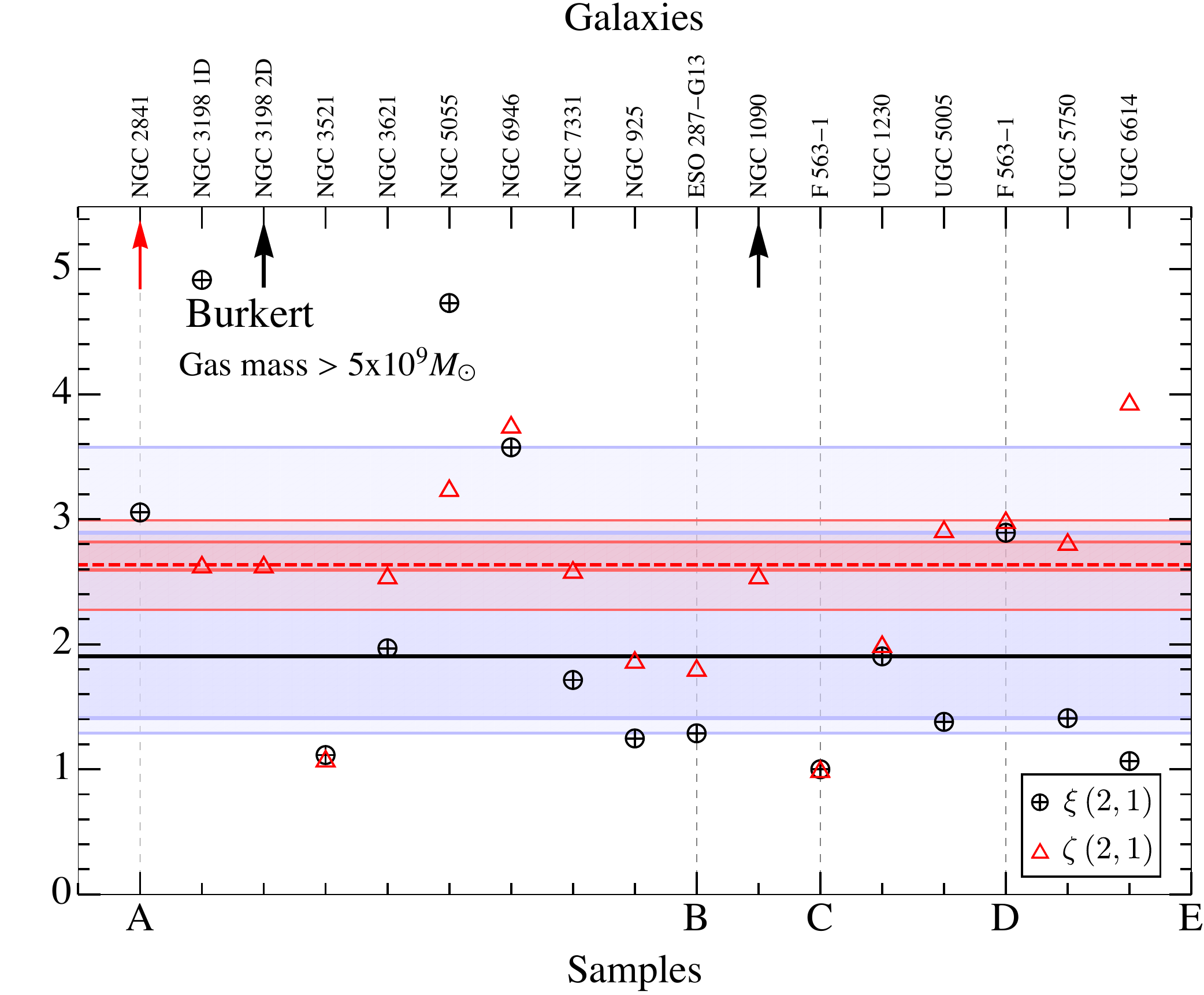}
    \end{subfigure}\hspace*{-0.61cm}
    \begin{subfigure}
	    \centering 
        \includegraphics[width=0.39\textwidth]{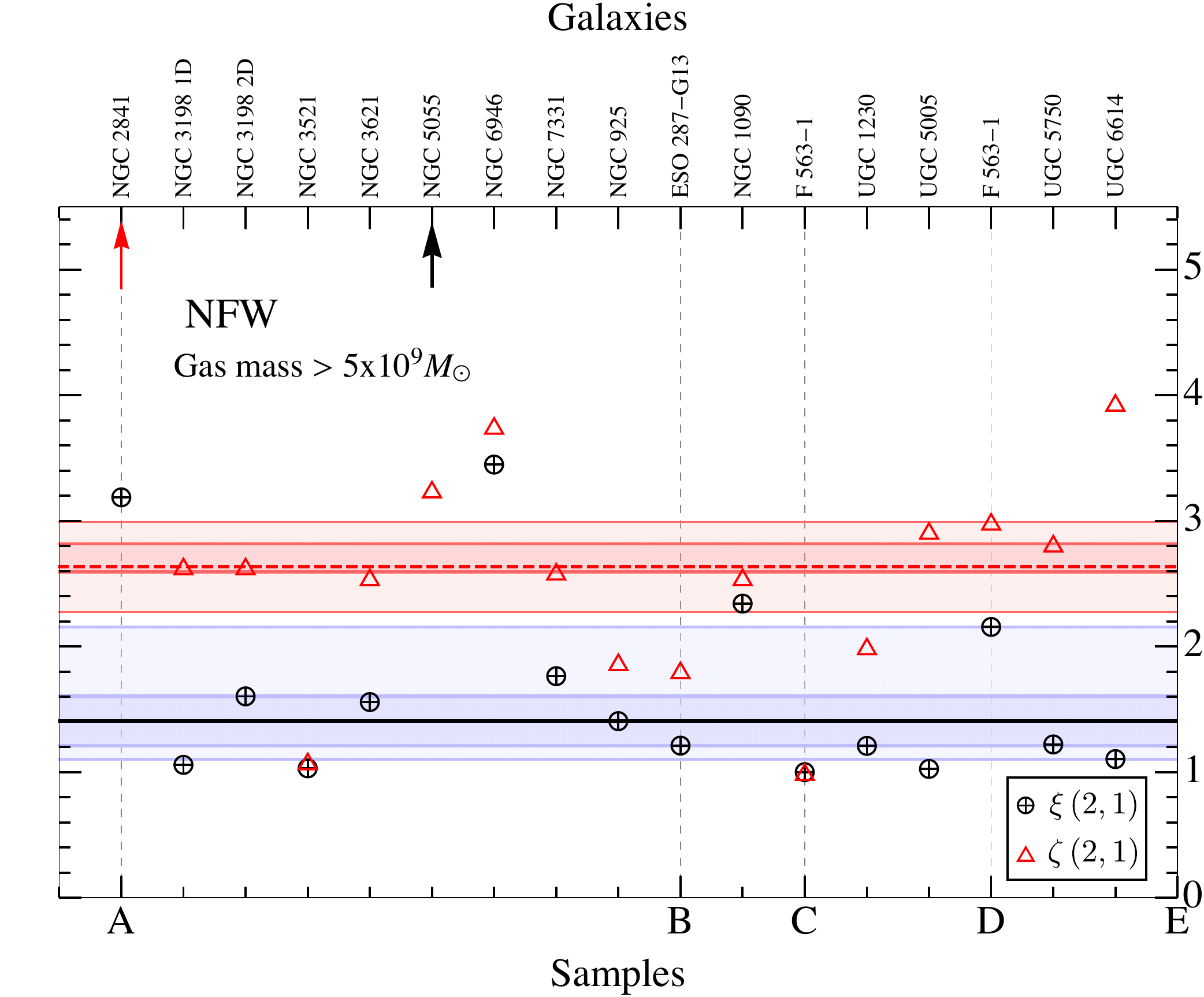}
    \end{subfigure}
    \begin{subfigure}
	    \centering 
        \includegraphics[trim = 0cm 0cm 0cm 0cm, clip=true, width=0.39\textwidth]{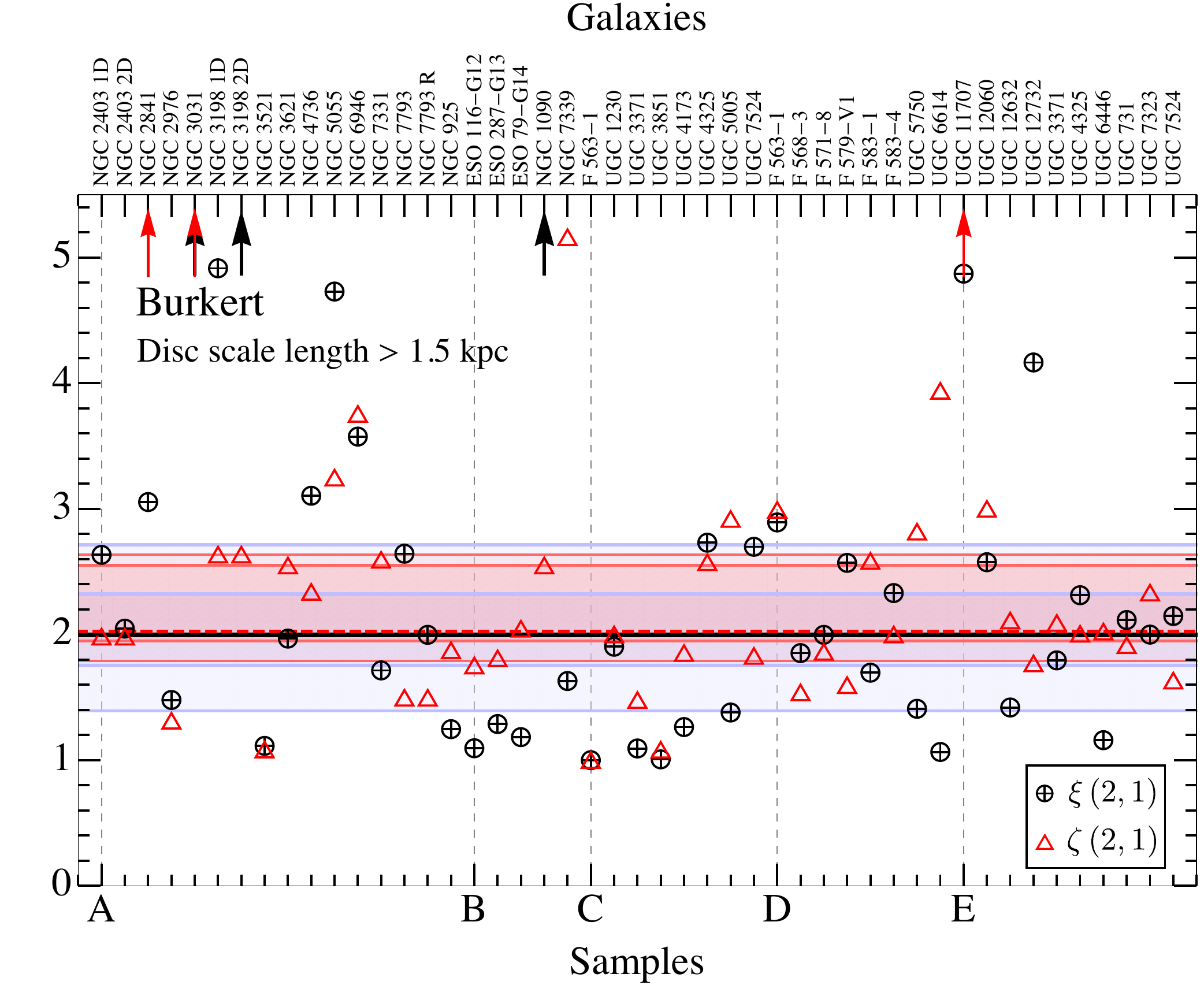}
    \end{subfigure} \hspace*{-0.67cm}
    \begin{subfigure}
	    \centering 
        \includegraphics[width=0.39\textwidth]{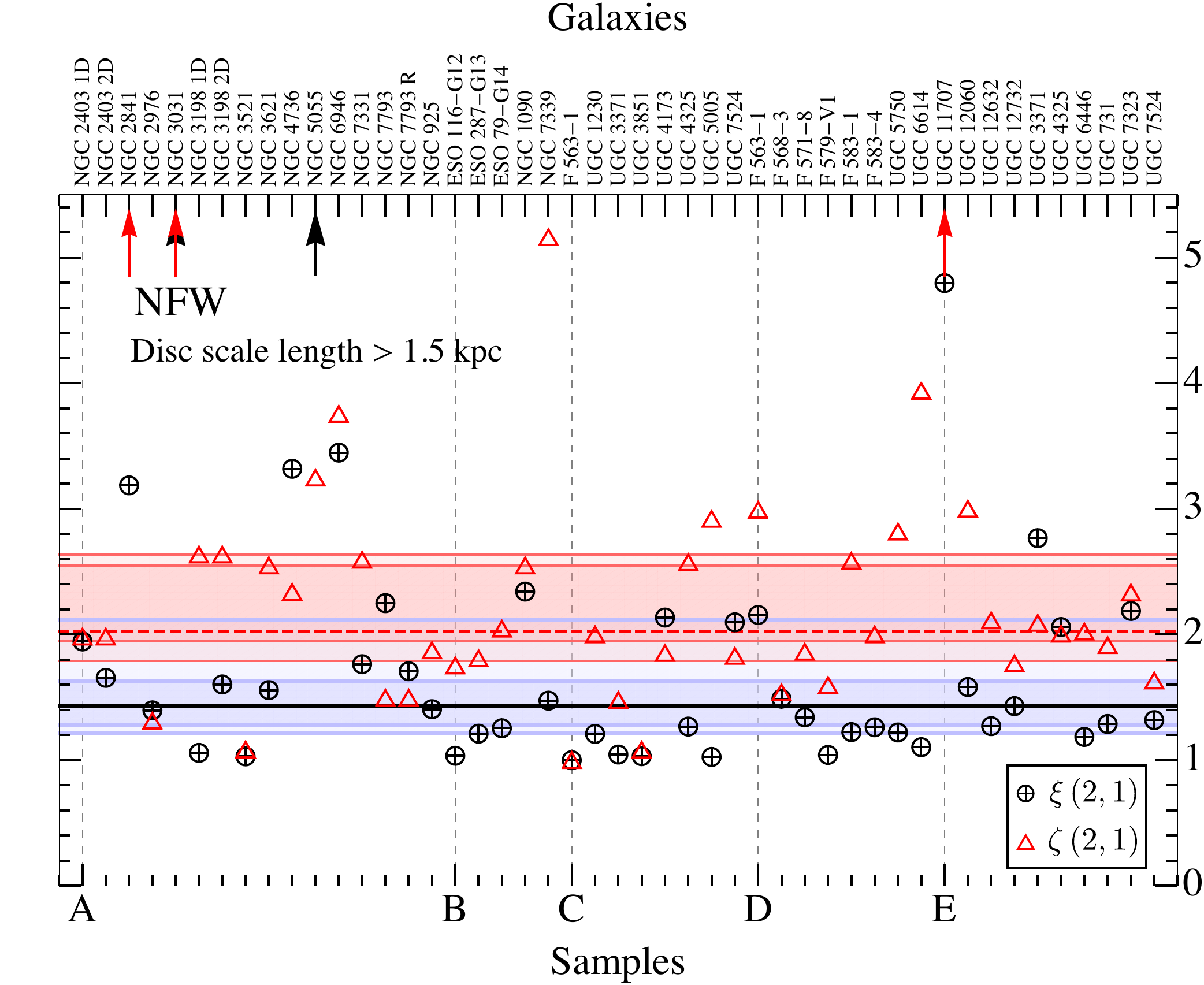}
    \end{subfigure}
	\hspace*{0.1cm}
	\begin{subfigure}
	    \centering 
        \includegraphics[trim = 0cm 0cm 0cm 0cm, clip=true, width=0.39\textwidth]{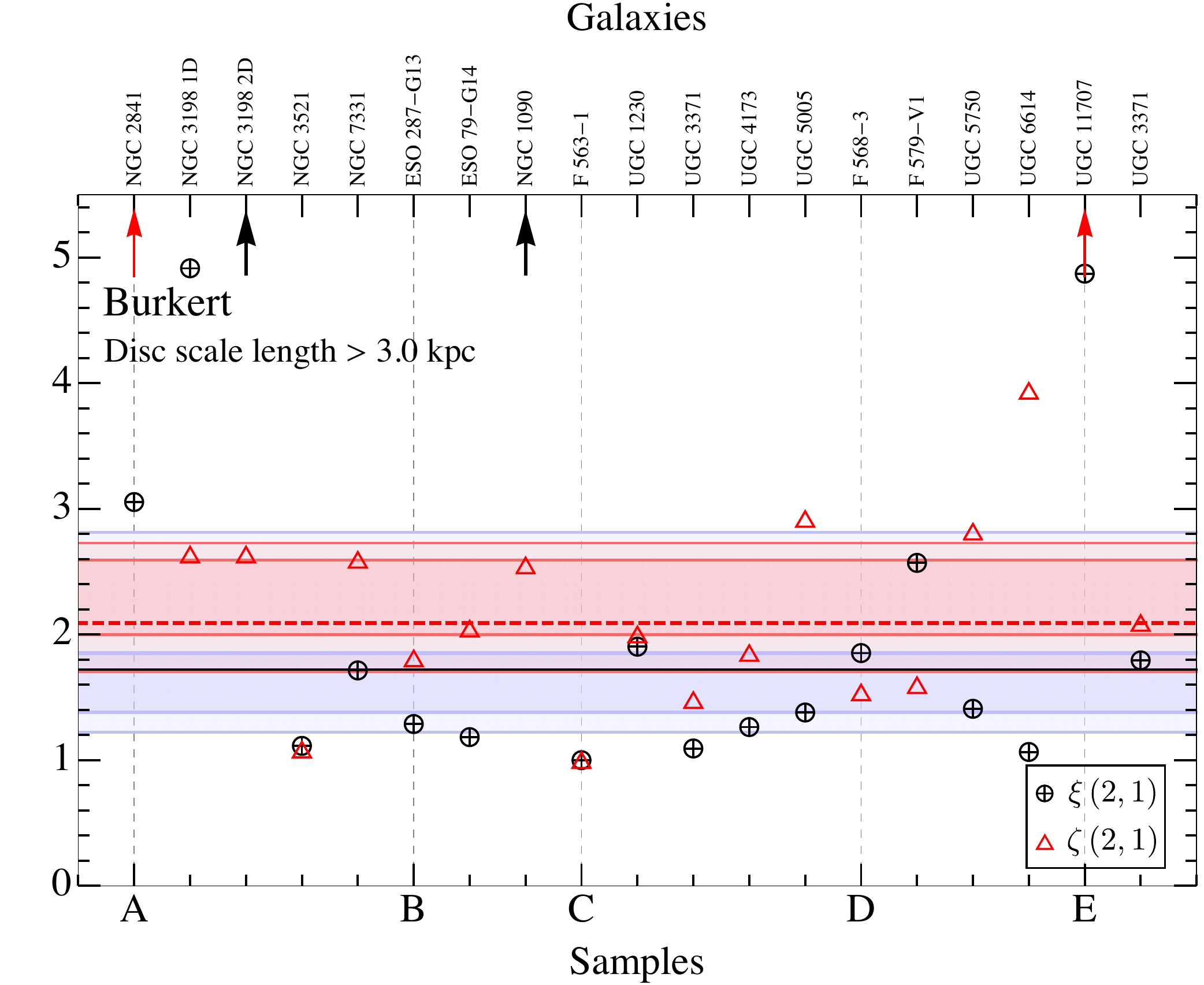}
    \end{subfigure} \hspace*{-0.67cm}
    \begin{subfigure}
	    \centering 
        \includegraphics[width=0.39\textwidth]{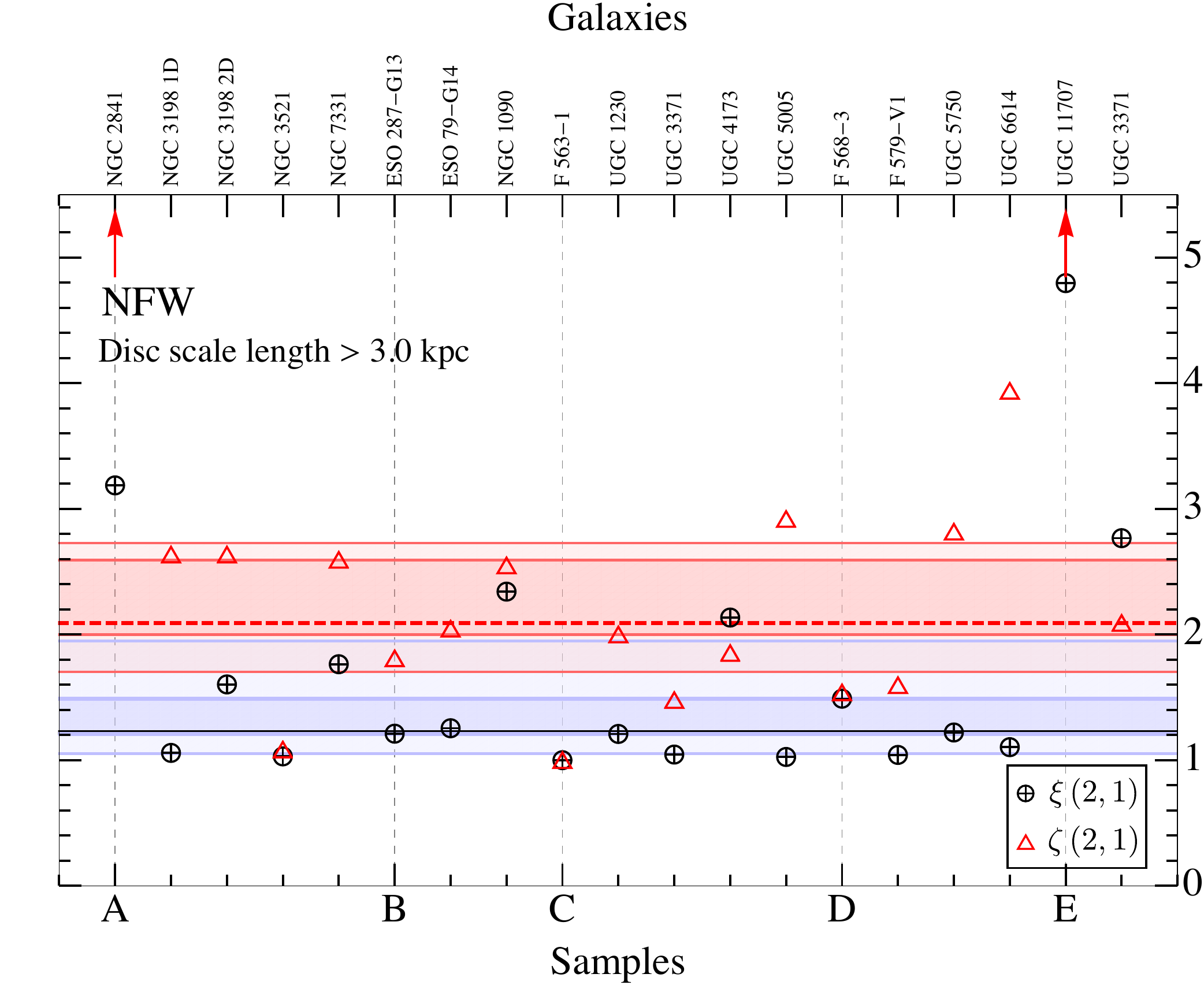}
    \end{subfigure}
    \caption{These plots show the values of $\xi(2,1), \zeta(2,1)$, their medians and their dispersions. The symbols follow the same conventions of Fig.~\ref{fig:bluered1}.
From top to bottom, the subsample relative to a given row is, respectively, ${\cal S}_\mscript{g1}$, ${\cal S}_\mscript{g2}$, ${\cal S}_\mscript{h1}$ and ${\cal S}_\mscript{h2}$.} \label{fig:bluered2}.
\end{figure*}

\begin{figure*}
    \begin{subfigure}
	    \centering 
        \includegraphics[trim = 0cm 0cm 0cm 0cm, clip=true, width=0.45\textwidth]{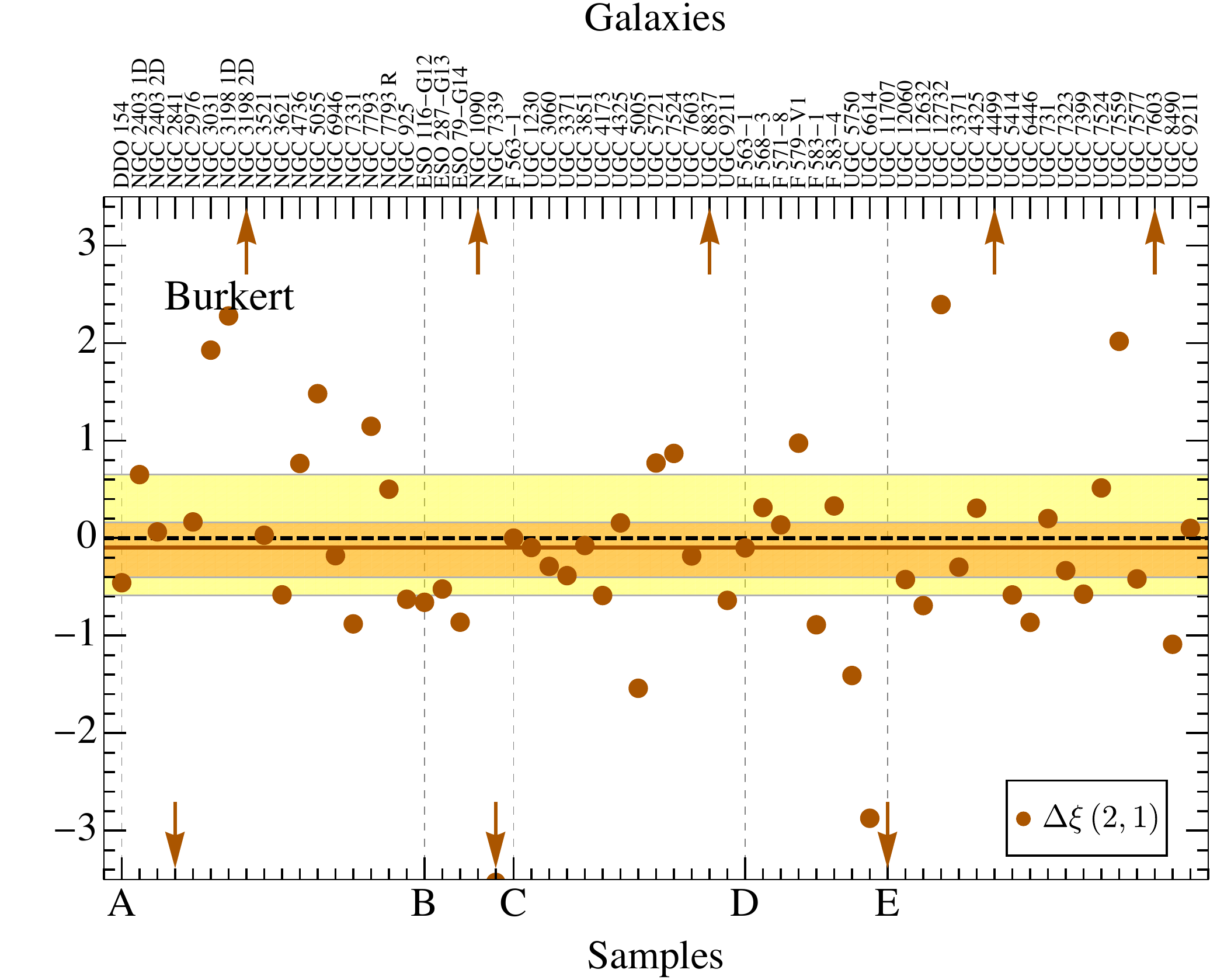}
   \end{subfigure} \hspace*{-0.75cm}
    \begin{subfigure}
	    \centering 
        \includegraphics[width=0.45\textwidth]{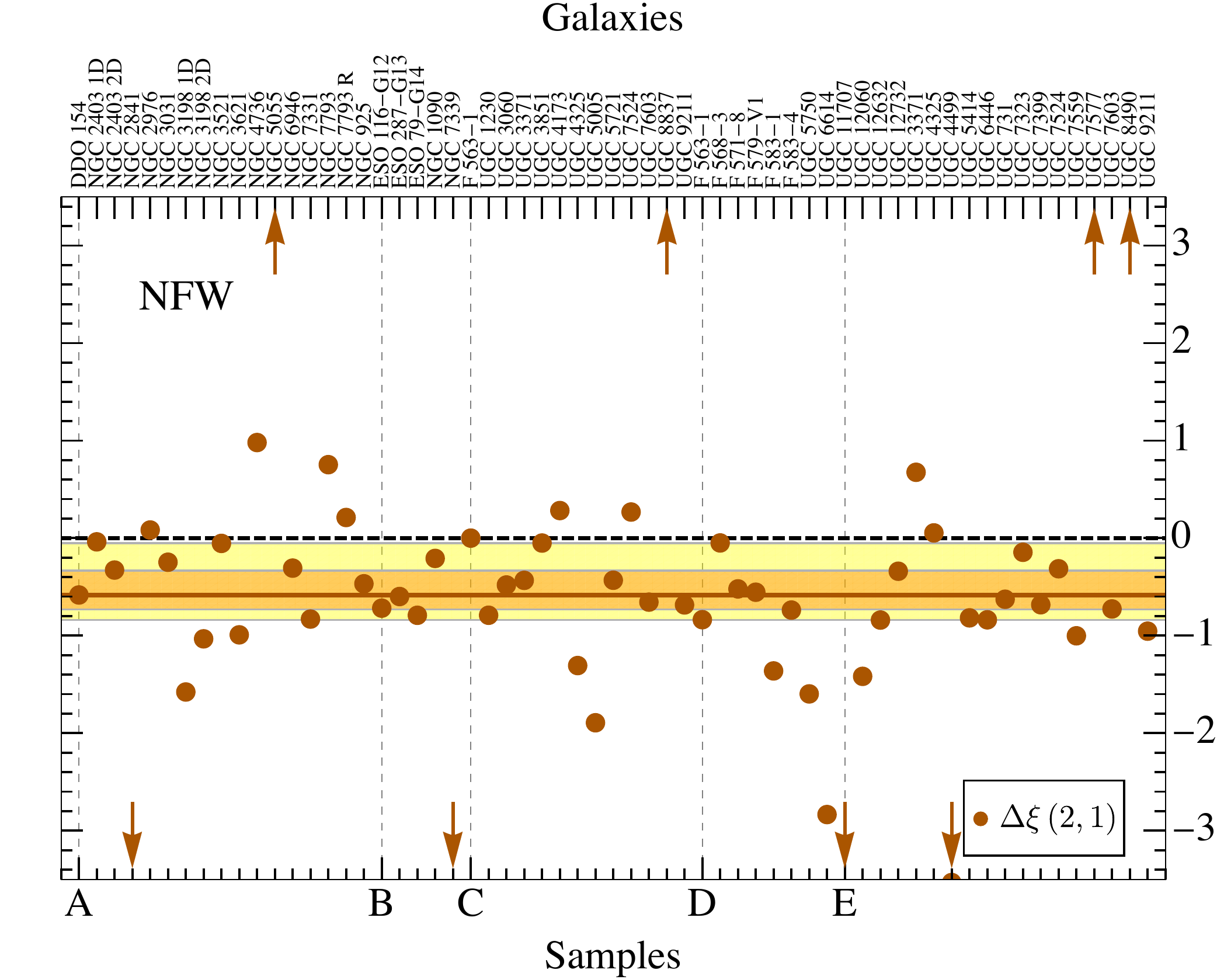}
    \end{subfigure}
    \begin{subfigure}
	    \centering 
        \includegraphics[trim = 0cm 0cm 0cm 0cm, clip=true, width=0.45\textwidth]{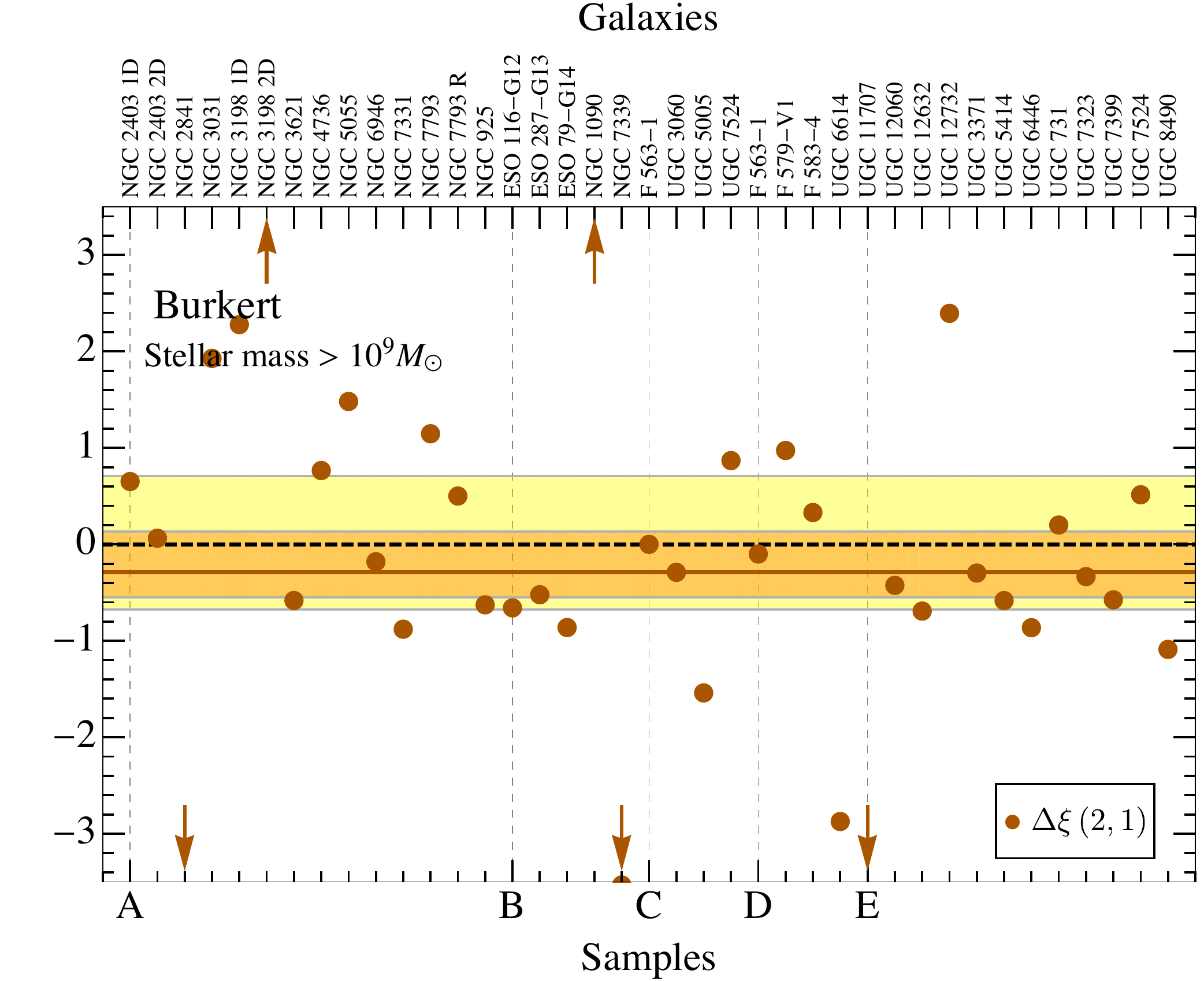}
    \end{subfigure} \hspace*{-0.75cm}
    \begin{subfigure}
	    \centering 
        \includegraphics[width=0.45\textwidth]{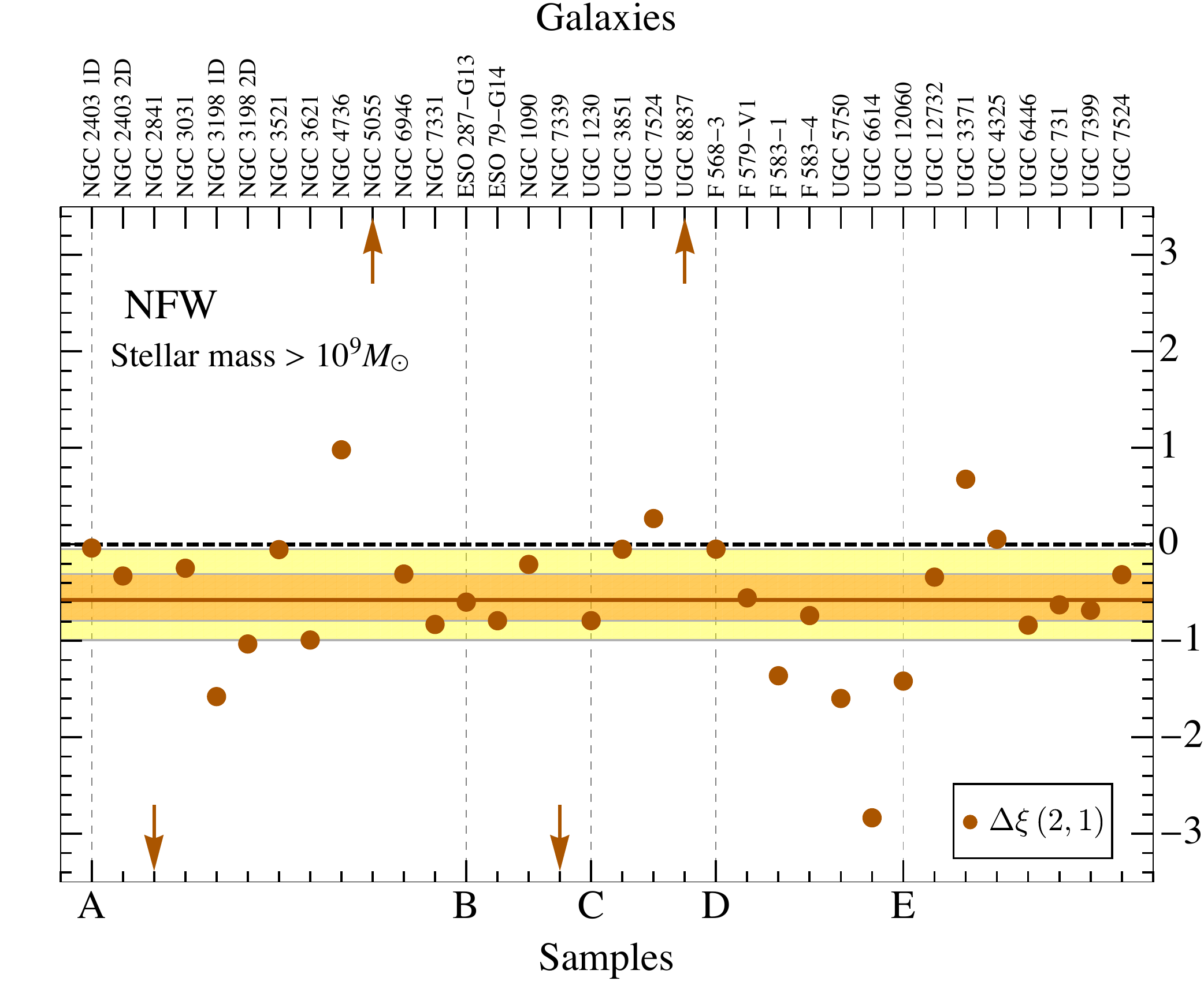}
    \end{subfigure}
	\hspace*{0.2cm}
    \begin{subfigure}
	    \centering 
        \includegraphics[trim = 0cm 0cm 0cm 0cm, clip=true, width=0.45\textwidth]{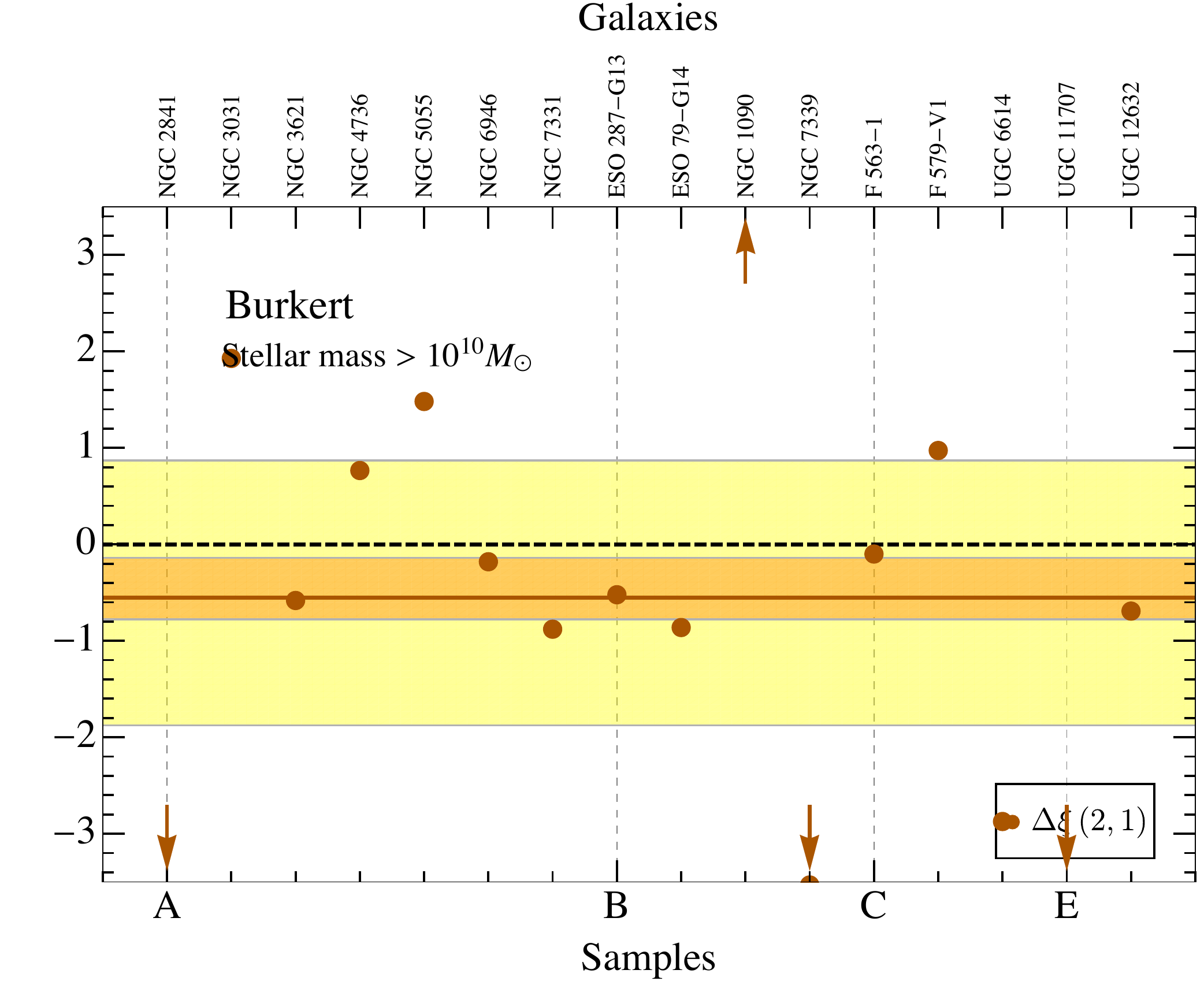}
    \end{subfigure} \hspace*{-0.75cm}
    \begin{subfigure}
	    \centering 
        \includegraphics[width=0.45\textwidth]{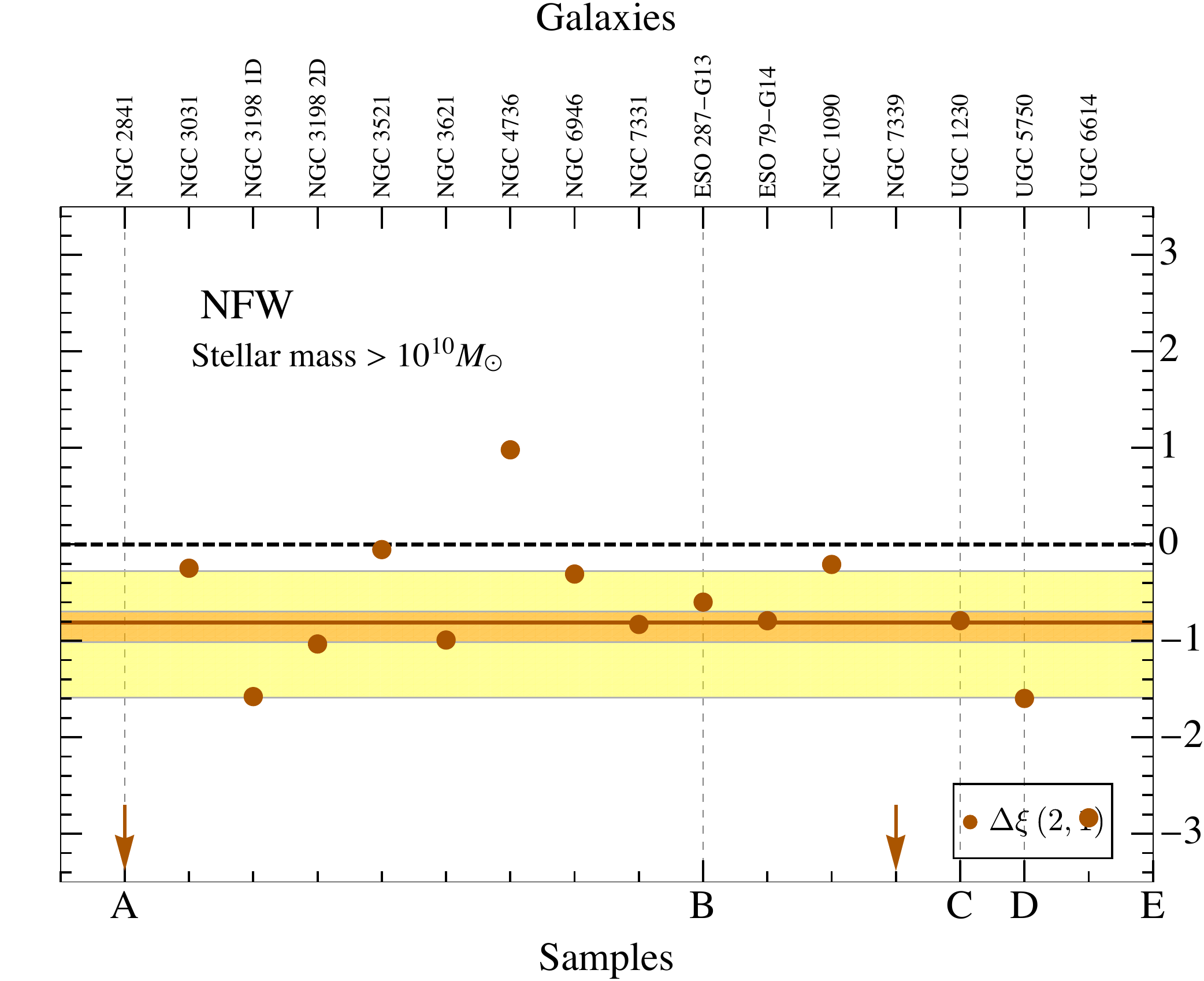}
    \end{subfigure}
    \caption{Plots that show the values of $\Delta\xi(2,1)$, its median and dispersion. The solid brown and the dashed black lines show respectively the value of $\langle \Delta \xi(2,1) \rangle$ and its expected value, i.e. zero. The lighter and darker yellow regions are  the dispersions derived from $\sigma^\pm_{50\%}(\Delta\xi(2,1))$ and $\sigma^\pm_{25\%}(\Delta \xi(2,1))$ respectively. See also Fig.~\ref{fig:bluered1}. These plots are consistent with $\langle \Delta \xi(2,1) \rangle \approx 0$ (i.e., homogeneous fit) for the Burkert profile and  $\langle \Delta \xi(2,1) \rangle < 0$ for the NFW profile.} \label{fig:yellow1}
\end{figure*}

\begin{figure*}
    \begin{subfigure}
	    \centering 
        \includegraphics[trim = 0cm 0cm 0cm 0cm, clip=true, width=0.39\textwidth]{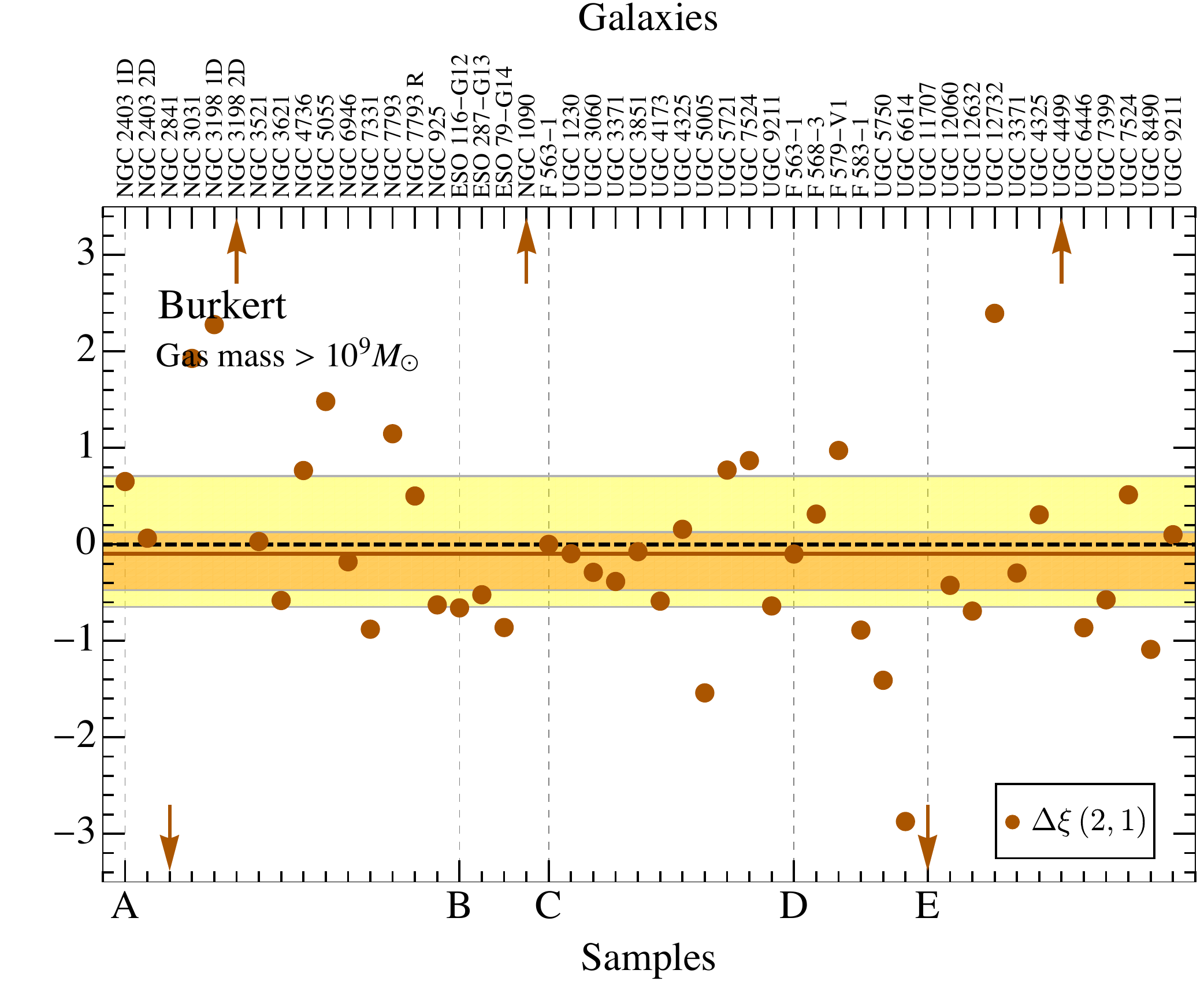}
   \end{subfigure} \hspace*{-0.67cm}
    \begin{subfigure}
	    \centering 
        \includegraphics[width=0.39\textwidth]{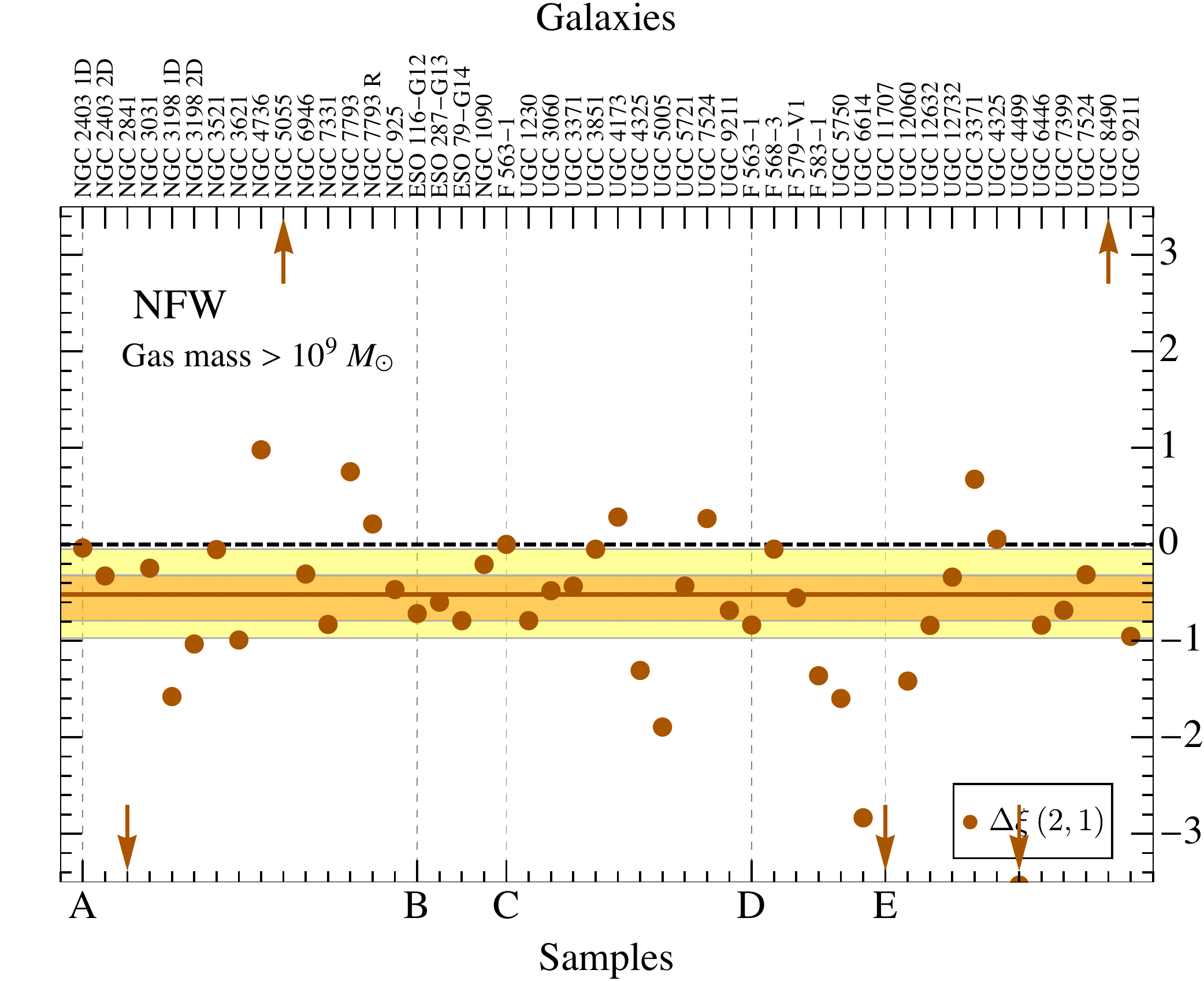}
    \end{subfigure}
    \begin{subfigure}
	    \centering 
        \includegraphics[trim = 0cm 0cm 0cm 0cm, clip=true, width=0.39\textwidth]{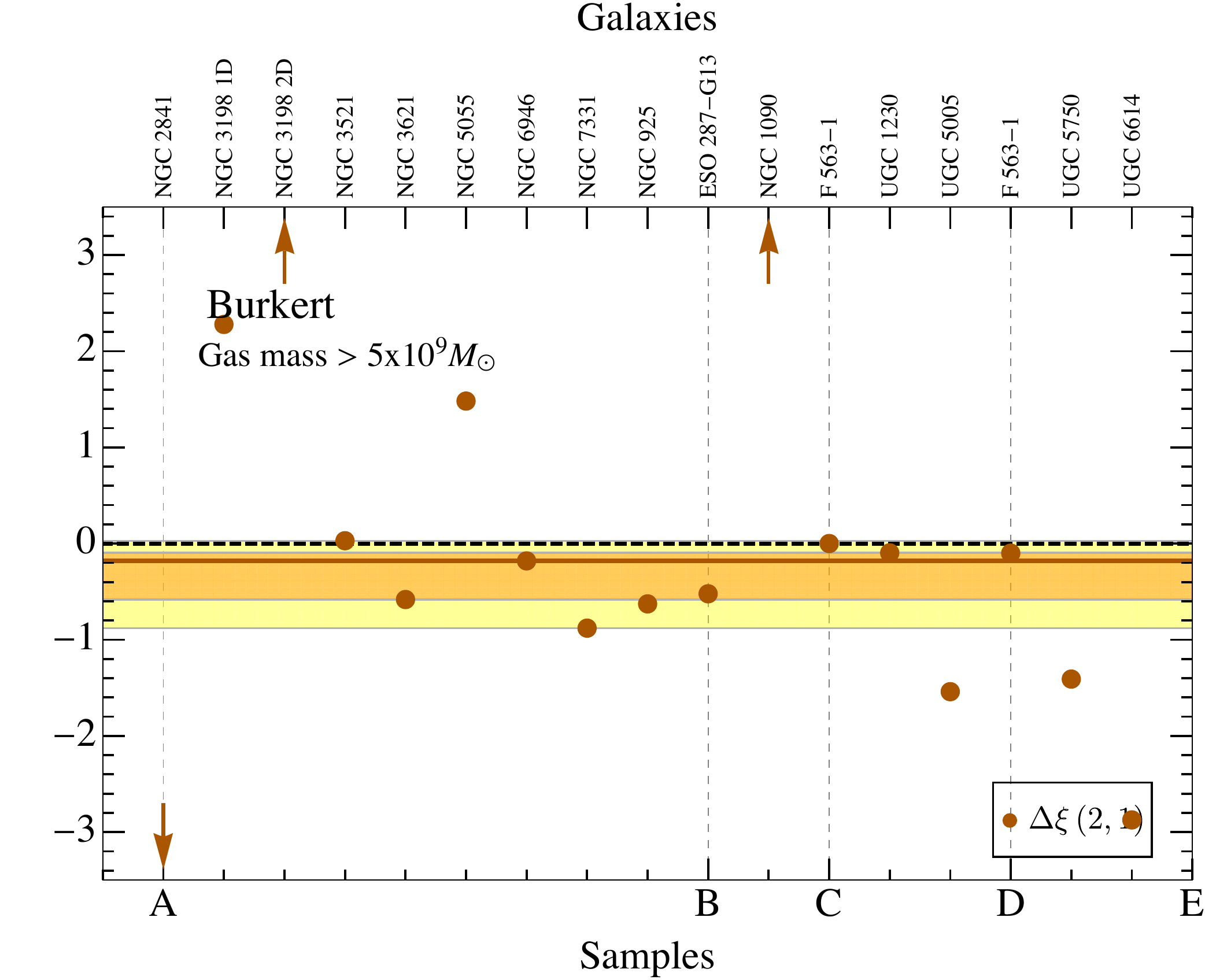}
    \end{subfigure} \hspace*{-0.68cm}
    \begin{subfigure}
	    \centering 
        \includegraphics[width=0.39\textwidth]{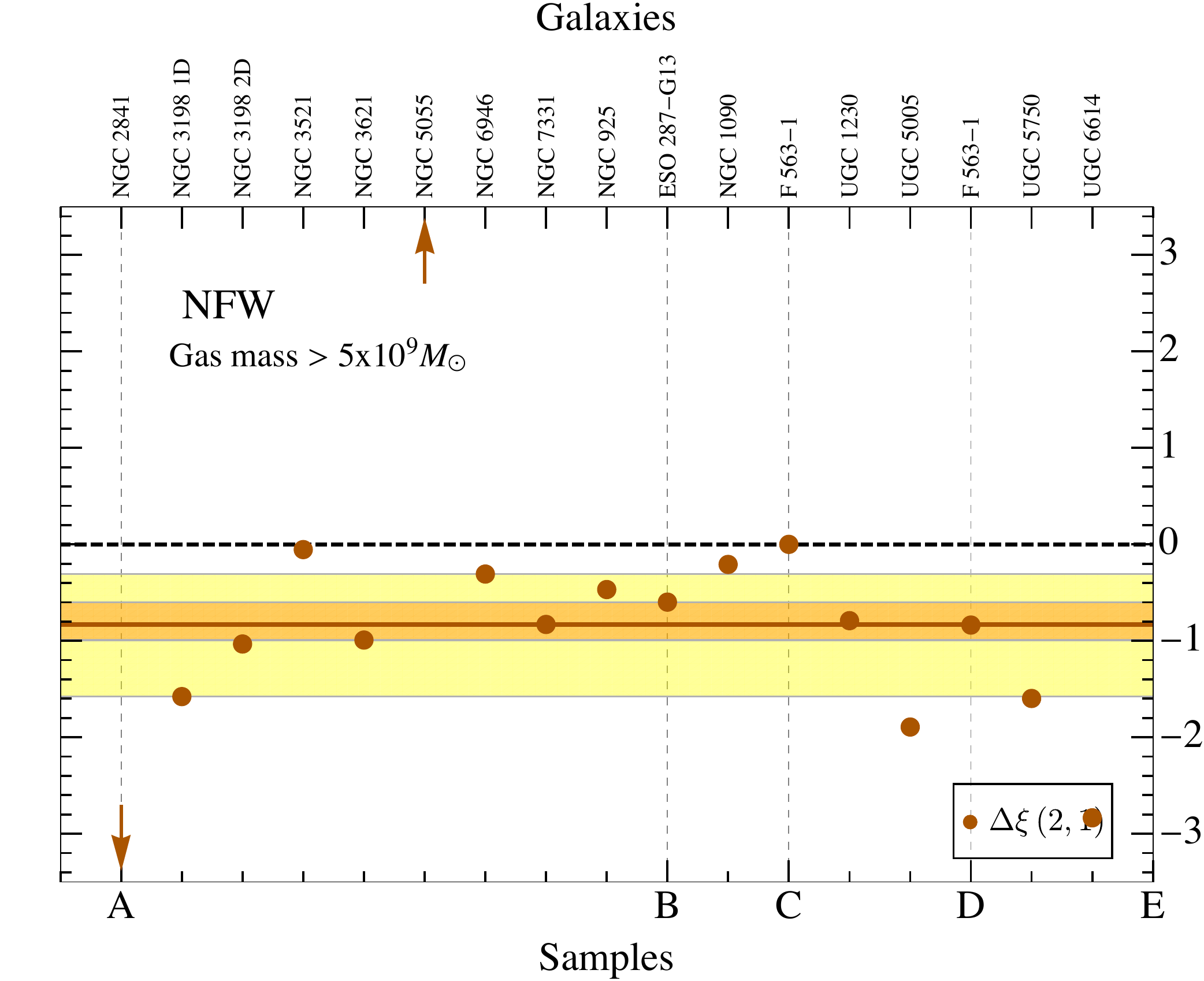}
    \end{subfigure}
    \begin{subfigure}
	    \centering 
        \includegraphics[trim = 0cm 0cm 0cm 0cm, clip=true, width=0.39\textwidth]{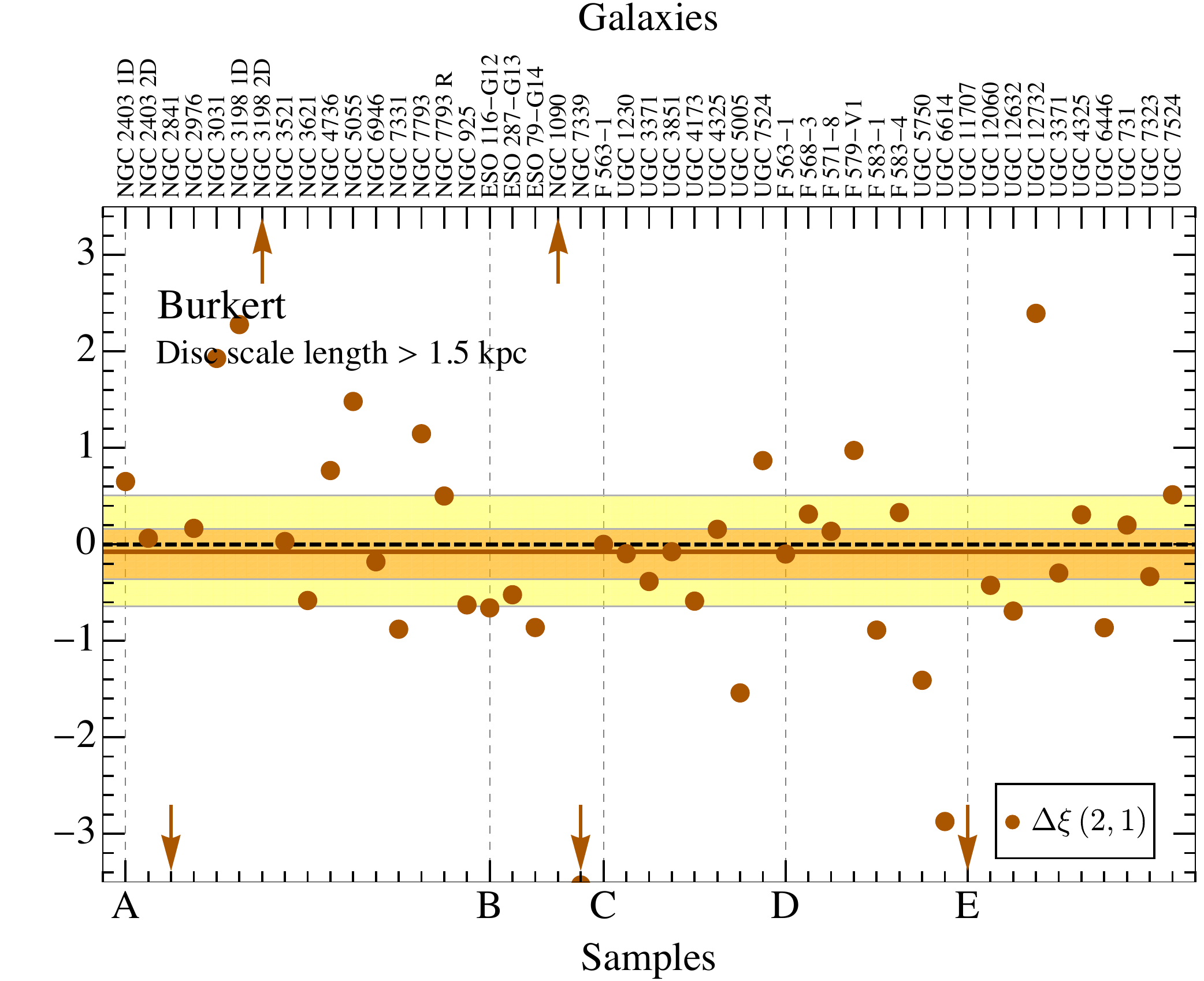}
    \end{subfigure} \hspace*{-0.67cm}
    \begin{subfigure}
	    \centering 
        \includegraphics[width=0.39\textwidth]{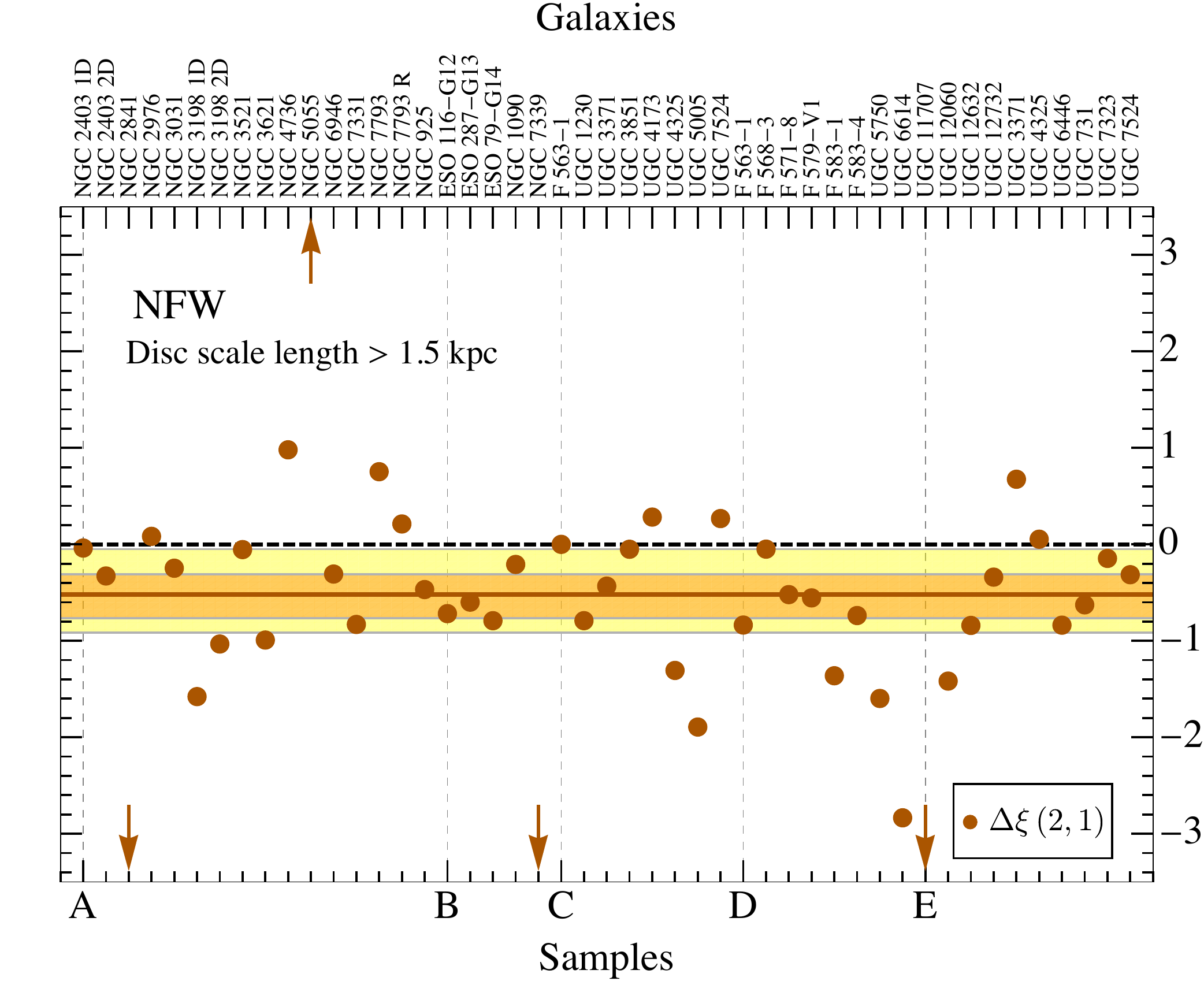}
    \end{subfigure}
    \hspace*{0.1cm}
	\begin{subfigure}
	    \centering 
        \includegraphics[trim = 0cm 0cm 0cm 0cm, clip=true, width=0.39\textwidth]{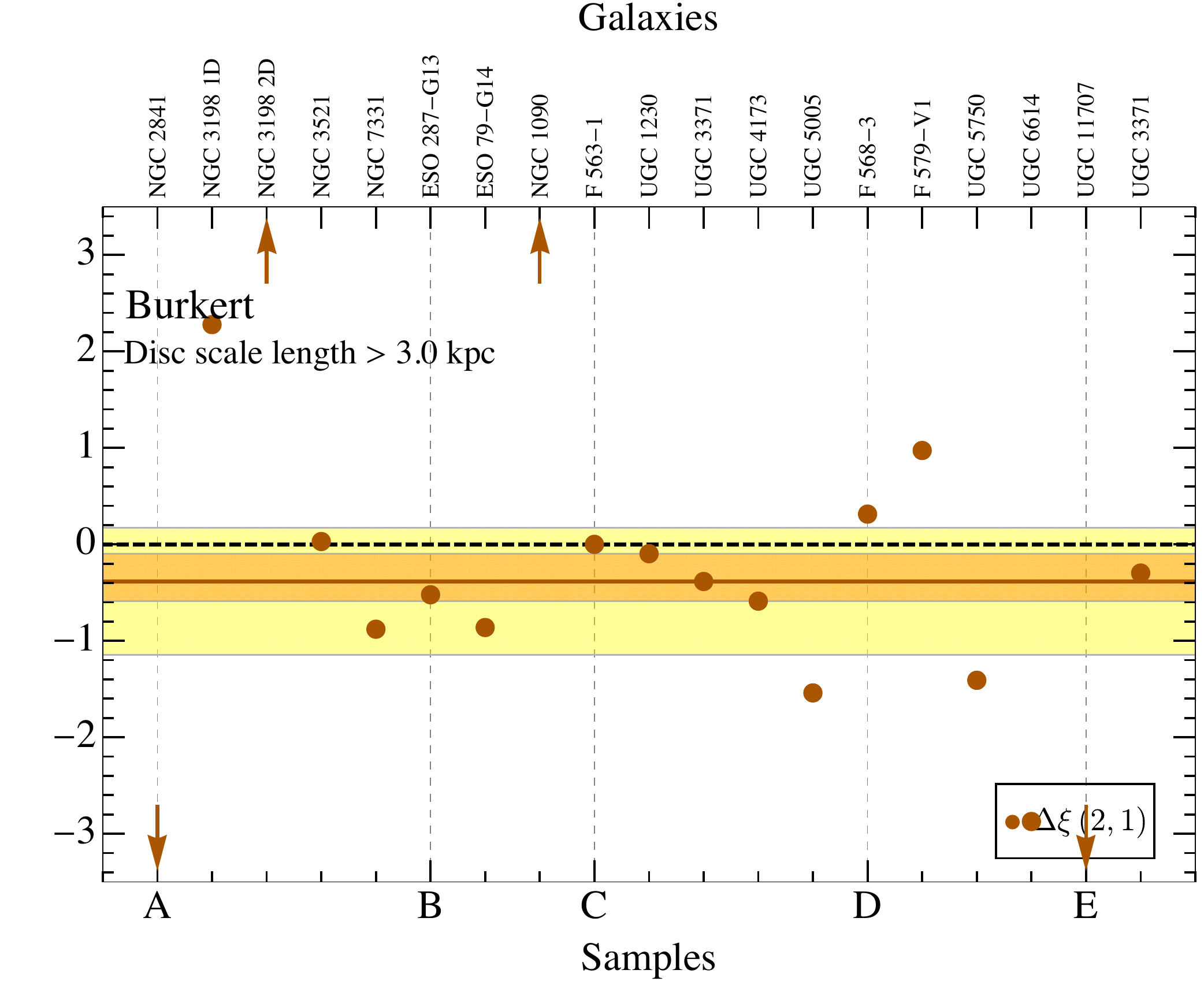}
    \end{subfigure} \hspace*{-0.67cm}
    \begin{subfigure}
	    \centering 
        \includegraphics[width=0.39\textwidth]{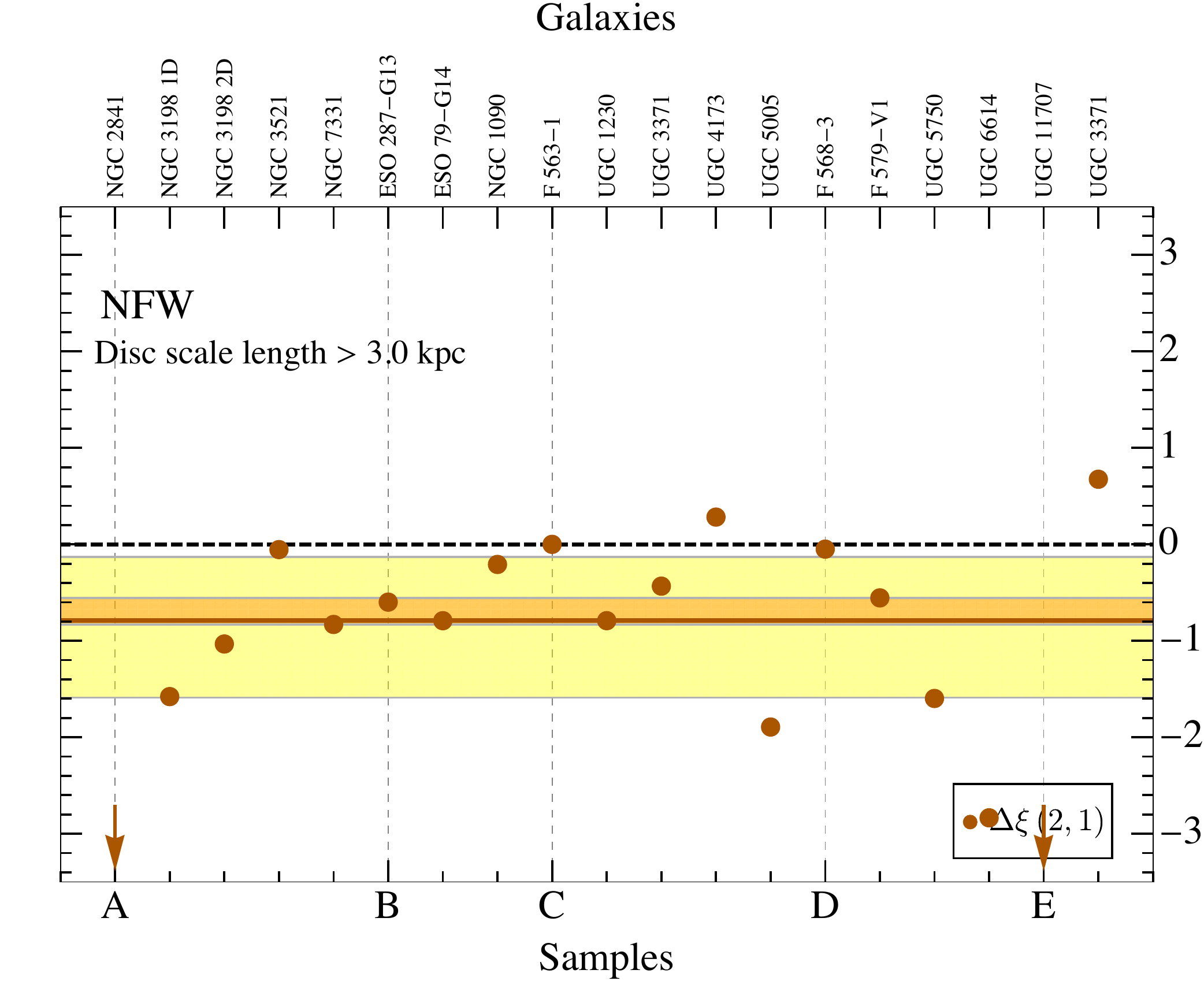}
    \end{subfigure}
    \caption{These plots show the values of $\Delta\xi(2,1)$, its median and dispersion. The symbols follow the same conventions of Fig.~\ref{fig:yellow1}. From top to bottom, the subsample relative to a given row is, respectively, ${\cal S}_\mscript{g1}$, ${\cal S}_\mscript{g2}$, ${\cal S}_\mscript{h1}$ and ${\cal S}_\mscript{h2}$.} \label{fig:yellow2}
\end{figure*}

Here we show in detail the plots of $\xi$, $\zeta$ and $\Delta \xi$ for all the subsamples considered in this work. These plots are in Figs. \ref{fig:bluered1}, \ref{fig:yellow1}, \ref{fig:bluered2} and  \ref{fig:yellow2}.

\section{The expected and the derived stellar mass-to-light ratios} \label{app:expectedY}

\begin{figure*}
    \begin{subfigure}
     	\centering 
        \includegraphics[trim = 0cm 0.00cm 0cm 0cm, clip=true, width=0.43\textwidth]{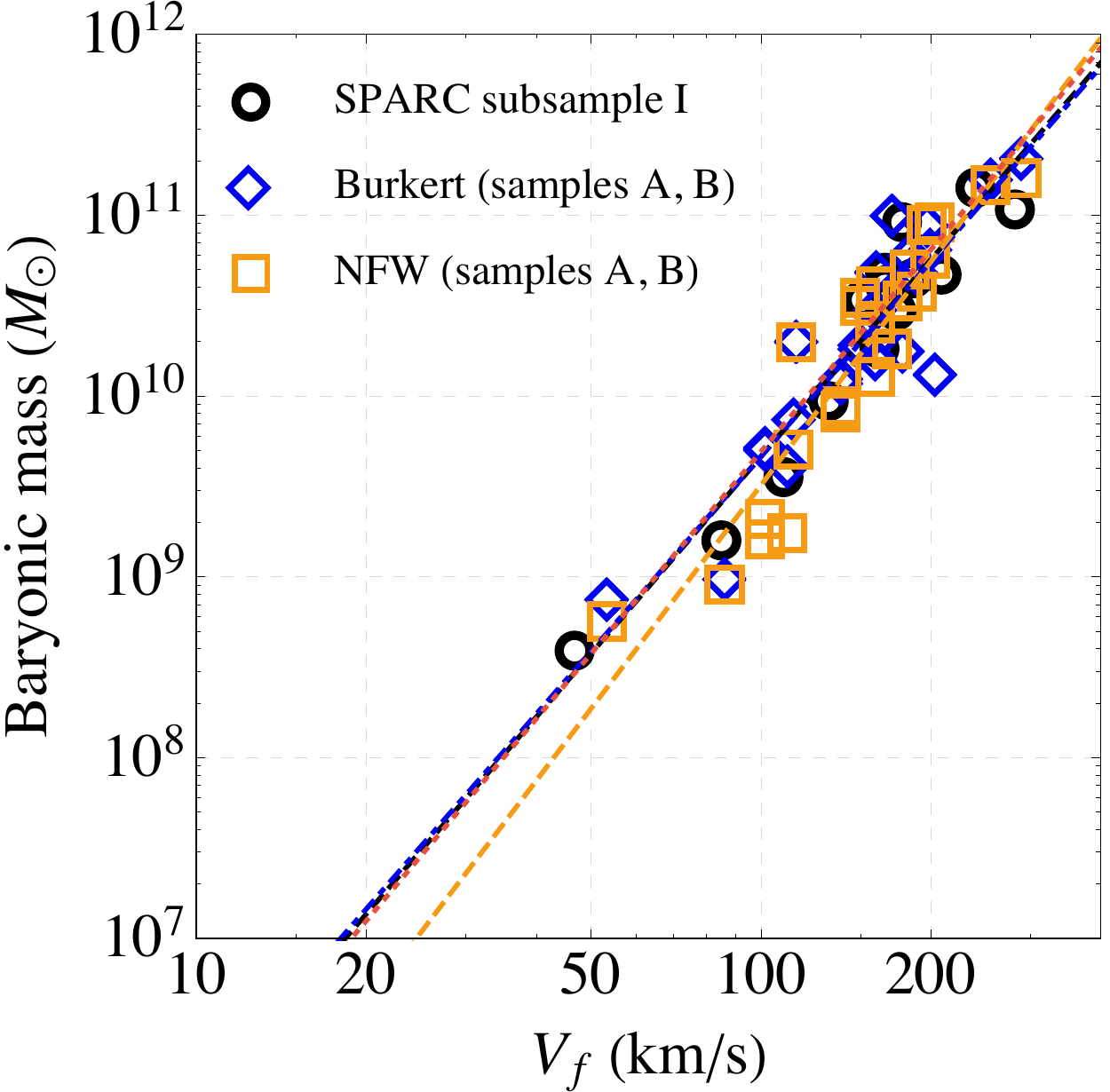} 
    \end{subfigure} \hspace*{0.5cm}
    \begin{subfigure}
     	\centering 
        \includegraphics[trim = 0.0cm 0cm 0cm 0.00cm, clip=true, width=0.43\textwidth]{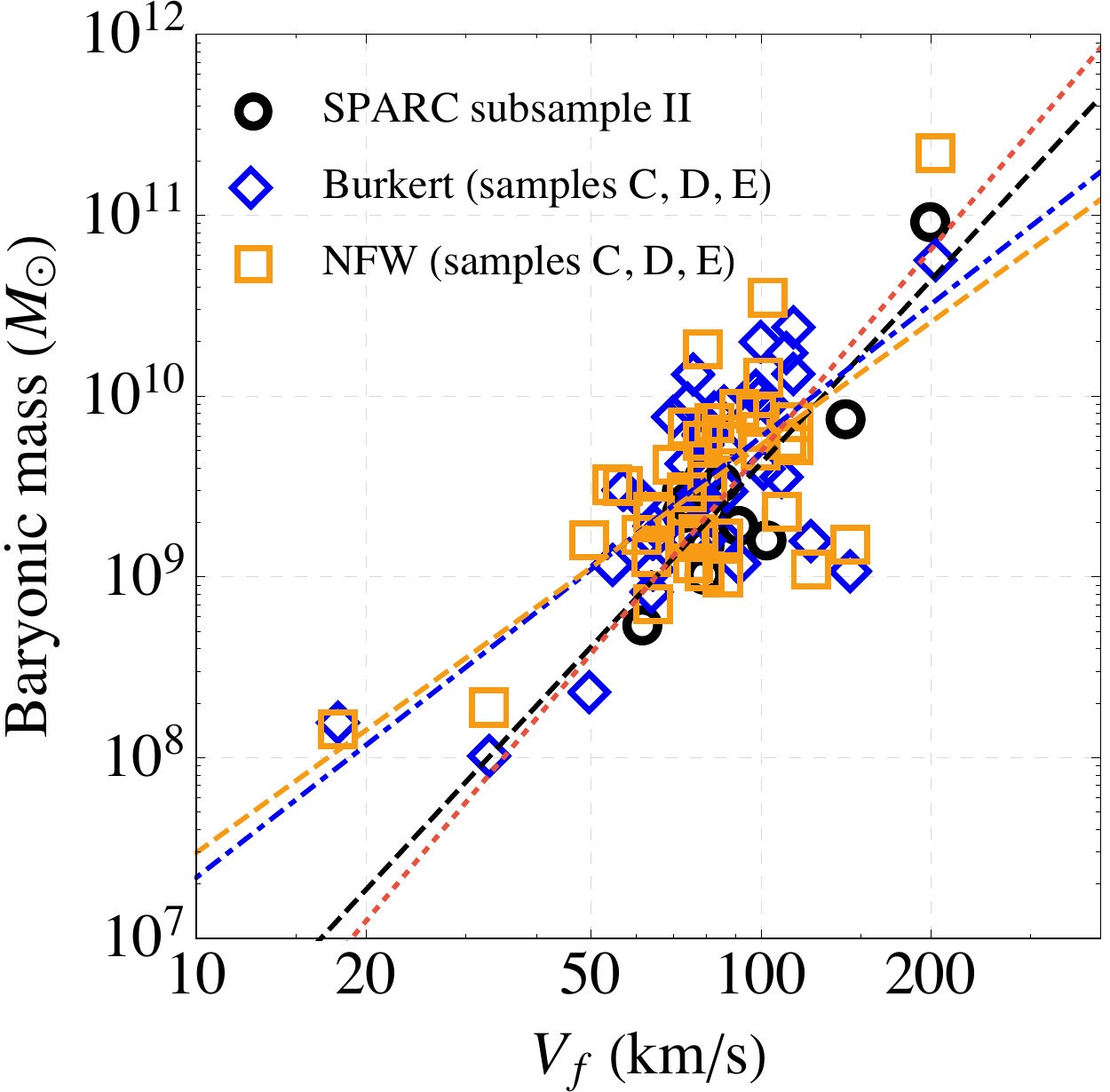}
    \end{subfigure} 
	\caption{A comparison between the  best fit $\Upsilon_*$, for the NFW and the Burkert profiles, and the BTFR. The left plot considers only the data from the samples A and B, and the data from SPARC that correspond to the same galaxies (subsample I). The right plot considers the samples C, D and E, together with the data from SPARC that correspond to the same galaxies (subsample II). The red dotted line shows the BTFR law from \citet{2016ApJ...816L..14L}, the dashed black line is the BTFR result considering only the SPARC data that appears in each of the plots above. The  dot-dashed blue and the dashed orange lines show the best straight lines that describe the BTFR inferred from the Burkert or the NFW halo respectively. There is good  agreement between the SPARC data and the derived values of $\Upsilon_*$ in this work for the Samples A and B, while for Samples C, D and E the dispersion is too large to infer the BTFR from the fitted values of $\Upsilon_*$. } \label{fig:BTF}
\end{figure*}

In this work, the stellar mass-to-light ratios ($\Upsilon_*$) were all derived from best fits from RC data. In this appendix we compare the derived values with the expected ones, and evaluate the consequences of changes on $\Upsilon_*$ for the results on $\xi$ and related quantities.

In general,  by comparing best fits that consider different dark matter profiles and use  $\Upsilon_*$ as a free parameter, one is testing the total combination of dark matter and the stellar component(s). If the derived values of  $\Upsilon_*$ are systematically reasonable for one of the dark matter models, but not for the other, this alone would be an evidence in favour of the first model. In this case  there would be a tension between the values of $\Upsilon_*$ that this model favours and the values of $\Upsilon_*$ that are expected to be physically viable (from stellar population synthesis models, dynamical arguments, or scaling laws like the Baryonic Tully-Fisher relation). If both the dark matter models lead to reasonable values of $\Upsilon_*$, then the comparison between the best fits results of each of the models is a comparison between these models.

The stellar components of the samples A and B are determined from infrared observations (with 3.6 $\mu$m wave length for Sample A and I-band for the Sample B). These samples include most of the  massive and luminous large galaxies that are considered in this work. Besides estimating values of $\Upsilon_*$ from stellar population synthesis models, the corresponding references  agree that there is significant uncertainty on $\Upsilon_*$, in part due to uncertainties on the stellar initial mass function (IMF), leading to uncertainties on $\Upsilon_*$ of about a factor two. Hence, as one of their approaches, the  $\Upsilon_*$ values are derived from best fit procedures. \citet{2008AJ....136.2648D} show that for some galaxies  the expected value of $\Upsilon_*$ leads to a reasonable dynamical picture, and the fitted values of $\Upsilon_*$ also agree with the latter; but there are also examples of some galaxies that show  tensions between the expected and the fitted values. It was found that the NFW profile favours the Kroupa IMF, while other profiles may favour different IMF's. 

Based on results from stellar population synthesis models \citep{2014AJ....148...77M, 2014ApJ...788..144M, 2014PASA...31...36S} and, also, on the minimization of the baryonic Tully-Fisher relation (BTFR) dispersion \citep{2016ApJ...816L..14L}, \citet{2016AJ....152..157L} consider the simplifying hypothesis that\footnote{See, however, \citet[][]{2016A&A...585A..17A,2016A&A...593A..39P}.} $\Upsilon_* = 0.5$ for all the stellar discs at 3.6 $\mu$m. Although the use of $\Upsilon_* = 0.5$ is too restrictive to be true for all galaxies, at least it is a reasonable starting point to study general properties of galaxies. Therefore, we compare our results on the inferred $\Upsilon_*$ values with those of the SPARC sample \citep{2016AJ....152..157L}. 

Some of the galaxies that constitute the SPARC sample can also be found in the samples A and B, and we use these, together with the complete SPARC results on the BTFR, in order to check  our results on $\Upsilon_*$. We will call ``SPARC subsample I'' the collection of the latter SPARC galaxies. These comparisons are performed in Fig.~\ref{fig:BTF}. It can be seen that both the NFW and the Burkert fits lead to BTFRs that are very close to that found from SPARC. 

Writing $M_b$ for the baryonic mass and $V_f$ for the final circular velocity  the BTFR  has the form,
\begin{equation} \label{eq:btfrelation}
	\log_{10} M_b = a \log_{10} V_f  + b.
\end{equation}
To be clear, the baryonic mass $M_b$ is defined as the total mass of gas (hydrogen and helium) plus the mass from the stellar components of each galaxy.  $V_f$ is essentially the observed circular velocity that is farthest from the galaxy center, and this is the definition used to generate the plots in Fig.~\ref{fig:BTF} for the NFW and Burkert data. \citet{2016ApJ...816L..14L} use a more robust variation for the definition for $V_f$, which in the end leads to small changes that are not relevant to the purposes of this appendix. This difference on the $V_f$, together with small differences on the RC data itself, is the reason that the SPARC data that appear in Fig.~\ref{fig:BTF} is slightly displaced in the $V_f$ axis for some galaxies. 

The best fit values for $a$ and $b$ read,:
\begin{eqnarray*}
	& a=3.71 \mbox{, } b = 2.27 & \mbox{: full SPARC sample} \\
	& a=3.62 \mbox{, } b=2.43 &\mbox{: SPARC subsample I}\\
	& a=3.58 \mbox{, } b=2.50 &\mbox{: Burkert for Samples A and B}\\
	& a=4.11 \mbox{, } b=1.29 &\mbox{: NFW for Samples A and B}.\\
\end{eqnarray*}
Although differences can  promptly be seen in the numbers above, in the range $20 < V_f/\mbox{(km/s)} < 300$ the corresponding lines are very close (see the left plot in Fig.~\ref{fig:BTF}), with three of them being  almost indistinguishable.

The situation with the stellar components of the samples C, D and E is clearly different. These samples are dominated by dwarf and LSB galaxies. These galaxies have observed RCs and stellar components that allow for large variations on $\Upsilon_*$.\footnote{This claim is supported by \citet{Swaters:2011yq}, and in particular by \citet{2016AJ....152..157L}. According to the latter, for the large luminous galaxies, $\Upsilon_*=0.5$ at [3.6] leads to stellar RCs close to maximal, while for the LSB and dwarfs with that same value for $\Upsilon_*$  much lower relative stellar contributions are found, such that dark matter commonly dominates at 2.2 $h$, (i.e., at the maximum of the stellar disc contribution to the RC).} The right hand side plot in Fig.~\ref{fig:BTF} shows a large dispersion on $\Upsilon_*$ for a given value of $V_f$. By considering the error bars on $\Upsilon_*$ derived from the fits, which are not small for these galaxies, the compatibility with the BTFR dispersion is improved. 

For the case of the Samples C, D, and E, the best fit for the BTFR parameters is not particularly meaningful, and does not show a robust systematic deviation from the standard BTFR, since the corresponding error on the $a$ and $b$  parameters (see eq. \ref{eq:btfrelation}) is  large. The distribution of the data in the plane $M_b \times V_f$ is essentially the same for both of the models, hence the large dispersion on $\Upsilon_*$ does not introduce a bias in favour of any one of the models.

It should be verified whether the  large dispersion in $\Upsilon_*$ for the samples C, D, and E  has impact on the results relative to the quantity $\xi$. Considering the figures on the $\xi$ and $\zeta$ results, Figs.~\ref{fig:bluered1}, \ref{fig:bluered2}, the large dispersion on $\Upsilon_*$ could at most increase the dispersion on the results of $\xi$, but without any effect on $\zeta$, since $\zeta$ only depends on the observational RC data. The dispersion of the $\xi$ data does not show any clear systematic increase between samples A and B, and the  samples C, D, and E. The same happens for Figs. \ref{fig:yellow1}, \ref{fig:yellow2}, where the dispersion on the $\Delta\xi$ data is essentially the same along the samples for a given model. Moreover, although most of the galaxies belong to the samples C, D and E, when considering the subsamples that select the most massive or large galaxies, the relative importance of the samples A and B is increased. Thus, our main results that concern the largest galaxies are specially robust to this issue. 

It would be interesting to analyse the data from the SPARC sample using the new methods here proposed, and considering different hypothesis on $\Upsilon_*$, which we plan to do in a future work.


\bibliographystyle{mnras} 
\bibliography{bibdavi2016c}{} 

\bsp
\label{lastpage}
\end{document}